\documentclass[twocolumn]{aastex631}

\usepackage{xspace}
\usepackage[version=4]{mhchem} 
\graphicspath{{figures/}}

\usepackage{aas_macros}
\usepackage[T1]{fontenc}
\usepackage[inline]{enumitem}
\usepackage{makecell}

\newcommand{\poseidon}{\texttt{POSEIDON}\xspace}
\newcommand{\eureka}{\texttt{Eureka!}\xspace}
\usepackage[hang,flushmargin]{footmisc}

\newcommand\teq{T$_{\rm{eq}}$ }

\hypersetup{linkcolor=violet,citecolor=teal}

\newcommand{\Goddard}{NASA Goddard Space Flight Center, 8800 Greenbelt Road, Greenbelt, MD 20771, USA}
\newcommand{\Carnegie}{Earth and Planets Laboratory, Carnegie Science, 5241 Broad Branch Road, NW, Washington, DC 20015, USA}
\newcommand{\Zurich}{Department of Astrophysics, University of Z\"urich, Winterthurerstrasse 190, 8057 Z\"urich, Switzerland}
\newcommand{\PSUAA}{Department of Astronomy \& Astrophysics, The Pennsylvania State University, 525 Davey Laboratory, University Park, PA 16802, USA}
\newcommand{\PSUCEHW}{Center for Exoplanets and Habitable Worlds, The Pennsylvania State University, 525 Davey Laboratory, University Park, PA 16802, USA}
\newcommand{\APL}{Johns Hopkins APL, 11100 Johns Hopkins Rd, Laurel, MD 20723, USA}

\begin{document}

\title{GEMS JWST: Hold on to your HATS(-6 b), a sub-solar metallicity giant planet with water, methane and ammonia in its atmosphere}

\author[0000-0001-6340-8220]{Giannina Guzm\'an Caloca}
\affiliation{Department of Astronomy, University of Maryland, College Park, MD 20742, USA}
\affiliation{\Goddard}

\author[0000-0003-4835-0619]{Caleb I. Ca\~nas}
\altaffiliation{NASA Postdoctoral Program Fellow}
\affiliation{Southeastern Universities Research Association, Washington, DC 20005, USA}
\affiliation{\Goddard}

\author[0000-0003-0354-0187]{Nicole L. Wallack}
\affil{\Carnegie}

\author[0000-0002-2739-1465]{Erin M. May}
\affiliation{\APL}

\author[0000-0002-8163-4608]{Shang-Min Tsai}
\affiliation{Institute of Astronomy \& Astrophysics, Academia Sinica (ASIAA), Taipei 10617, Taiwan}

\author[0000-0002-8278-8377]{Simon M\"{u}ller}
\affiliation{\Zurich}

\author[0000-0001-5555-2652]{Ravit Helled}
\affiliation{\Zurich}

\author[0000-0001-8401-4300]{Shubham Kanodia}
\affiliation{\Carnegie}

\author[0000-0002-0746-1980]{Jacob Lustig-Yaeger}
\affiliation{\APL}

\author[0000-0001-8020-7121]{Knicole D. Col\'on}
\affil{\Goddard}

\author[0000-0002-1483-8811]{Ian Czekala} 
\affiliation{School of Physics \& Astronomy, University of St. Andrews, North Haugh, St. Andrews KY16 9SS, UK}

\author[0000-0003-1439-2781]{Megan Delamer}
\affiliation{\PSUAA}
\affiliation{\PSUCEHW}

\author[0000-0002-8518-9601]{Peter Gao}
\affiliation{\Carnegie}

\author[0000-0002-7127-7643]{Te Han}
\affil{Department of Physics \& Astronomy, The University of California, Irvine, Irvine, CA 92697, USA}

\author[0000-0002-2990-7613]{Jessica Libby-Roberts}
\affiliation{\PSUAA}
\affiliation{\PSUCEHW}
\affiliation{Department of Physics and Astronomy, University of Tampa, Tampa, FL 33606, USA}

\author[0000-0001-9596-7983]{Suvrath Mahadevan}
\affiliation{\PSUAA}
\affiliation{\PSUCEHW}

\author[0000-0002-4487-5533]{Anjali A. A. Piette}
\affiliation{School of Physics and Astronomy, University of Birmingham, Edgbaston, Birmingham B15 2TT, UK}

\author[0000-0001-7409-5688]{Gu\dh mundur Stef\'ansson}
\affil{Astrophysics \& Space Institute, Schmidt Sciences, New York, NY 10011, USA}
\affil{Anton Pannekoek Institute for Astronomy, University of Amsterdam, Science Park 904, 1098 XH Amsterdam, The Netherlands}

\author[0000-0002-7352-7941]{Kevin B. Stevenson}
\affiliation{\APL}

\correspondingauthor{Giannina Guzm\'an Caloca}
\email{gguzmanc@umd.edu}

\begin{abstract}

HATS-6 b is one of several recently discovered Giant Exoplanets orbiting M-dwarf Stars (GEMS) and is part of a JWST survey that aims to compare bulk and atmospheric properties of these rare planets against their FGK star counterparts. HATS-6 b is a warm ($\mathrm{T_{eq}}\sim700$ K), Saturn-mass ($M_p\sim0.3~\mathrm{M_J}$), Jupiter-radius ($R_p\sim1~\mathrm{R_J}$) planet that transits its star every $\sim$ 3 days. In this study, we present the transmission spectrum of HATS-6 b obtained with two transits using the PRISM mode of JWST Near Infrared Spectrograph (NIRSpec), spanning a wavelength range of $0.6-5.3$ \textmu{}m. Analyzing these JWST observations using an iterative approach between forward modeling and free chemistry retrievals, we derive a low metallicity ($\log\mathrm{[M/H]}=-1.99^{+0.2}_{-0.2}$) sub-solar C/O ($\log\mathrm{[C/O]=-0.46^{+0.2}_{-0.2}}$) atmosphere, and find strong evidence for H$_2$O, CH$_4$, and NH$_3$ at volume mixing ratios (in $\log[X]$) of $-4.88_{-0.24}^{+0.25}$, $-5.38_{-0.19}^{+0.18}$, and $-6.03_{-0.19}^{+0.18}$, respectively. We consistently retrieve a significantly lower $\mathrm{T_{eq}}$ than predicted from the orbital configuration of HATS-6 b, which was impervious to any data reduction and retrieval choices, suggesting a non-zero bond albedo. Our planetary interior models retrieve bulk metallicities three orders of magnitude larger than our retrieved atmospheric metallicity, also suggesting that the atmosphere is not well-mixed. We find an excess feature around 3 \textmu{}m, and expand on possible explanations for this, such as the presence of HCN or hydrocarbons like C$_2$H$_4$. Yet, due to the degeneracies present for hydrocarbon features in this wavelength region, we do not draw any conclusions about the excess feature and instead encourage further observations and follow-up of this intriguing target.

\end{abstract}


\section{Introduction} \label{sec:intro}

In our galaxy, M-dwarfs are the most common stars \citep{henry_solar_2006, winters_solar_2015, reyle_10_2021}. Despite their ubiquitous hosts, the population of \uline{G}iant \uline{E}xoplanets orbiting \uline{M}-dwarf \uline{S}tars ($R_p>8~\mathrm{R_\oplus}$, henceforth GEMS) is known to be rare \citep[e.g.,][]{2023MNRAS.521.3663B,2023AJ....165...17G,2025arXiv250908773G}. Under the theory of core accretion, M-dwarfs are predicted to rarely form such planets because of the low disk masses and long orbital timescales which impede the efficient formation of massive planetary cores ($\sim$ 10 M$_\oplus$) required to activate runaway gas accretion \citep[e.g.,][]{Mizuno1980,Stevenson1982,Pollack1996,Ikoma2000,laughlin_core_2004, burn_new_2021}. Furthermore, there is observational evidence through radial velocity and transit surveys that the occurrence rates of giant planets decrease with decreasing stellar mass \citep[e.g.,][]{endl_exploring_2006, johnson_new_2007,Bowler2010,Johnson2010,gaidos_understanding_2013, bonfils_2013, montet_trends_2014, sabotta_2021,Beleznay2022}. 

Recent discoveries of short period ($P<10$ days) GEMS form part of a growing population that challenges our current theories of formation \citep[e.g.,][]{kanodia_toi-5205b_2023, delamer_toi-4201_2024,Bryant2025,Kanodia2025,angeli_2025,OBrien2025}. Some of these short period GEMS represent an extreme of planetary formation due to their high planet-to-star mass ratios, which would require more dust mass to be present in the protoplanetary disk than typically estimated for M-dwarfs \citep[e.g.,][]{morales_giant_2019, quirrenbach_carmenes_2022,Hotnisky2025}. Although conclusions from planetary formation studies are highly dependent on the models used and assumptions made \citep[e.g.,][]{movshovitz_2010, venturini_2020, kessler_2023}, and despite the discovery of close-in giant planets around FGK stars inspiring new theories of planetary formation and migration \citep[e.g.,][]{dawson_origins_2018, fortney_hot_2021}, GEMS still continue to pose a challenge for formation theories, especially during the protoplanetary disk phase. Furthermore, in contrast to their well-studied hot Jupiter counterparts, GEMS have equilibrium temperatures $<$ 1000 K due to the lower masses and temperatures of their M-dwarf hosts and may not display inflated atmospheres as a result \citep{hot_jupiter_inflation}. With temperatures $<$ 1000 K, the dominant carbon-bearing species transitions from CO to CH$_4$ \citep{Fortney2020}. Additionally for these cooler temperatures, nitrogen species, such as NH$_3$, are expected at higher abundances than their hot Jupiter counterparts \citep{nitrogen_bulk}. For this reason, we expect the atmospheres of GEMS to be distinct from the canonical hot Jupiter population.

Studying the atmospheres of GEMS and comparing the results to those of giant planets orbiting FGK stars, including those in our own Solar System, will provide invaluable insight into this unique population of planets. \cite{kanodia_searching_2024} detailed a survey design for the discovery of new GEMS and predicted the number of GEMS needed to draw sample-level statistical conclusions regarding their bulk properties. To complement this, \cite{canas_gems_2025} highlighted the GEMS JWST survey design and motivation for the Cycle 2 JWST program GO 3171 \citep[][]{GEMS_proposal}. In this work, we present an in-depth atmospheric characterization of HATS-6 b, the second target observed as part of the GEMS JWST survey. HATS-6 b is a short-period ($P\sim$ 3 days), warm ($\mathrm{T_{eq}}\sim$ 700 K), Saturn-mass and Jupiter-radius planet ($M_p\sim0.3~\mathrm{M_J}$, $R_p\sim1~\mathrm{R_J}$, $\rho_p\sim0.39~\mathrm{g~cm^{-3}}$) orbiting a $\sim$ 0.6~M$_\odot$ M1V star \citep{Hartman2015}. 

The system parameters for HATS-6 b are summarized in detail in \autoref{tab:sysparams}. Section \S\ref{sec:observation} details our observations;   \S\ref{sec:datareduction} describes the data reductions, including an investigation of spot crossing events (\S\ref{subsection:spots}). We characterize the atmosphere of HATS-6 b utilizing both forward models, shown in \S\ref{sec:forwardmodel} and Bayesian retrievals described in \S\ref{sec:retrievals}. We then present a detailed discussion on the results of our characterization studies in \S\ref{sec:discussion}. HATS-6 b has also been the focus of recent theoretical global circulation model (GCM) studies \citep{kiefer_why_2024, kiefer_under_2024}, and we discuss our results in relation to these simulations.


\section{Observations}\label{sec:observation}
On 2024 January 15 and January 21 UT, we observed two transits of HATS-6 b (GO 3171; observation numbers 6 and 7) using the JWST NIRSpec PRISM in Bright Object Time Series (BOTS) mode \citep{Birkmann2022,Espinoza2023}. The NRSRAPID readout pattern and the SUB32 array were utilized to observe the host star ($J=12.05$) for target acquisition. All science observations were performed using the NRS1 detector with the SUB512 subarray ($32\times512$ pixels) and consisted of 17850 integrations with 4 groups per integration. Exposures lasted $\sim$5.6 hrs, providing a median observational cadence of 1.13 s. No observation saturated, as they remained below the conservative threshold for non-linearity effects used in \citet{carter_benchmark}. The transit duration was $\sim$ 2 hours with a $\sim$1.5 hour pre-transit baseline and a 1 hour post-transit baseline for each observation. We independently reduced the observations using \texttt{ExoTiC-JEDI} \citep{Alderson2022} and \texttt{Eureka!} \citep{Bell2022} and describe the reduction for each pipeline in \S\ref{sec:datareduction}.

\startlongtable
\begin{deluxetable*}{lhcccccc}
\tabletypesize{\tiny}
\tablecaption{System parameters for HATS-6 b \label{tab:sysparams}}
\tablehead{\colhead{Name} &
\nocolhead{placeholder} &
\colhead{Units} &
\colhead{Prior$^a$} &
\colhead{Joint Spot Fit} &
\colhead{Zero Spot Fit} &
\colhead{Source} &
}
\startdata
\sidehead{Stellar parameters:}
~~~Stellar Mass ($M_\star$) & & $\mathrm{M_\odot}$ & \nodata & $0.59\pm0.03$ & \nodata & 1\\
~~~Stellar Radius ($R_\star$) & & $\mathrm{R_\odot}$ & \nodata & $0.60\pm0.03$ & \nodata & 1\\
~~~Effective Temperature ($T_{\mathrm{eff}}$) & & K & \nodata & $3791\pm94$ & \nodata & 1\\
~~~Surface Gravity ($\log g_\star$) & & dex & \nodata & $4.65\pm0.05$ & \nodata & 1\\
~~~Metallicity ($\mathrm{[Fe/H]}$) & & dex & \nodata & $0.36\pm0.10$ & \nodata & 1\\
\hline
\sidehead{Fitted transit parameters:}
~~~Period ($P$) & & day & $\mathcal{N}(3.325272,0.1)$ & $3.325271 \pm 0.000005$ &  $3.325264 \pm 0.000003$  & This work\\
~~~Time of mid-transit ($T_0$) & & $\mathrm{BJD_{TDB}}$ & $\mathcal{N}(2460331.460000,0.1)$ & $2460331.4608961740 \pm 0.0000094$ & $2460331.4608793380 \pm 0.0000043$ & This work\\
~~~Eccentricity ($e$) & & \nodata & Fixed & 0 & 0 & This work\\
~~~Argument of periastron ($\omega$) & & deg & Fixed & 90 & 90 & This work\\
~~~Scaled radius ($R_p/R_\star$) & & \nodata & $\mathcal{U}(0,1)$ & $0.1849_{-0.0002}^{+0.0002}$ & $0.1823_{-0.0006}^{+0.0006}$ & This work\\
~~~Scaled semi-major axis ($a/R_\star$) & & \nodata & $\mathcal{N}(13.65,1.0)$ & $13.725_{-0.008}^{+0.008}$ & $13.623_{-0.022}^{+0.026}$& This work\\
~~~Impact parameter ($b$) & & \nodata & $\mathcal{U}(0,1)$ & $0.402 \pm 0.001$ & $0.443 \pm 0.003$ & This work\\
\hline
\sidehead{Derived system parameters:}
~~~Inclination ($i$) & & deg & \nodata & $88.202_{-5.38}^{+5.38}$ & $88.136_{-5.98}^{+4.76}$ & This work\\
~~~Semi-major axis ($a$) & & au & \nodata & $0.036 \pm 0.01$ & $0.038 \pm 0.01$ & This work\\
\hline
\sidehead{Planetary Parameters:}
~~~Planet Radius ($R_p$) & & $\mathrm{R_J}$ & \nodata & $1.08 \pm 0.06$ & \nodata & This work\\
~~~Planet Mass ($M_p$) & & $\mathrm{M_\oplus}$ & \nodata & $101\pm22$ & \nodata & 1\\
~~~Planet \teq (K) & & $\mathrm{M_\oplus}$ & \nodata & $713\pm5$ & \nodata  & 1\\
\enddata
\tablecomments{System parameters were derived from the white light curve fits with \texttt{juliet} \citep{espinoza_juliet_2019} for our main \textbf{Eureka1} reduction. We adopted the orbital parameter values from our joint spot fit due to reduced RMS in the white light curves.} 
\tablenotetext{a}{Normal priors with a mean of $X$ and standard deviation of $Y$ are denoted as $\mathcal{N}(X,Y)$. \\ Uniform priors between a lower limit of $X$ and upper limit of $Y$ are denoted as $\mathcal{U}(X,Y)$.}
\tablerefs{1) \cite{tianjun_mdwarf}}
\end{deluxetable*}


\section{Independent Reductions}\label{sec:datareduction}

The uncalibrated data files (\texttt{*uncal.fits}) were downloaded from the Barbara A. Mikulski Archive for Space Telescopes (MAST).\footnote{\url{https://mast.stsci.edu/portal/Mashup/Clients/Mast/Portal.html}} The download consisted of each visit exposure broken down into 4 segment files. The raw files were processed using two independent pipelines described in the following sections.

\subsection{ExoTiC-JEDI}\label{subsection:exotic}
For our first independent reduction, we used the Exoplanet Timeseries Characterisation - JWST Extraction and Diagnostics Investigator pipeline (\texttt{ExoTiC-JEDI})\footnote{\url{https://github.com/Exo-TiC/ExoTiC-JEDI}} \citep{Alderson2022}. The first two stages of the pipeline are wrappers to the \texttt{jwst} pipeline (v1.11.4 with CRDS v11.17.25, mask file \texttt{jwst\_nirspec\_mask\_0074.fits}, and CRDS context \texttt{jwst\_1293.pmap}) with minimal adjustments to the default settings. For Stage 1, we skipped both the \texttt{superbias} and \texttt{jump} steps. The latter, designed to flag outliers due to anomalies such as cosmic rays, has proven ineffective for observations with a small number of groups \citep[e.g.,][]{Rustamkulov2023}. We also skipped the \texttt{ExoTiC-JEDI} \texttt{custom\_bias} step. This step creates a custom pseudo-bias image using the median of the first group and was designed to reduce systematics in the data between the NRS1 and NRS2 detectors, yet NIRSpec PRISM only uses the NRS1 detector. Furthermore, we identified a trace-adjacent hot pixel in column 125, which introduced correlated noise to the extracted light curve for that column. First, we applied a ``DO\_NOT\_USE'' mask to the hot pixel. We then attempted to replace the hot pixel with the median value of the surrounding pixels. Despite these attempts to correct the hot pixel, the correlated noise persisted. In the end, we chose to mask that column entirely during the spectral fitting stage. 

Prior to ramp fitting in Stage 1, ExoTiC-JEDI includes a custom ``de-stripping'' stage, where the median background value is subtracted column-by-column at the group level. In order to subtract the background without removing any of the astrophysical signal, pixels within 10 times the Full-Width-at-Half-Maximum (FWHM) from the trace center were masked. This custom step removes the background and helps combat correlated noise at the detector readout level \citep[such as ``$1/f$ noise'' ;][]{Birkmann2022, Alderson2022, Rustamkulov2023}. For Stage 2, we ran the following steps: (1) \texttt{assign\_wcs}, (2) \texttt{extract\_2d}, (3) \texttt{srctype}, and (4) \texttt{wavecorr} directly from the \texttt{jwst} pipeline. For more detailed information about these pipeline steps, see the JWST documentation \citep{2016jdox.rept......}. For the \texttt{extract\_2d} step, we modified the \texttt{jwst} default values for the wavelength range and slit height using the \texttt{assign\_wcs} function. We updated these values because the default values from \texttt{jwst} were truncating the detector sub-array and introducing wavelength offsets between our reductions. Before extraction in Stage 3, we used data-quality flags to mark saturated, dead, bad, low quantum efficiency, and no gain pixels. We replaced said pixels with the median value of the 5 pixels surrounding it. We also replaced outliers both spatially and temporally. Spatially, we applied a rejection threshold of 5$\sigma$ deviation from the median using the median of the surrounding 20 columns for pixel replacement. Temporally, we applied a 10$\sigma$ rejection threshold using the median of that pixel in the surrounding 10 integrations (one integration $\sim$ 1 s). 

For spectral extraction, we used a box aperture that extended to 6 times the FWHM ($\sim$0.7 pixels) on each side of the trace for a total aperture of 8.4 pixels. We also set a buffer of 4 pixels from the upper and lower aperture edges for further 1/f and background noise correction. ExoTiC-JEDI corrects for this noise by subtracting column-by-column the median of the unilluminated region. In order to extract the final 1D light curves, ExoTiC-JEDI follows the steps for optimal spectral extraction outlined by \cite{Horne1986}. The pipeline also realigns any positional shifts determined by cross-correlation, and removes any outliers utilizing a 4$\sigma$ rejection value. We extracted light curves at this stage both at the pixel-level resolution (470 spectroscopic channels) and spatially binned to a predetermined wavelength grid of 40 nm bins for a total of 120 spectroscopic channels between $0.6-5.3$ \textmu{}m.

\subsection{Eureka!}\label{subsection:eureka}
\eureka{} \citep{Bell2022} (v1.0) is an end-to-end data reduction pipeline for JWST time series observations. It allows for the processing of level 0 \texttt{uncal} data files, which are the raw uncalibrated data products from single exposures. The first two stages of \eureka{} serve primarily as a wrapper for the \texttt{jwst} \citep{bushouse2023} pipeline. In Stage 1, we used a custom group level background subtraction step (GLBS) that accounted for background and striping due to 1/f noise. For each group, we subtracted the median of the top and bottom 7 pixels in each column, after removing any outliers using a rejection threshold of 3$\sigma$. \eureka's Stage 1 also includes a custom reference pixel correction step to account for the lack of true reference pixels in NIRSpec PRISM's 512 subarray. This step used a Row-by-row, Odd-Even By Amplifier (ROEBA) correction \citep[e.g., as used by the \texttt{tshirt} pipeline in][]{Rustamkulov2023}. Using the top/bottom 8 rows as our reference rows, we subtracted the median of the even reference rows from all even rows, and the median of the odd reference rows from all odd rows. For Stage 2 of \eureka, we made a few modifications to the default steps in the \texttt{jwst} pipeline. We did not run the flat field correction step nor the photometric calibration step (\texttt{photom}), used to convert to absolute flux units, as these add additional noise due to the uncertainty of the flux calibration. This is a standard method for exoplanet observations, where we are only concerned with relative measurements \citep[i.e.][]{Rustamkulov2023, erin_NIRSpec_skip}.

Stage 3 of \eureka{} extracts the 1D stellar spectra. We first performed a second round of background subtraction, using a background exclusion region of 9 pixels, such that all pixels more than 9 away from the center of the spectrum are included in the background calculation. We again subtracted the median of each column, with an outlier threshold of 5 $\sigma$. The spectral extraction used the optimal extraction algorithm \citep{Horne1986} with an aperture half width of 3 pixels on either side of the central pixel (e.g., 7 total pixels). We rejected outliers more than 10$\sigma$ away from the median when constructing the median frame for optimal extraction, and outliers more than 10$\sigma$ away from the median frame when performing optimal extraction.

Stage 4 of \eureka{} assembles the white-light curve and spectroscopic light curves. The white-light curve was summed from data spanning $0.6-5.3$ \textmu{}m and corrected for outliers using a 4 $\sigma$ threshold on the rolling median with a 10 integration box-car width. We iterated through 5 rounds of outlier removal. The spectroscopic light curves were also binned between $0.6 - 5.3$ \textmu{}m in 40 nm bin widths (4 pixel bins), resulting in 120 spectroscopic light curves. The same outlier rejection thresholds were applied. 

\subsection{Choosing our final reduction}\label{subsection:final_reduction}

We performed two independent \eureka{} reductions, with two different \texttt{jwst} pipeline and \texttt{CRDS} configurations to test whether any notable differences arose from using independent configurations on the same exact pipeline. The second \eureka{} reduction had the following differences: (i) Stage 1 employed a GLBS using the top/bottom 8 rows, (ii) Stage 3 adopted an aperture half width of 4 pixels with a background exclusion region of 7 pixels, an outlier rejection threshold of $7\sigma$ when generating the median, and a rejection threshold of $7\sigma$ when computing the spatial profile and for optimal extraction, (iii) Stage 4 calculated the rolling median using a width of 15 pixels and a rejection threshold of $3\sigma$, and (iv) For Stage 4, we generated limb darkening coefficients with the \texttt{ExoTiC-LD} \citep{Grant2022, Grant2024} package for a quadratic limb darkening law using the MPS2 stellar grid \citep{Kostogryz2022} on the same wavelength grid as the extracted light curves. For the MPS2 grid limb darkening calculations we adopted stellar parameters from \cite{Hartman2015}. 

Aside from these configuration choices, the methods described above remained the same for both reductions. The first reduction (henceforth known as \textbf{Eureka1}) was reduced at the binned (120 channels) and pixel-level resolution (470 channels), whilst the second reduction (henceforth \textbf{Eureka2}) was only reduced at the binned resolution. 

We adopted the \textbf{``Eureka1''} (\texttt{jwst} v1.18.0 and CRDS v11.17.21) reduction at the pixel-level resolution as our main reduction for analysis since it uses a newer version of the JWST pipeline, and a limb darkening prescription that did not rely on pre-computed grids (see \autoref{appendix:limbdark}). The \eureka{} binned-then-fit reduction with a grid limb darkening prescription will henceforth be known as \textbf{``Eureka2''} (\texttt{jwst} v1.9.4 and CRDS v11.16.19), and the ExoTiC-JEDI reduction described in \S\ref{subsection:exotic} will be known as \textbf{``ExoticJEDI''} (\texttt{jwst} v1.11.4 and CRDS v11.17.25). Finally, we adopt the \textbf{Eureka1} reduction over \textbf{ExoticJEDI} due to the lower scatter in the white-light curve, which had a Mean Absolute Deviation (MAD) of $\sim$ 490 ppm for \textbf{Eureka1} vs 650 ppm for \textbf{ExoticJEDI}. We attributed this difference in MAD to the larger aperture used for the \textbf{ExoticJEDI} reductions. The differences between each pipeline are detailed in \autoref{fig:data-reduction-differences}.


\begin{deluxetable*}{lccc}
\tabletypesize{\small}
\tablecaption{Summary of reduction choices for HATS-6 JWST observations. \label{fig:data-reduction-differences}}
\tablehead{
&
\colhead{\textbf{Eureka1}}&
\colhead{\textbf{Eureka2}}&
\colhead{\textbf{ExoticJEDI}}
}
\startdata
~~\texttt{jwst} version & \textbf{1.18.0} & \textbf{1.9.4} & \textbf{1.11.4} \\
~~CRDS version & 11.17.21 & 11.16.19 & 11.17.25 \\
~~Limb darkening prescription & Kipping 2013 & MPS2 Stellar Grid & Kipping 2013 \\
~~Limb darkening priors adopted & Uniform & Gaussian & Uniform \\
~~Systematic prescription & Quadratic polynomial & Linear ramp & Quadratic polynomial \\
~~Sampler for light curve fits & \texttt{dynesty} & \texttt{emcee} & \texttt{dynesty} \\
~~Spot-fitting routine & \texttt{spotrod} & \texttt{fleck} & \texttt{spotrod} \\
~~Binning routine & pixel-level and binned extraction & binned extraction & pixel-level and binned extraction
\enddata
\tablerefs{CRDS: \cite{GREENFIELD201641}, \texttt{dynesty}: \cite{Speagle2020}, \texttt{emcee}: \cite{emcee2013}, \texttt{fleck}: \cite{Morris2022}, \texttt{jwst}: \cite{bushouse2023}, Kipping 2013: \cite{kipping_efficient_2013}, MPS2 Stellar Grid: \cite{Kostogryz2022}, \texttt{spotrod}: \cite{spotrod2014}}
\tablecomments{We summarize the biggest differences in our three pipeline reductions. Due to lower scatter in the white light curve, we adopt \textbf{Eureka1} as our final reduction. We note that for our \textbf{Eureka1} reductions, joint white-light curve fitting was performed with \texttt{juliet}, while spectral fitting was performed with \eureka{} due to available computational resources.}
\end{deluxetable*}

\subsection{Light Curve Fitting}\label{subsec:lc-fit}

\subsubsection{Investigating Spot Crossings}\label{subsection:spots}

During our initial white-light curve fitting, we noted some residual structure in both transits. Due to the high cadence nature of the data, some visible bumps, albeit small, were seen exclusively in the transiting part of the white light curve residuals. This structure was present across reductions and independent of noise and outlier treatment as well as aperture choices. Since this could be a potential indication of spot crossing events, we followed the same investigation steps as in \cite{canas_gems_2025}. Spot crossings were investigated using two different code suites: \texttt{spotrod} \citep{spotrod2014} and \texttt{fleck} \citep{Morris2022}. 

One main difference between the aforementioned spot modeling codes is that \texttt{fleck} does not wrap spots on the limb, making it suboptimal for fitting spots that sit close to the stellar limb. \cite{2026ApJ..1003L..45M} describe some of the limitations when modeling large spots with \texttt{fleck} and note that there may be bright artifacts in the transit model if an ``overhang'' in a limb spot intersects the transit chord. Furthermore, our implementation of \texttt{fleck} sampled the position along a 3D sphere using latitude and longitude, making sampling inefficient without tight informed priors on the spot positions. In contrast, the \texttt{spotrod} input requires a spot center using a projected planetary coordinate system. This allowed for sampling the center from a unit disk with uninformative uniform priors. By doing this, we do introduce degeneracies in spot position along the vertical axis (i.e. mirror images produce similar likelihoods), yet it is not expected that spot position will affect the spectral analysis as much as other spot parameters, such as spot contrast. The computational efficiency of our implementation of \texttt{spotrod} also allowed us to (i) simultaneously model both visits and (ii) explore the possibility of a greater number of spots than we could with \texttt{fleck}.

For the \texttt{fleck} implementation in our \textbf{Eureka2} reduction, we integrated \texttt{fleck} into the \eureka{} light curve fitting module.\footnote{Spot modeling with \texttt{fleck} is now a default functionality of \texttt{Eureka!}.} For both visits, we considered between 0 and 3 spots, inclusively. By comparing the Bayesian Information Criterion \citep[BIC;][]{Schwarz1978} and Akaike Information Criterion \citep[AIC;][]{Akaike1974}, for visit 1 we found a statistical preference for 1 spot, while for visit 2 we found a statistical preference for 0 spots. 

For the \texttt{spotrod} implementation on our \textbf{Eureka1} reduction, we modeled from 0 to 5 spots\footnote{Our \texttt{spotrod} implementation and sampling method allowed us to explore more spots in each visit when compared to \texttt{fleck} because the convergence time with \texttt{spotrod} was quick and thus compatible with the available computational resources.} for each visit, inclusively. We did so by integrating \texttt{spotrod} into the light curve fitting software \texttt{juliet} \citep{espinoza_juliet_2019}. Both the AIC and BIC revealed a statistical preference for 1 spot in visit 1 and 3 spots in visit 2. The next best-fitting model favored 0 spots for both visits for both criteria, which would make our second visit results consistent with \texttt{fleck} results. The best-fitting models there had $\Delta$BIC $>10$ and $\Delta$AIC > $16$ that strongly favored a spot fit over our non-spotted fit. Our non-spotted fit presented higher Bayesian evidence (a Bayes factor of ln B$_{0,13}$ = 5) overall. Formally, the model with the highest Bayesian evidence had 5 spots in visit 1 and 1 spot in visit 2, a configuration heavily disfavored by the AIC and BIC ($\Delta\mathrm{AIC}_{51,13}\sim28$, $\Delta\mathrm{BIC}_{51,13}\sim61$). We note that the Bayesian evidence relies on the fully marginalized likelihood (e.g., integrated over parameter space) while the AIC and BIC are approximations of the evidence that penalize models \citep{Akaike1974,Schwarz1978} based on complexity (in this case, number of spots). Here, we adopt the best fit based on the AIC/BIC metrics given the degeneracy in the spot parameters \citep[e.g.,][]{spotrod2014}. 

\begin{figure*}
\epsscale{0.8}
\plotone{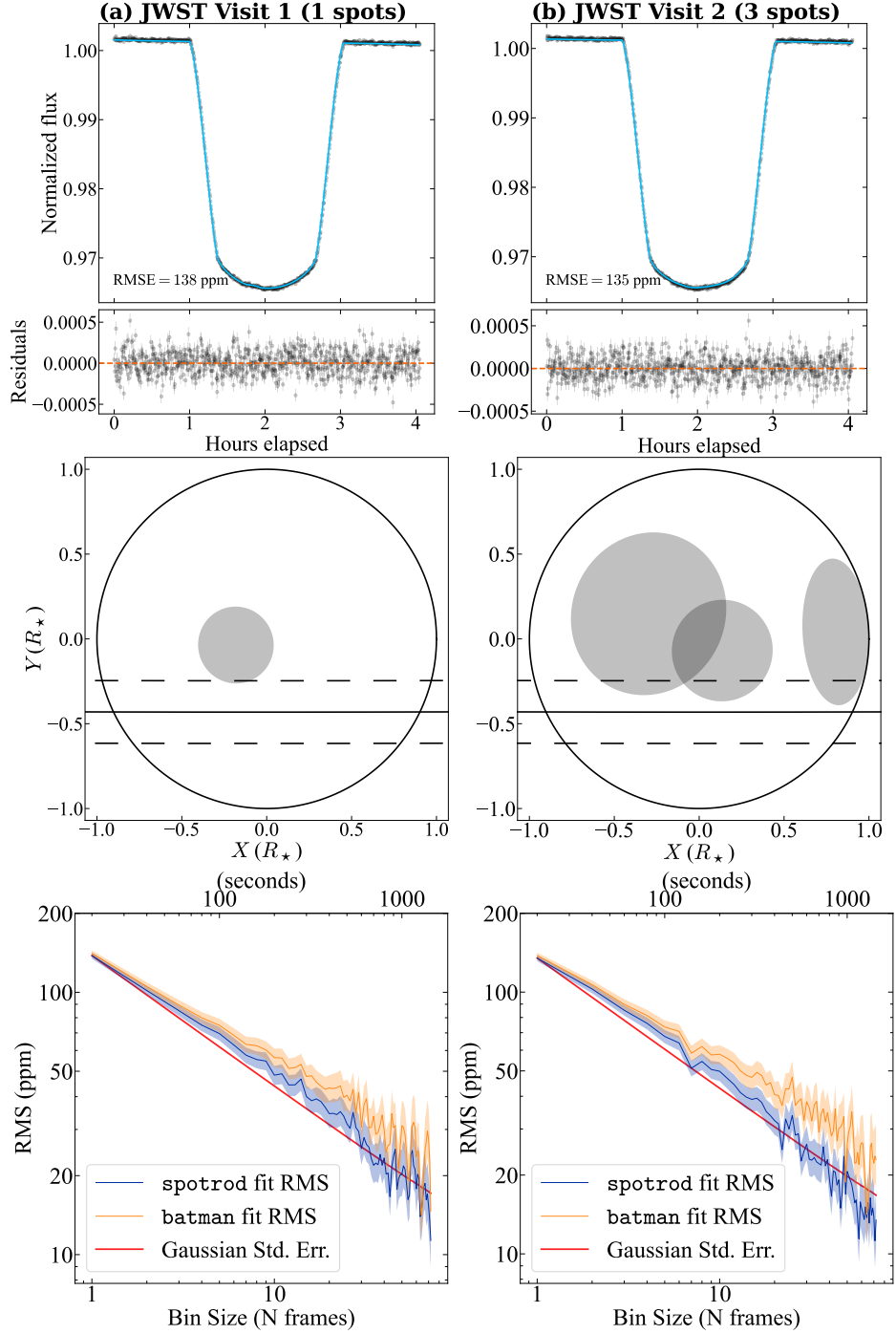}
\caption{Panels (a) and (b) show the JWST NIRSpec PRISM white light curves produced using \texttt{Eureka1} after binning to a cadence of 20 s. \textbf{Top row}: The data along with the best-fitting model (solid line) with the residuals below. \textbf{Middle row}: The stellar surface for the adopted spot configuration. The solid line is the position of the transit chord while dashed lines reflect the planet radius. The shading of the spots is arbitrary and is not meant to convey the spot flux ratio. \textbf{Bottom row}: The RMS for each visit for the \texttt{spotrod} model (blue) and spot-free model (orange) data. Gaussian white noise is shown as a red solid line. For our final reduction, we show that fitting for spots spectroscopically does not produce any significant differences in the final transmission spectra (see \autoref{appendix:spot_chrome}). Because the wavelength-dependent transit depths generally demonstrated no sensitivity to spot crossings, we proceed with a hybrid approach in which we hold the orbital parameters constant from the spotted fit, but do not fit for the spots spectroscopically. This decision was made in order to avoid any overfitting due to higher model complexity and any degeneracies with limb darkening due to the spot on the stellar limb.}
\label{fig:wlc}
\end{figure*}

We attributed the differences in our results from \texttt{fleck} and 
\texttt{spotrod} to differences in sampling and spot limb wrapping between the codes. Furthermore, differences in the limb darkening treatment of each reduction could also contribute to degeneracies in the spot-fitting results. In \S\ref{subsec:spectralfit}, we summarize the effects that our selected spot configurations have on the finalized transmission spectra. Given the limitations on sampling and spot wrapping around the stellar limbs for \texttt{fleck}, we move forward with our \texttt{spotrod} results for our analysis. \autoref{fig:wlc} shows our best-fit spot configurations along with the white light curves.

The retrieved spots have radii that are larger than the planet ($r_p/R_\star\sim0.18$ compared with $0.2\lesssim~r_{\mathrm{spot}}/R_\star\lesssim0.3$) but appear to be consistent with the size of spots observed on M dwarf using JWST \citep[e.g.,][]{canas_gems_2025,2026AJ....171..294A,2026ApJ..1003L..45M}. We adopt the best fitting spot configuration but note that there exists a degeneracy between the spot radius and spot contrast such that smaller spots can also provide similar light curves \citep[e.g., a larger contrast for smaller spots;][]{2010A&A...510A..25S,2026AJ....171..344M}. We also note that our spot modeling with both \texttt{fleck} and \texttt{spotrod} assume a simplified model in which spots are circles with uniform limb darkening, ignoring the effects of complex spot structures and differential limb-darkening that will be important for large spots \citep[e.g.,][]{2012MNRAS.427.2487K,Morris2022,2026A&A...706A.281S}. 

\subsubsection{White-light Curve Fitting}\label{subsec:whitelight}

For the binned-then-fit \textbf{Eureka2} reductions, we tested both a linear and a quadratic temporal ramp. We also tested two different quadratic limb darkening prescriptions: fixed pre-calculated values from MPS2, and Gaussian priors centered at those same pre-calculated values that allow limb darkening parameters to float. We further explored a preference for linear temporal ramps and fixed limb darkening. We fit visit 1 and visit 2 independently, letting R$_{p}$/R$_{s}$, a/R$_{s}$, inclination, and mid-transit time be free parameters, in addition to the spot contrast, radius, latitude, and longitude for visit 1 using \texttt{fleck}. All fitting for \texttt{Eureka2} is done using \texttt{emcee} \citep{emcee2013}. 

For the pixel-level (\textbf{Eureka1} and \textbf{ExoticJEDI}) reductions, we excised the transit baseline and fit a polynomial to it using \texttt{scipy.curve\_fit} \citep{Virtanen2020}, which favored a quadratic temporal ramp. We chose the Kipping 2013 \citep{kipping_efficient_2013} prescription for quadratic limb darkening coefficients, sampling the limb darkening parameters using a wide uninformative prior. We fit visit 1 and visit 2 independently with R$_{p}$/R$_{s}$, a/R$_{s}$, inclination, and mid-transit time as free parameters, in addition to the spot contrast, radius, and spot position as free parameters for our spot fits, done with \texttt{spotrod} for \textbf{Eureka1} and \textbf{ExoticJEDI}. The posteriors for all fits to these reductions were sampled with \texttt{dynesty} \citep{Speagle2020} using dynamic nested sampling with a convergence criterion of $\Delta \ln Z=0.01$. \autoref{tab:sysparams} summarizes our mean best-fit orbital parameters.

\subsubsection{Spectral Light Curve Fitting}\label{subsec:spectralfit}

For the spectroscopic light curve fitting, we fixed the orbital parameters to the best-fit values for each reduction. In the \textbf{Eureka2} reductions, we gave the quadratic limb darkening coefficients tight Gaussian priors centered on the values calculated from the MPS2 stellar grid \citep{Kostogryz2022} (as detailed in \S\ref{subsection:eureka}). In contrast, for the \textbf{Eureka1} and \textbf{ExoticJEDI} reductions, we allowed the limb darkening coefficients to float along wide uniform priors using the \citet{kipping_efficient_2013} prescription. Limb darkening prescriptions have been extensively studied for the effect they have on extracted abundances \citep[e.g.,][]{coulombe_biases_2024, keers_reliable_2024}. Therefore, we adopted different limb darkening prescriptions for one of our reductions to observe the impact on the resulting spectra. We also modeled the transit spectra with different limb darkening model grids and found that the spectra were most affected by these differences blueward of 3 \textmu{}m. We also note that none of the stellar grids were able to fully capture the complexities present in our spectra (see \autoref{appendix:limbdark}), and therefore we proceeded with the Kipping 2013 \citep{kipping_efficient_2013} free limb darkening prescription.

We also adopted another difference in our spectroscopic light-curve fitting by defining two spectral resolutions: pixel-level and binned. ``Pixel-level'' refers to the entire native 470 channel resolution, while ``binned'' refers to a wavelength grid with 40nm bin widths (4 pixel bins), resulting in 120 spectroscopic channels. We used both of these resolutions during our reductions. The two binning strategies we tested were the following: ``bin-then-fit'' v.s. ``fit-then-bin''. For ``bin-then-fit'' we binned the data down to 40nm bin widths before fitting the spectral channels, resulting in 120 spectroscopic channels. For ``fit-then-bin'' we fit the spectroscopic light curves at native resolution, resulting in 470 spectroscopic channels that can then be binned down to any desired wavelength grid. 

Binning before spectroscopic fitting leads to less computationally intensive fitting, but it may also hide important outliers that bias the resulting spectra. Fitting the light curves in native resolution first may also lead to smaller uncertainties in the extracted spectra \citep{coulombe_biases_2024}. Testing these approaches with the same reduction (\textbf{ExoticJEDI}) we find that the ``fit-then-bin'' v.s. ``bin-then-fit'' approaches led to some differences in the resulting spectra, especially redder than 4 \textmu{}m and when a light curve with high correlated noise was not masked in the binning stage (\autoref{appendix:bin}). We believe the differences in the redder end may be due to the bin-then-fit approach absorbing some of the high scatter that happens at those wavelengths in the pixel-level spectrum. Regardless, the resulting spectra remained within 2 $\sigma$ of each other and therefore shouldn't produce significant differences in the general conclusions. Therefore, we caution readers to employ the ``fit-then-bin'' approach wherever possible, especially if differences in visits are expected. In order to avoid any biases from outliers and maximize the amount of information we can extract from the spectra, we proceed with a pixel-level resolution spectral fit as our final reduction, and utilize this dataset for our final analysis in \S\ref{subsec:retrieval_results}.

We also investigated the chromatic effect of spot crossings with \texttt{spotrod} by performing spectral fits for our best-fit spot configurations (see \S\ref{subsection:spots}). We do this by holding all spot parameters constant except the spot contrast. Despite our white light curves highly favoring a spot crossing configuration, we do not find notable differences in the resulting spectra between our non-spotted and spotted configurations (see \autoref{appendix:spot_chrome}). The same spots that were evident in the white light curve were not resolved spectroscopically with confidence. Because the wavelength-dependent transit depths generally demonstrated no sensitivity to spot crossings, we adopted the simplest model without spots. This was done in order to avoid any overfitting due to higher model complexity and any degeneracies with limb darkening since one of our spots is in the limb. 

Finally, we investigated the possibility of limb asymmetries in our data given recent evidence of possible limb asymmetries in other gas giants \citep{macdonald_why_2020, murphy_evidence_2024}. We used both \texttt{catwoman} \citep{jones_catwoman_2020, espinoza_constraining_2021} and \texttt{harmonica} \citep{grant2022transmission} for our analysis. \texttt{catwoman} models limb asymmetries as two semi-circles of different sizes, while \texttt{harmonica} models the effective radius as a function of angle from the planet's orbital velocity vector (called a transmission string). First, we fit the white-light curve with both, allowing the semi-circle radius values to float in the case of \texttt{catwoman} and the first-order term (a$_1$) to float in the case of \texttt{harmonica} in order to allow for equatorial bulging. In both cases, there was some slight evidence of bulging in the evening limb of the planet, yet when performing spectroscopic fits with both \texttt{catwoman} and \texttt{harmonica}, we found a high negative correlation between the resulting limb spectra (see \autoref{appendix:limb_diff}) for all models and visits. Since our limbs are highly correlated, any features that arise in the limb spectra are not independent from each other, and often a perfect mirror of each other. This artifact makes any result from our limb asymmetry models an unreliable measure of anything physical happening in the atmosphere of HATS-6 b. Therefore, we once again proceed with our simplest model, assuming that no limb asymmetries are present.

\begin{figure*}
\plotone{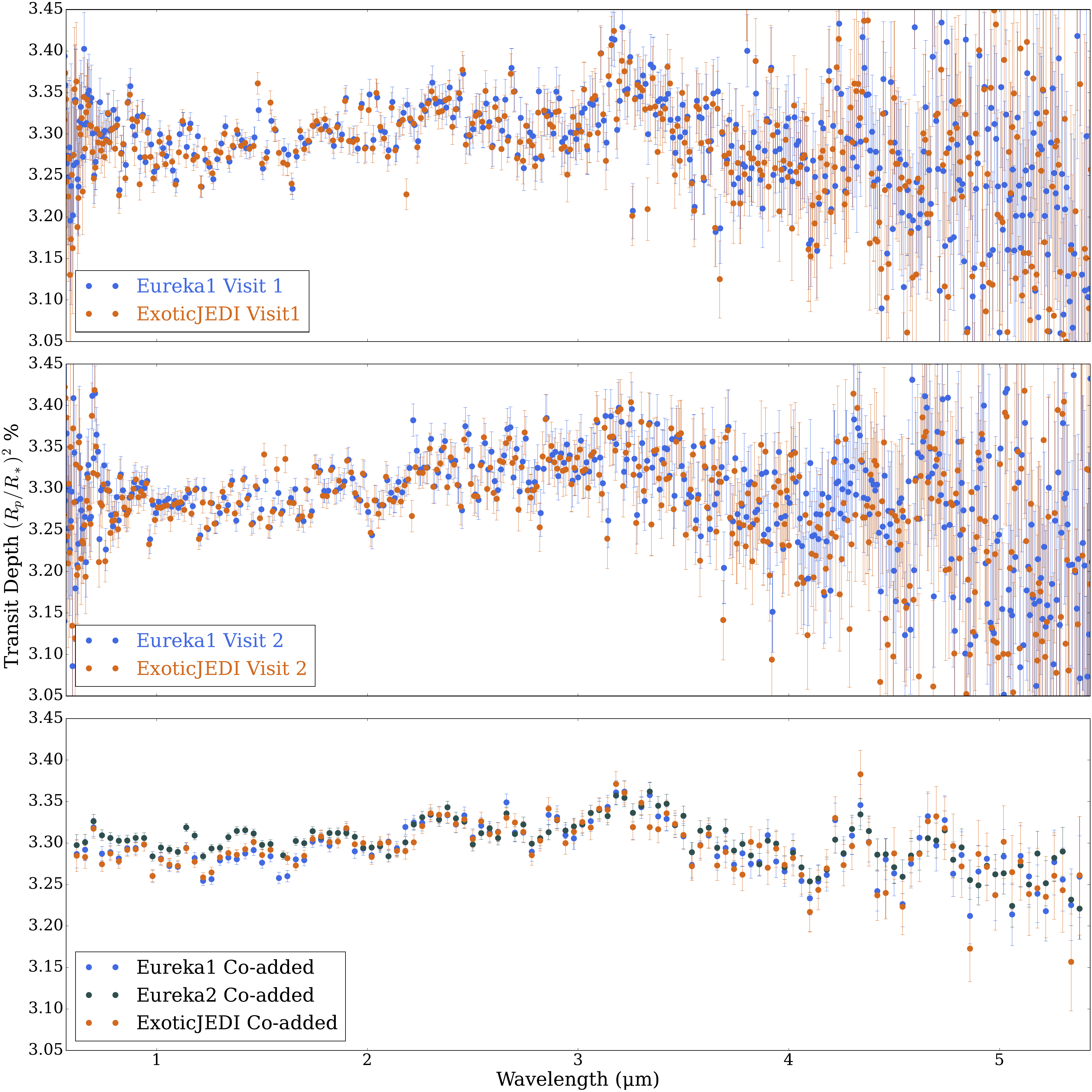}
\caption{The finalized, full-resolution transmission spectra shown by visit in the first two panels using \textbf{Eureka1} and \textbf{ExoticJEDI} reductions. The lower panel shows the co-added transmission spectrum for each of the three reductions: \textbf{Eureka1}, \textbf{Eureka2}, and \textbf{ExoticJEDI}. We find a generally good agreement between all three reductions, on average within 2 $\sigma$ of each other, despite a significant difference in limb-darkening and systematic treatment in the \textbf{Eureka2} reduction, which could explain its slightly higher slope on the bluer end of the spectrum. The binned spectra have 120 spectral channels with 40nm bins. A comparison of the visits for each reduction is shown in \autoref{fig:transmission_by_code}.}
\label{fig:transmission}
\end{figure*}

\subsection{Finalized Transmission Spectra}\label{subsec:spectra}

The finalized transmission spectra are shown in \autoref{fig:transmission} (see more in  \autoref{appendix:compare_visits}). Each reduction was done independently with the reduction choices outlined in \autoref{fig:data-reduction-differences}. We chose \textbf{Eureka1} as our primary reduction due to lower scatter in the data. Despite the different reduction choices made, there is generally good agreement in the data, and the biggest discrepancies can be traced back to limb darkening prescriptions and aperture choices. The co-added spectra were calculated only using the binned-then-fit version of our reductions because the two visits had different wavelength solutions, making a weighted average for our pixel-level spectra unreliable. As a result, there is no co-added pixel-level reduction for this target and any mention of co-added data henceforth implies that it is also binned. We ultimately employ the pixel-level visit-by-visit spectra to make our final decisions on model choice in order to avoid any bias from the binned approach (see \S\ref{subsection:results}) and only use the coadded binned data to verify and synthesize our results. Finally, we run the final model on the co-added binned spectra to coalesce our results and report those along with the results for each single visit (see \S\ref{subsec:retrieval_results}). 


\section{Atmospheric Forward Modeling}\label{sec:forwardmodel}

\subsection{Chemical Equilibrium with PICASO}\label{subsec:PICASO}
In order to explore how the atmosphere of HATS-6 b compared to the models of gas giants in chemical equilibrium, we leveraged the capabilities of PICASO \citep[v3.2.2][]{batalha_exoplanet_2019, mukherjee_picaso_2023}. PICASO is an open-source modeling package that allows the user to explore one-dimensional radiative-convective thermochemical equilibrium (RCTE) models for a given set of grid parameters. PICASO utilizes a correlated-k opacity database that is publicly available \citep[see][]{lupu_correlated_2021} and uses abundances given by equilibrium chemistry from the Sonora Bobcat models for gas giants \citep{marley_sonora_2021}. For our RCTE chemical equilibrium grid, we adopted the pressure-temperature profile based on  \cite{guillot_radiative_2010} as the initial approximation for the climate model iterations. We adopted a wide range of intrinsic temperatures that are predicted for temperate giant planets \citep{thorngren_intrinsic_2019}, as well as a clear atmosphere. The model grid consists of the following parameters: 

\begin{enumerate}[label=(\roman*)]
    \item 13 metallicities  | 
    [M/H] = $-1.0$, $-0.7$, $-0.5$, $-0.3$, 0.0, 0.3, 0.5, 0.7, 1.0, 1.3, 1.5, 1.7, 2.0 dex
    \item 6 carbon-to-oxygen ratios\footnote{These are multiplicative factors of the protosolar abundance of $\mathrm{[C/O]}_\odot=0.457$ from \cite{Lodders2010}.} |
    C/O = 0.25, 0.5, 1.0, 1.5, 2.0, 2.5
    \item 5 intrinsic temperatures | 
    T$_{int}$ = 50, 100, 200, 300, 400 K
    \item 3 heat redistribution factors\footnote{This represents the fractional contribution of stellar radiation; see Equation 20 in \cite{mukherjee_picaso_2023}.} |
    $r_{st}$ = 0.25, 0.50, 0.75
\end{enumerate}

The resulting grid consisted of a total of 1170 clear-atmosphere models. Transmission spectra were calculated with PICASO from the RCTE model grid using a resampled opacity database of R $\sim$ 60,000 \citep{batalha_resampled_2020}. Once the RCTE transmission spectra grid was calculated, we fit our native-resolution visits separately by using a $\chi_\nu^2$ grid search algorithm implemented in PICASO. We fit from $1.5-5.3$ \textmu{}m in order to mitigate any model discrepancies due to TLS or aerosol contamination, since we do not forward model those effects. We explore haze and stellar contamination in our free chemistry retrievals (\S\ref{subsection:freechem}) instead. For our RCTE grid, both visits were best fit with the same combination of parameters and yielded goodness-of-fit criteria as follows: $\chi_\nu^2 = 1.523$ for visit 1 and $\chi_\nu^2 = 1.378$ for visit 2. The best-fit model had an intrinsic temperature (T$_{int}$) of 50 K, a heat redistribution factor ($r_{st}$) of 0.25, a metallicity ([M/H]) of $-1.0$ dex, and a sub-solar Carbon-to-Oxygen ratio (C/O) of 0.25 multiplicative to solar (equivalent to a C/O of 0.1145). The best-fit model sits at the lower metallicity and lower C/O limits of the k-correlated opacities from the Sonora Bobcat models such that we could not exclude lower atmospheric metallicity and C/O ratios for HATS-6 b. 

\begin{deluxetable*}{lc|c|cc}
\tabletypesize{\small}
\tablecaption{Parameters for the best-fitting models derived from \texttt{PICASO} (RCTE) and \texttt{VULCAN} (disequilibrium chemistry) on our visit by visit data. \label{tab:picasovulcan}}
\tablehead{
\colhead{Parameter} &
\multicolumn{1}{c|}{Units} & 
\colhead{RCTE$^\dagger$} & 
\multicolumn{2}{c}{Disequilibrium Chemistry} 
}
\startdata
~~Atmospheric metallicity ($\mathrm{[M/H]}$) & dex & $-1.0$ & $-1.0$ & $-1.0$ \\
~~Atmospheric carbon-to-oxygen ratio (C/O)$^\ddag$ & \nodata   & 0.25  & 0.25 & 0.25 \\
~~Intrinsic Temperature ($T_\mathrm{int}$) & $K$   & 50    & 50  & 50 \\
~~Heat redistribution factor ($r_{st}$) & \nodata  & 0.25   & 0.25  & 0.25 \\
~~Eddy diffusive coefficient ($K_{zz}$) & dex & \nodata & 7  & 9 \\
\hline
~~ Reduced chi-squared ($\chi^2_\nu$) Visit 1& \nodata   & 1.523 & 1.485 & 1.486 \\
~~ Reduced chi-squared ($\chi^2_\nu$) Visit 2& \nodata   & 1.378 & 1.333 & 1.334 \\
\enddata
\tablenotetext{\dagger}{Parameters that are empty in a given column were not applicable to the specific model.}
\tablenotetext{\ddag}{The C/O in this table is the value reported by the Sonora Bobcat models \citep{marley_sonora_2021} and represents multiples of $\mathrm{[C/O]}_\odot=0.457$ based on the protosolar abundances from \cite{Lodders2010}.}
\tablecomments{The \texttt{PICASO} grid parameters are described in \S\ref{subsec:PICASO} while the \texttt{VULCAN} grid parameters are described in \S\ref{subsection:vulcan}. Each column in the disequilibrium chemistry section represents the best-fit combination for independent eddy diffusive coefficients ($K_{zz}$).}
\end{deluxetable*}

\begin{figure*}
\plotone{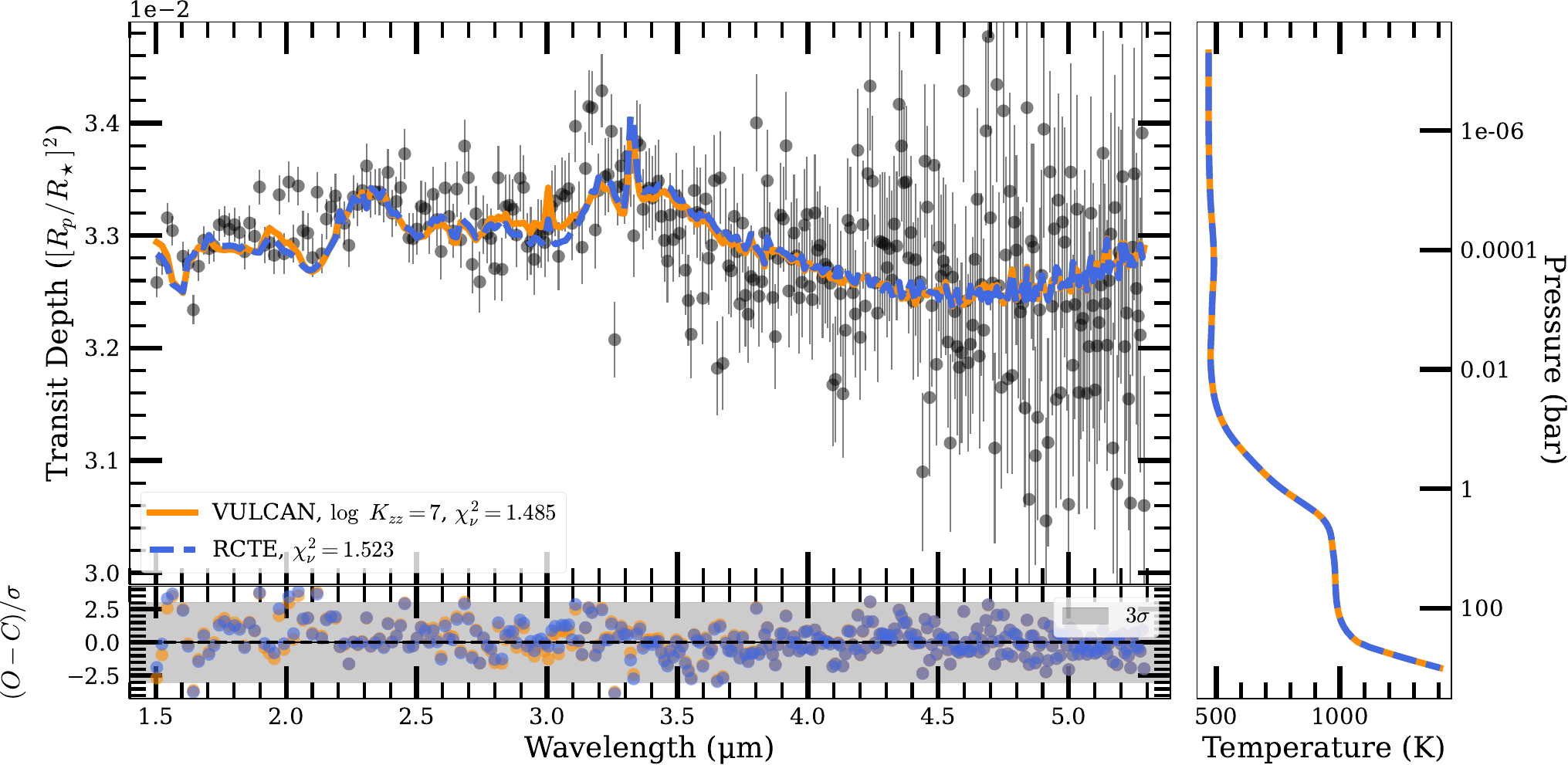}
\caption{\textbf{Left:} The best-fit clear-atmosphere RCTE and VULCAN forward models against the visit 1 \textbf{Eureka1} reduction. \textbf{Right:} The respective pressure-temperature profiles of each model. Even though chemical disequilibrium with VULCAN provides a better fit to the data, the resulting pressure-temperature profiles are virtually the same.}
\label{fig:picaso_best1}
\end{figure*}

\begin{figure*}
\plotone{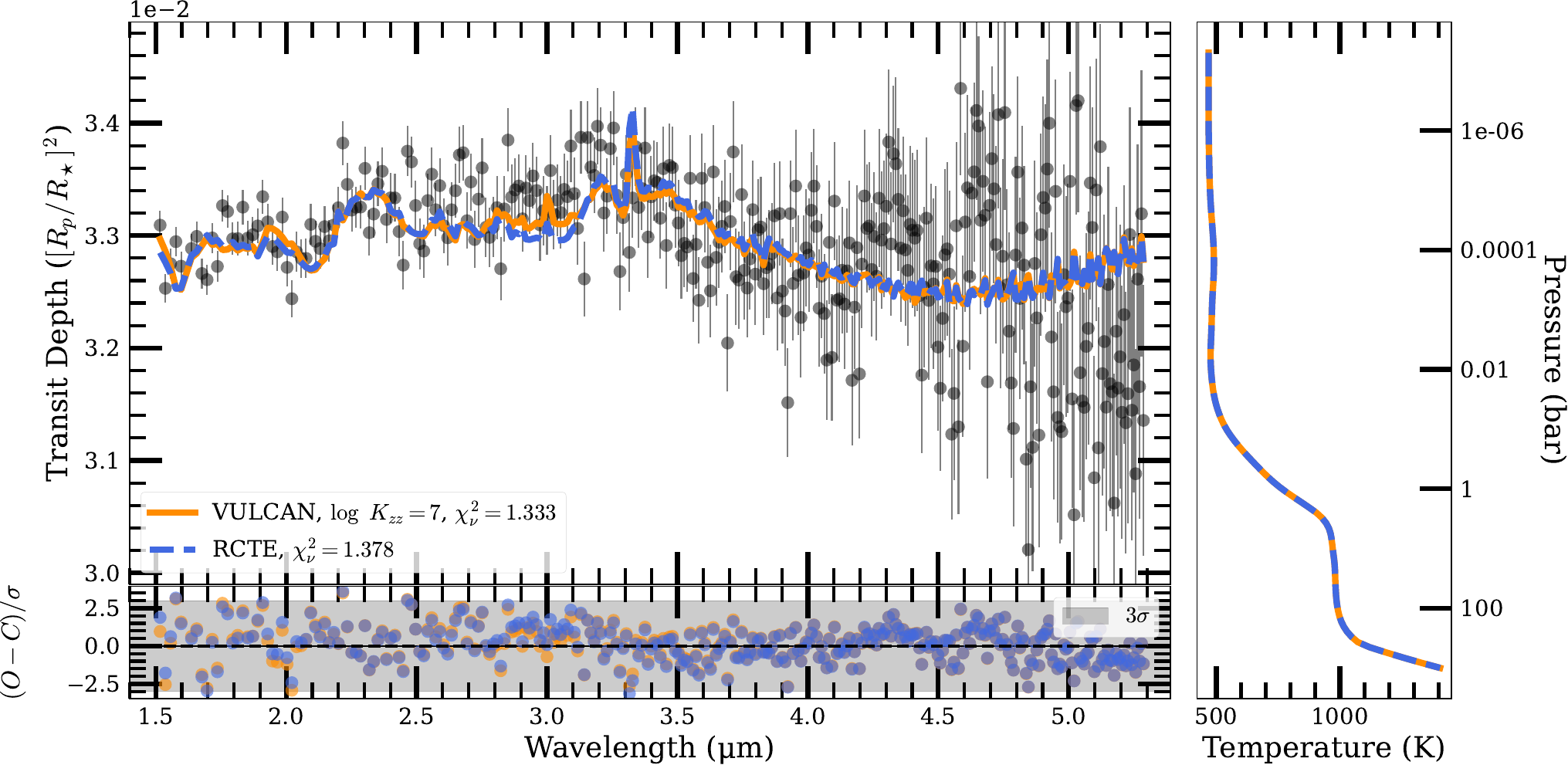}
\caption{\textbf{Left:} The best-fit clear-atmosphere RCTE and VULCAN forward models against the visit 2 \textbf{Eureka1} reduction. \textbf{Right:} The respective pressure-temperature profiles of each model. Even though chemical disequilibrium with VULCAN provides a better fit to the data, the resulting pressure-temperature profiles are virtually the same.}
\label{fig:picaso_best2}
\end{figure*}

\subsection{Disequilibrium Processes with VULCAN}\label{subsection:vulcan}

In order to expand beyond the limitations of our chemical equilibrium models, we also model the effect of disequilibrium processes using the photochemical model VULCAN \citep{tsai_vulcan_2017, tsai_comparative_2021}. We adopted the temperature profiles obtained from our RCTE grid and evolved the atmospheric composition from the initial equilibrium abundances set by \texttt{FastChem} \citep{stock_fastchem_2018}. We used the S–N–C–H–O photochemical network \footnote{\url{https://github.com/exoclime/VULCAN/blob/master/thermo/ SNCHO_photo_network_2024.txt}} and the UV spectrum of GJ 436 \citep[M3V;][]{muscles1,muscles2,muscles3} as a proxy for the stellar UV spectrum of HATS-6 (M1V). Previous papers have shown that different M star analogues produce very similar results for this sort of modeling \citep{renyu2025}. Following a similar setup to \cite{canas_gems_2025}, we investigated the same atmospheric parameter space as our RCTE models, along with two different strengths of vertical mixing ($K_{zz}=10^7$ and 10$^9$ cm$^{-2}$ s$^{-1}$).

Once we obtained the abundance profiles from VULCAN, we generated synthetic spectra as described in \S\ref{subsec:PICASO}. We then fit the grid of disequilibrium chemistry transmission spectra using the same methods described in \S\ref{subsec:PICASO}. Similar to our chemical equilibrium fits there was a single best-fit parameter combination for both visits, with varying goodness-of-fit: $\chi_\nu^2 = 1.485$ for visit 1 and $\chi_\nu^2 = 1.333$ for visit 2. The spectral fits are shown in \autoref{fig:picaso_best1} and \autoref{fig:picaso_best2}. The best-fit model had an intrinsic temperature (T$_{int}$) of 50 K, a heat redistribution factor ($r_{st}$) of 0.25, a metallicity ([M/H]) of $-1.0$ dex, a Carbon-to-Oxygen ratio (C/O) of 0.25 relative to solar, and a vertical mixing strength factor (K$_{zz}$) of 10$^7$ cm/s$^2$. For both chemical equilibrium and disequilibrium fits, HATS-6 b consistently trended towards a metal-poor and sub-solar C/O atmosphere. In both cases, metallicity and C/O hit the grid lower limits, meaning we could not exclude lower metallicity and C/O ratios. We test for two different Eddy diffusion coefficients ($K_{zz}$) for our disequilibrium models, and find that the difference in goodness of fit between our first and second best-fit models minimal (see \autoref{tab:picasovulcan}). Finally, although our VULCAN implementation only yielded minor improvements to the fit compared to the RCTE fit, the consideration of disequilibrium processes allowed us to explore more physically motivated chemical networks during our analysis of free-chemistry retrievals presented in \S\ref{sec:retrievals}.

\subsection{Cloud contributions with virga}\label{subsection:virga}

With the goal of thoroughly exploring any physical processes that might affect our spectra, we implemented the cloud modeling suite \texttt{virga} and assumed spherical cloud particles (\texttt{virga} v1). Using a database of refractive indices \citep{batalha_refractive_2020}, \texttt{virga} allows users to explore bulk H$_2$ atmospheres using the cloud prescription from \cite{ackerman_precipitating_2001}. We first ran condensation curves in \texttt{virga} as a way to identify which condensate species were readily applicable to HATS-6 b. In order to identify which condensates could be present in the atmosphere of HATS-6 b, we assumed an isotherm at 700 K (i.e., \autoref{fig:picaso_best1} and \autoref{fig:picaso_best2}) within our pressures probed, along with a subsolar atmospheric metallicity, and found that the only species whose condensation curve crossed our isotherm within the \texttt{virga} framework was ZnS \citep{querry_optical_1998}. We also tested a slightly lower temperature ($\sim500$ K) that was more representative of our final results, and found no new condensate species that crossed the lower temperature isotherm.

We took our  best-fit chemical equilibrium model and post-processed it with \texttt{virga} at R $\sim$ of 60,000. We found no quantifiable difference between the original spectra and the virga post-processed spectra. We then tested the same post-processing comparison at R $\sim$ 15,000 \citep{batalha_resampled_2025} to extend the models to 15 \textmu{}m, in case any differences between the models were more apparent redward of 5.3 \textmu{}m. Finding no differences between the wavelength-extended post-processed models, we isolated the cloud-top pressures in the calculated \texttt{virga} model and found that the ZnS cloud-top pressure sat at much higher pressures than we are able to probe with our transmission spectra for our assumed K$_{zz}$ of 10$^7$ (from our best-fit VULCAN models). Therefore, we concluded that none of the currently available condensate species in \texttt{virga} had a quantifiable effect on our data.


\section{Atmospheric Retrievals} \label{sec:retrievals}

We performed Bayesian atmospheric retrievals \citep[e.g.,][]{madhusudhan_atmospheric_2018} assuming both equilibrium and free chemistry to further interpret our spectra. This inverse-modeling approach allows us to explore a full range of models that may be consistent with our data, along with uncertainties on the retrieved model parameters. We first run a set of chemical equilibrium retrievals, which we summarize in \S\ref{subsection:chemeq_results}. Then, we perform an extensive set of exploratory retrievals on our \textbf{Eureka1} binned co-added reductions. The purpose of the exploratory retrievals is to get a general sense of what chemistry and atmospheric models work best for our data. We summarize these exploratory retrievals in \S\ref{subsection:coaaddposeidon} - \S\ref{subsection:Tpriors}. Finally, informed by our results from our exploratory retrievals, we choose a canonical suite of models that is broadly representative of our exploratory results. We also physically inform our final chemistry using our forward models and run this canonical suite of models on our final visit-by-visit native resolution reductions (\textbf{Eureka1}) in order to avoid any potential biases from binning down our data. We use these independent visit-by-visit fits to choose our final model (see \S\ref{subsection:visitbyvist}-\S\ref{subsection:results}). Finally, we run the final model with another retrieval suite to test the robustness of our results (\S\ref{subsection:taurex}), and also run it on the binned co-added data in order to coalesce our results and report a single value (\S\ref{subsec:retrieval_results}). We verify that the general conclusions are not impacted by the choice of data reduction, and report to the reader our results for both our native resolution visit-by-visit fits and our binned co-added fits. Our final retrieval results are summarized in \S\ref{subsec:retrieval_results}.

\subsection{Chemical Equilibrium Retrievals}\label{subsection:chemeq}

We first explored bulk atmospheric properties for HATS-6 b using chemical equilibrium retrievals with three different retrieval suites: \texttt{PLATON}, \texttt{POSEIDON}, and \texttt{TauREx}, which are described in the following subsections. For all chemical equilibrium retrievals, an isothermal temperature-pressure profile was adopted. All chemical equilibrium fits were run on the binned coadded data. We did not include the effects of ions or condensation equilibrium in any of our equilibrium chemistry calculations. The results for chemical equilibrium retrievals are summarized in \S\ref{subsection:chemeq_results}. 

\subsubsection{Chemical Equilibrium with PLATON}\label{subsubsection:platon}

We explored chemical equilibrium models with the retrieval suite PLanetary Atmospheric Tool for Observer Noobs \citep[\texttt{PLATON} v6.1;][]{Zhang2019,zhang_platon_2020,2025AJ....169...38Z}. This version of \texttt{PLATON} uses \texttt{FastChem} \citep{stock_fastchem_2018} to calculate the chemical equilibrium abundances between a pre-set grid of $-2\leq\log\mathrm{[M/H]}\leq3$ and $0.01\leq \mathrm{C/O}\leq2$. The \texttt{PLATON} prescription uses a cross-section opacity database with a resolution of R $\sim$ 20,000, and we adopted the default list of chemical species for chemical equilibrium calculations. We also fit for an error multiplier (a constant by which all errors are multiplied) to test whether the inclusion of an error multiplier significantly improves our fits and find that our results are not improved or affected by the inclusion of this error multiplier. The temperature-pressure profile was modeled with 13 atmospheric levels that covered pressures of $10^{-9}-10^3$ bar, and we adopted a reference pressure of 1 bar. We sampled the parameter space using the importance nested sampling algorithm, \texttt{MultiNest} \citep{multinest}, through the \texttt{PyMultiNest} interface \citep{2014A&A...564A.125B,2016ascl.soft06005B} with $N=1000$ live points and a convergence criterion of $\Delta\ln Z=0.5$. 

\subsubsection{Chemical Equilibrium with \texttt{POSEIDON}}\label{subsection:poseidon}

We also explored chemical equilibrium models with \texttt{POSEIDON} \citep{macdonald_hd_2017,macdonald_poseidon_2023}. \texttt{POSEIDON} also uses \texttt{FastChem} for the calculation of the equilibrium chemistry abundances in a grid that spans a pre-set range of $-1\leq\log\mathrm{[M/H]}\leq4$ and $0.2\leq \mathrm{C/O}\leq2$. We adopted a cross-sectional opacity resolution of R $\sim$ 20,000 for the radiative transfer calculations and included all species in the \texttt{FastChem} grid available for chemical equilibrium calculations in the \texttt{POSEIDON} code suite. The temperature-pressure profile was modeled with 42 atmospheric levels that covered pressures of $10^{-6}-10^2$ bar, and we adopted a reference pressure of 1 bar. We performed the retrievals on the co-added binned spectrum using \texttt{PyMultiNest} with $N=1000$ live points and a convergence criterion of $\Delta \ln Z=1$.

\subsubsection{Chemical Equilibrium with TauREx}\label{subsection:ggchem}
We also performed chemical equilibrium retrievals with the package Tau Retrieval for Exoplanets \citep[\texttt{TauREx} v3.2.4;][]{taurex1, taurex2}. We used the \texttt{GGChem} \citep[Gleich-Gewichts-Chemie;][]{ggchem} plugin\footnote{\url{https://github.com/ucl-exoplanets/GGchem}} \citep{taurex2} for the calculations of the chemical equilibrium abundances, assuming a solar chemical abundance, an isothermal profile, and a clear or cloudy atmosphere. The inclusion of \texttt{GGChem} allowed us to probe wider priors for metallicity and C/O than we did for the other retrieval suites, which were restricted to pre-computed grids. We included all available species containing the elements H, He, C, N, O, P, S for the calculations provided by \texttt{TauREx} \citep{taurexopacity} or available on \texttt{ExoMol} \citep{exomolop}.  The temperature-pressure profile was modeled with 300 atmospheric levels that covered pressures of $10^{-7}-10^3$ bar, and the reference pressure was 1000 bar. The fits to the co-added spectrum used log-uniform priors of $\mathcal{LU}(-10,5)$ for both $\mathrm{\log[M/H]}$ and $\mathrm{\log[\mathrm{C/O}]}$. We sampled the opacities at a spectral resolution of R $\gtrsim15,000$ for radiative transfer calculations. We sampled the posteriors using \texttt{MultiNest} with $N=1000$ live points and a convergence criterion of $\Delta\ln Z=1$. 

\subsubsection{Chemical Equilibrium Retrieval Results}\label{subsection:chemeq_results}

\begin{figure*}
\epsscale{1.17}
\plotone{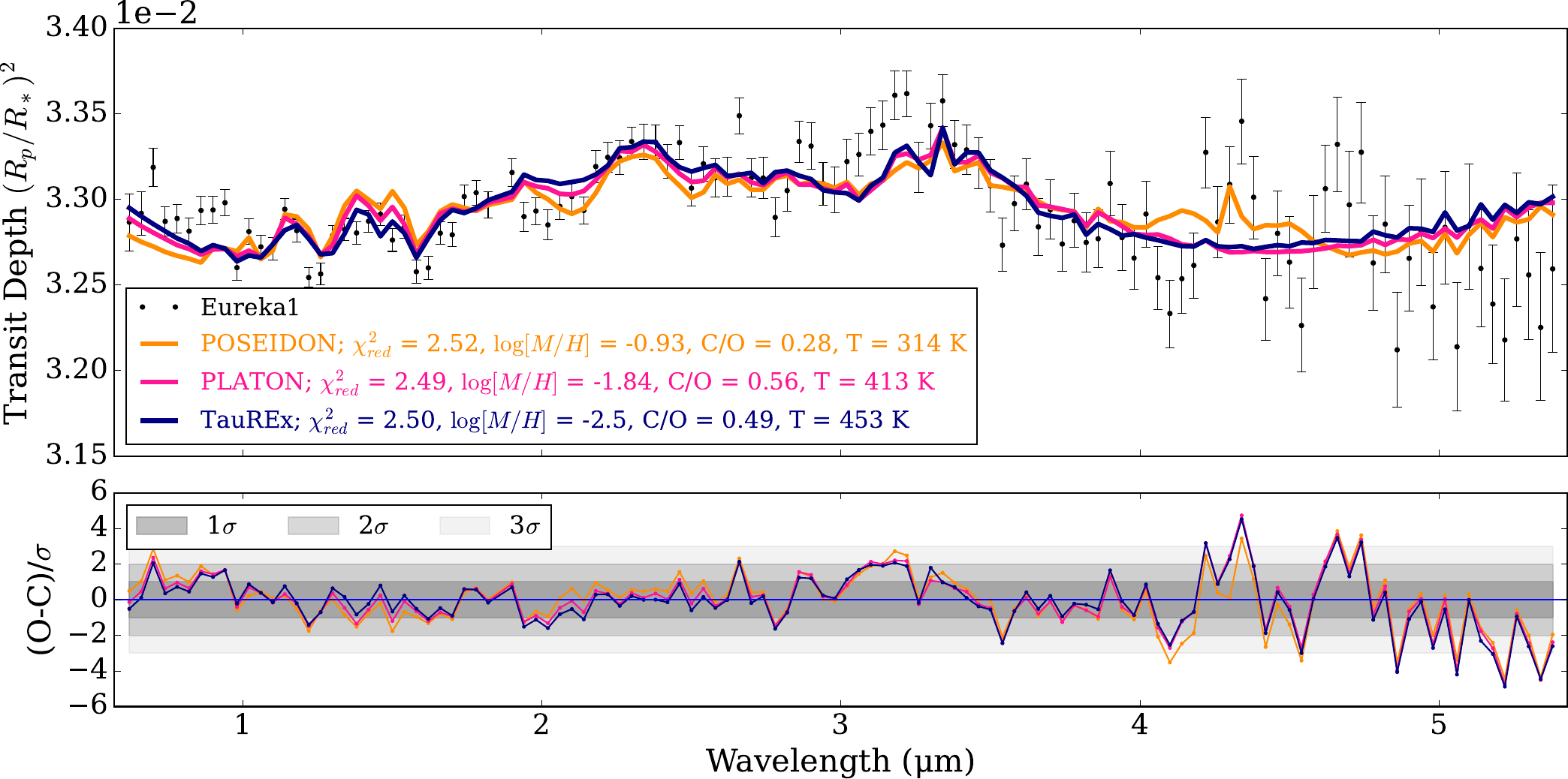}
\caption{The best-fit chemical equilibrium models to our co-added data performed with three different retrieval suites. For all chemical equilibrium fits, the models produced poor fits to several features on the spectrum, namely near 3 \textmu{}m and between $4-5$ \textmu{}m region. For the 4.3 micron region, the differences between features are due to a combination of differences in the retrieval framework (namely prior constraints) and chemical networks used (\texttt{FastChem} vs \texttt{GGChem}) for each retrieval suite.}
\label{fig:chemeq_fits}
\end{figure*}

\begin{figure*}
\plotone{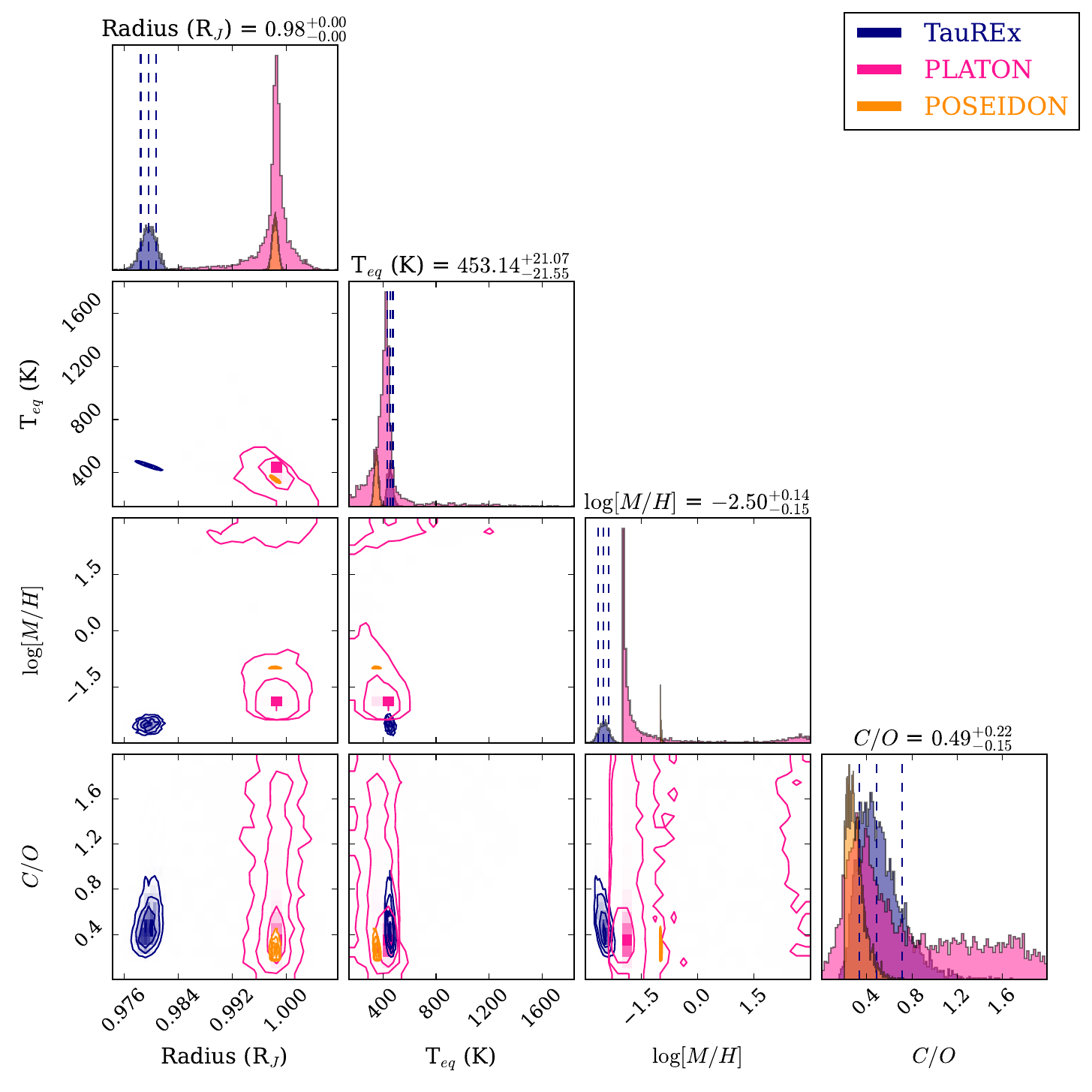}
\caption{The retrieved posteriors for our clear-atmosphere chemical equilibrium retrievals with three different retrieval suites. The captions at the top and the vertical dashed lines reflect the 16th, 50th, and 84th percentiles from the TauREx retrievals. By default, \texttt{TauREx} reports the planetary radius at the bottom of the atmosphere ($10^3$ bar) while \texttt{POSEIDON} and \texttt{PLATON} report the radius at 1 bar, which is why the distributions do not overlap. The corresponding best-fit models are shown in \autoref{fig:chemeq_fits}.}
\label{fig:chemeq_posteriors}
\end{figure*}

\autoref{fig:chemeq_fits} displays our chemical equilibrium best-fit models for the co-added \textbf{Eureka1} spectrum and \autoref{fig:chemeq_posteriors} shows the corresponding posteriors. The chemical equilibrium retrievals show a worse chi-squared than our forward models. These retrievals also trended towards subsolar atmospheric metallicity (log[M/H] $<$ -1), and near-solar Carbon-to-Oxygen ratios (C/O $\sim$ 0.2-0.4). Despite good agreement between the models of different retrieval suites, the chemical equilibrium retrievals highlighted a persistent artifact that permeated the retrieval analysis of this target: a consistently low retrieved temperature. Although the equilibrium temperature of HATS-6 b is expected to be $710-740$ K from photometric analysis, which assumes efficient heat transport and a zero bond albedo, the retrieved temperatures ranged from 300--450 K for all equilibrium chemistry retrievals, which is not a negligible difference \citep[i.e.,]{macdonald_why_2020}. If we still assume efficient and uniform heat redistribution, these retrieved temperatures would yield a bond albedo close to that of Venus ($>$ 0.7), which would point toward the presence of clouds and hazes in the atmosphere of HATS-6 b. The best fitting chemical equilibrium model ($\chi^2_\nu$ = 2.18) also pointed towards the need for clouds and hazes to better explain the data. In order to test if our results were a product of inefficient sampling we also used \texttt{POSEIDON} to run a chemical equilibrium retrieval with a more informed Gaussian prior around the expected 700 K T$_{eq}$ for the clear atmosphere fit. This exercise yielded both a high metallicity (log[M/H] $\sim$ 2.6) and a high Carbon-to-Oxygen ratio (C/O $\sim$ 1.89) that was pushing on prior bounds, a stark contrast to our retrievals with uninformed uniform priors. We note that the model with a Gaussian prior on temperature also had a worse fit to the data ($\chi^2_\nu$ = 2.48; ln Z = 2095.84) compared to the equivalent clear atmosphere uninformed prior fit ($\chi^2_\nu$ = 2.27; ln Z = 2134.26).

\subsection{Free Chemistry Retrievals with \texttt{POSEIDON}}\label{subsection:freechem}

The assumption of thermochemical equilibrium may bias the retrieved atmospheric parameters \citep[e.g.,][]{taurex2}. To explore non-equilibrium conditions in the atmosphere of HATS-6 b, we performed a thorough set of exploratory retrievals for which we allowed individual chemical abundances to vary during inverse modeling. For this analysis we utilized \texttt{POSEIDON}. For all retrievals we adopted an isothermal temperature-pressure profile unless stated otherwise (i.e. the temperature-pressure profiles employed in \S\ref{subsection:Tpriors}). We performed our first iteration of retrievals on the binned coadded data in order to efficiently explore the parameter space and determine what physical prescriptions were most likely for our atmosphere. We also utilized this exploratory retrieval approach to test which parameters were driving certain results, such as the low retrieved T$_{eq}$. With the information gleaned from these exploratory retrievals on the co-added binned data, we built a comprehensive set of canonical models (\S\ref{subsection:results}) that we run for our final analysis on our visit-by-visit reductions. We extract our conclusions from the visit-by-visit results, and check them against a final binned co-added reduction run in order to show which conclusions are robust across visits. 

\subsubsection{Co-added Full Chemistry Spectral Retrievals with \texttt{POSEIDON}}\label{subsection:coaaddposeidon}

We began our free chemistry retrieval analysis by assuming `full chemistry' composition for our initial fits. The \emph{full chemistry} models included opacities for the following species: H$_2$O \citep{h2o}, CH$_4$ \citep{ch4}, CO$_2$ \citep{co2}, CO \citep{co}, SO$_2$ \citep{so2}, NH$_3$ \citep{nh3}, HCN \citep{hcn}, C$_2$H$_2$ \citep{c2h2}, H$_2$S \citep{h2s}, PH$_3$ \citep{ph3}, CS \citep{CS}, CS$_2$, C$_2$H$_4$, and C$_2$H$_6$ \citep[all from:][]{hitrancia}. They also included collision-induced absorption (CIA) from H$_2$–H$_2$, H$_2$–He, H$_2$–CH$_4$, CO$_2$–H$_2$, CO$_2$–CO$_2$, and CO$_2$–CH$_4$ \citep{hitrancia}. Furthermore, we assumed a bulk H$_2$ and He composition reflective of their cosmochemical abundances (N$_{\rm{He}}$/N$_{\rm{H_2}}$ = 0.17) and kept this fixed. We allowed the logarithmic mixing ratio of each of the listed species to float freely within prior bounds [$\mathcal{U}(-14.00, -1.00)$] while maintaining evenly mixed vertical gas abundances. 

Along with a clear atmosphere, we also adopted an identical chemical configuration while applying different permutations of the cloud and haze parameters to investigate their contributions to the retrieval fits. Clouds and hazes were modeled using the haze/cloud-deck parameterization in \citet{macdonald_hd_2017}. We also model the Transit Light Source effect (TLS) \citep{rackham_transit_2018} by parameterizing the stellar contamination component with synthetic stellar spectra using a spot coverage fraction and spot temperature using PHOENIX synthetic spectra \citep{phoenix}, generated using the \texttt{pyMSG}\footnote{\url{https://github.com/rhdtownsend/msg}} software package \citep[v1.3;][]{2023JOSS....8.4602T}, to model the dilution effects on the spectra. We included a `patchy cloud' fit which also allowed more flexibility than a simple gray cloud-deck. For all exploratory retrievals, we adopt 1000 live points (N), an opacity resolution of R $\sim$ 20000, and a convergence criterion of $\Delta\ln Z$ = 1.0. The posteriors and results for all \texttt{POSEIDON} retrievals, along with the \textbf{Eureka1} spectra, are available on Zenodo \citep{guzman_caloca_2026_21500595}.\footnote{\url{https://doi.org/10.5281/zenodo.21500595}}  

\begin{figure*}
\plotone{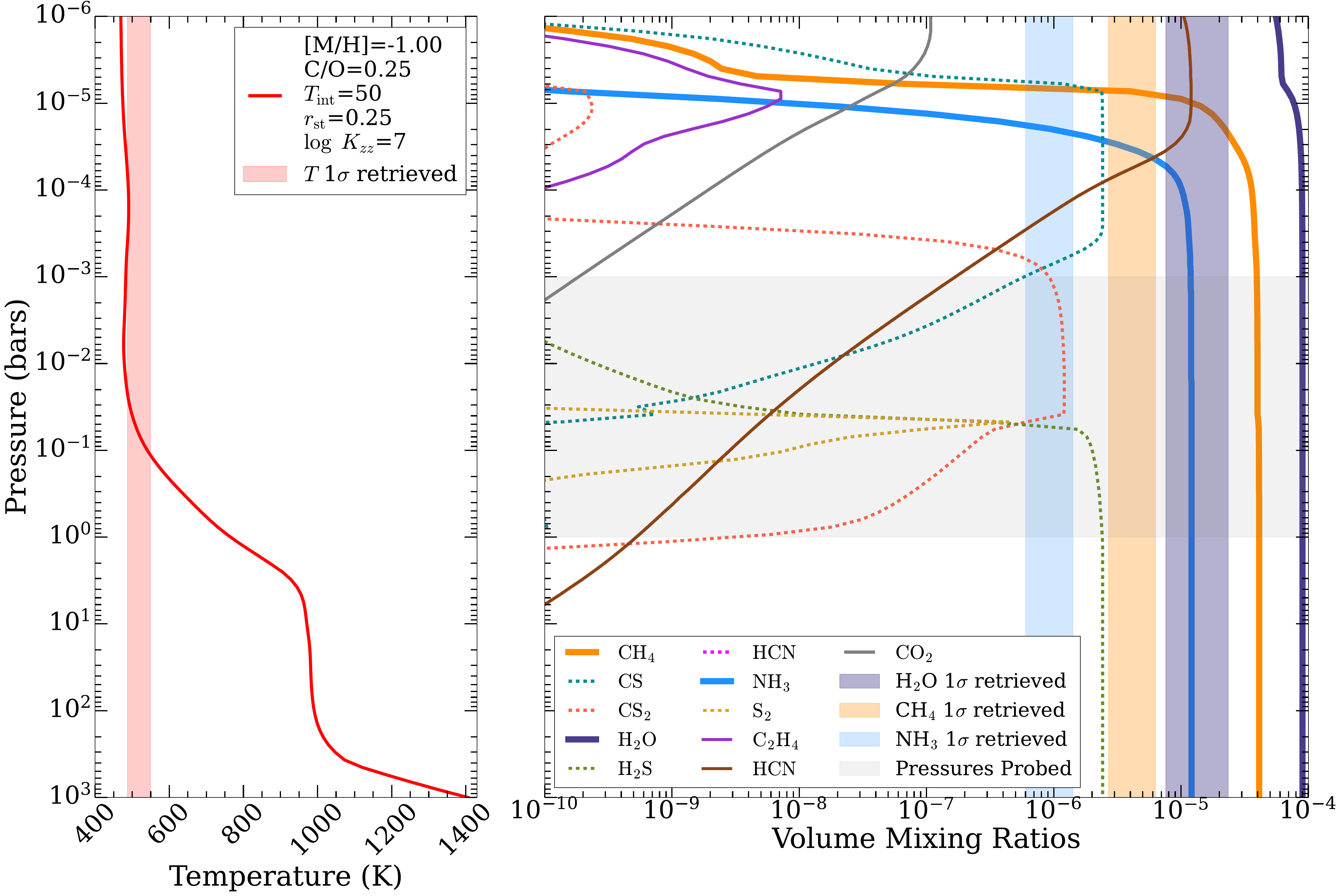}
\caption{Our best-fit VULCAN chemical profiles.  We highlight the 1$\sigma$ retrieved planetary temperature and abundances from our final retrieval model with \poseidon (M2.5; see \autoref{tab:retrievals_final}) for our best-constrained species (H$_2$O, CH$_4$, and NH$_3$) and show the abundances for these species as solid lines. For ease of comparison against forward model expectations from VULCAN, we also highlight the abundance profiles of C$_2$H$_4$ and HCN as solid lines. These species were part of follow-up retrieval tests that yielded tentative VMRs of: $\log\mathrm{C_2H_4}=-6.20^{+0.49}_{-2.09}$ and $\log\mathrm{HCN}=-5.71^{+0.34}_{-0.39}$ (see \S\ref{section:feature}).}
\label{fig:TPprofile}
\end{figure*}

The results of our initial exploratory free chemistry retrievals on the co-added spectra presented the same persistent artifact as our chemical equilibrium retrievals (see \autoref{appendix:full_chemistry}): a significantly lower retrieved \teq. On average, our fits had $\mathrm{T_{eq}}=300-400$ K retrieved temperatures. We also found robust constraints on H$_2$O, CH$_4$, and NH$_3$, and the average retrieved mixing ratios for these species are consistent with what was predicted from our \texttt{VULCAN} forward models. In addition, we retrieved an unusually high abundance of more complex aliphatic hydrocarbons, especially ethene (C$_2$H$_4$). For some configurations, ethene was retrieved at abundances higher than methane. Methane is the simplest aliphatic hydrocarbon and is predicted to be the dominant hydrocarbon and much more abundant in a \teq $\sim$ 700 K atmosphere than other hydrocarbons, like C$_2$H$_4$. Furthermore, comparing against our \texttt{VULCAN} best-fit abundances, the abundances retrieved for these aliphatic hydrocarbons were unexpectedly high even when taking into account photochemical disequilibrium processes (see \autoref{fig:TPprofile}). This begged the questions: \textit{what was the driver of our retrieved low \teq in our free chemistry fits, even when considering clouds and hazes, and why were we retrieving such high abundances of complex aliphatic hydrocarbons compared to model predictions?}

\subsubsection{Wavelength Cutoff Tests and Species Contributions with \texttt{POSEIDON}}\label{subsection:wavecut}

As a means to pinpoint what part of our spectra might be driving the temperature down and identify where the strongest C$_2$H$_4$ feature was, we also ran a retrieval wavelength cutoff test \citep[see][]{canas_gems_2025}. In these tests, we fit small segments of the spectra in order to isolate which wavelength regions are responsible for various effects in our retrieved results. We ran these tests on a clear atmosphere, \emph{full chemistry} model. From these tests, we concluded that our retrievals were anchoring onto a feature in the data redward of 2.5 \textmu{}m and a further look into the chemical contributions of our best-fit models confirmed that there was a significant absorption feature between $3-3.5$ \textmu{}m that was best fit by C$_2$H$_4$. Due to the proximity with the methane absorption feature at $\sim 3.0-3.5$ \textmu{}m, we also performed a retrieval only for H$_2$O and CH$_4$ to investigate whether higher abundances of CH$_4$ in the absence of C$_2$H$_4$ could also explain the feature around $3-3.3$ \textmu{}m. The result still trended to a lower temperature (\teq=350 K) and displayed a worse fit ($\chi^2_\nu$ = 2.45) than the run including C$_2$H$_4$ ($\chi^2_\nu$ = 1.97), with difficulty matching the region around 3 \textmu{}m. We concluded that the methane contribution retrieved in our simple chemistry run was not sufficient to explain the data between $3.0-3.3$ \textmu{}m, and we discuss this further in \S\ref{section:feature}.

\subsubsection{Testing Different Temperature Priors with \texttt{POSEIDON} retrievals}\label{subsection:Tpriors}

Another approach we took was to adjust our priors on the temperature in several ways: (i) an informative uniform prior centered at 713 K ($\pm$ 100 K); $\chi^2_\nu$ = 3.01 and ln Z = 770, (ii) a Gaussian prior ($\sigma$ $\pm$ 100 K) around 713 K; $\chi^2_\nu$ = 3.06 and ln Z = 768, (iii) a fixed temperature retrieval; $\chi^2_\nu$ = 3.09 and ln Z = 768, (iv) fitting for a Guillot \citep{guillot_radiative_2010} temperature-pressure (T-P) profile; $\chi^2_\nu$ = 1.94 and ln Z = 822, and (v) fitting for a Madhusudhan $\&$ Seager \citep{madandseager} T-P profile; $\chi^2_\nu$ = 2.07 and ln Z = 821. We ran these tests assuming a clear atmosphere with \emph{full chemistry}. Finally, we ran both a Guillot T-P profile with a gray cloud deck; $\chi^2_\nu$ = 1.98 and ln Z = 823, and a Madhusudhan $\&$ Seager T-P profile with a gray cloud deck as well; $\chi^2_\nu$ = 2.04 and ln Z = 822.

All of these tests resulted in worse fits to the data compared to our original isothermal run, except for the cloudy Guillot and Madhusudhan $\&$ Seager temperature-pressure profile tests, which yielded slightly better evidences (ln $B_{1,2} \simeq 1.2-1.4 $) when comparing to the original fit. Yet, the abundances derived were indistinguishable from the original isothermal fit. This was because the best-fit T-P profiles were mostly isothermal in the regions of the atmosphere probed by our transmission spectra (\autoref{appendix:guillot} and \autoref{appendix:pressure}). In the end these more complex T-P profile fits yielded only a very slight statistical preference that we attribute to more flexibility in the model, but the physical insight gleaned from the best-fit models was indistinguishable from our isothermal fits, and we therefore decide to continue with our isothermal model.

In the case of the other informative prior tests, they produced inconsistent and nonphysical results. This included CO and C$_2$H$_4$ mixing ratios corresponding to 1\% of the atmosphere as well as the planet radius hitting our higher prior bounds (R$_{p,ref} \sim$ 1.15). From these tests, we concluded that more informed priors on temperature or a more flexible thermal profile did not resolve any degeneracies and instead exacerbated other limitations in the retrievals. 

\subsubsection{Visit-by-visit retrievals with \texttt{POSEIDON}}\label{subsection:visitbyvist}

Our analysis first focused on the \textbf{Eureka1} binned co-added spectra under the assumption the planet atmosphere was stable over the two visits, but while co-adding our spectra allowed us to run a plethora of retrieval tests to gain a general understanding of our data, we also noted that there was unique structure within each visit. Some of this differing structure was especially present around the contentious 3 \textmu{}m  region that we identified during our chemical contribution analysis. Therefore, we performed visit-by-visit retrievals on each individual pixel-level resolution visit for our final retrievals, in order to allow comparison across visits. For \emph{full chemistry} clear atmosphere fits, the difference across visits was not significantly notable. For retrievals including other physical parameters (i.e. clouds), we did notice significant visit-by-visit variations in the retrieved values for those parametric properties. We therefore proceeded with a visit-by-visit approach for our final set of retrievals. 

\subsection{Canonical Model Suite Selection}\label{subsection:results}

For our final canonical retrieval suite, we adopted $N=1000$ live points, an opacity grid resolution of R $\sim$ 60,000, and a convergence criterion $\Delta\ln Z=1$. In order to test the detection robustness of the chemical species that were physically informed by our VULCAN models, we also restricted the chemical species to the species identified as the most abundant within the pressures probed (see \autoref{fig:TPprofile}) by using a VMR cutoff of 10$^{-7}$ (H$_2$O, CH$_4$, CO$_2$, CO, N$_2$, NH$_3$, CS$_2$, S, S$_2$, NS, CS$_2$, CS, and H$_2$S). We further restricted the chemical species in our final retrievals to those with available line lists in \texttt{HITRAN} \citep{hitrancia} and \texttt{ExoMol} \citep{exomolop} for application in retrieval suites at our disposal. These models, which we refer to as \emph{``Sans Chemistry''}, consisted of the following species: H$_2$O, CH$_4$, CO$_2$, CO, NH$_3$, CS, CS$_2$, and H$_2$S. CS$_2$ was not only present in our forward models, but has also recently been observed in other giant planets \citep{trianta_wasp80b, dai_vtau, zhang_cs2}. 

From our free chemistry retrievals with \poseidon (see \S\ref{subsection:freechem}), we noticed that adding physical complexity to our atmosphere (in the form of clouds, haze, or TLS) generally improved the fits and brought our equilibrium temperature closer to what is expected. We therefore also adopt clouds, haze, and TLS as part of our canonical suite of models to test any evidence of these in the data. Our final retrieval suite, performed on each individual visit, is summarized in \autoref{tab:RetrievalSetup}. For verification of our final  model fits, we utilized \texttt{TauREx}.

\begin{deluxetable*}{l|l|l}[t]
\tablewidth{0.97\textwidth}
\tabletypesize{\small}
\tablecaption{\texttt{POSEIDON} Final Retrieval Model Suite  \label{tab:RetrievalSetup}} 
\tablehead{
\colhead{Model Name} & \colhead{Chemistry} & \colhead{Bulk Properties}
}
\startdata
M1: Full Chemistry   & \makecell{H$_2$O, CH$_4$, CO$_2$, CO, SO$_2$, NH$_3$, \\ HCN, C$_2$H$_2$, C$_2$H$_4$, C$_2$H$_6$, H$_2$S, PH$_3$, CS, CS$_2$}       & Clear Atmosphere \\ \hline
M2.1: Sans Chemistry  & \makecell{H$_2$O, CH$_4$, CO$_2$, CO, NH$_3$, H$_2$S, \\ CS, CS$_2$} & Clear Atmosphere \\ \hline
M2.2: Clouds       & \makecell{H$_2$O, CH$_4$, CO$_2$, CO, NH$_3$, H$_2$S \\ CS, CS$_2$}   & \textbf{MacMad17} Gray Cloud Deck \\ \hline
M2.3: Haze    & \makecell{H$_2$O, CH$_4$, CO$_2$, CO, NH$_3$, H$_2$S \\ CS, CS$_2$}   & Parametric Rayleigh Scattering \\ \hline
M2.4: Clouds+TLS    & \makecell{H$_2$O, CH$_4$, CO$_2$, CO, NH$_3$, H$_2$S \\ CS, CS$_2$}    & \makecell[l]{\textbf{MacMad17} Gray Cloud Deck \\ TLS prescription} \\ \hline
M2.5: Clouds+Haze   & \makecell{H$_2$O, CH$_4$, CO$_2$, CO, NH$_3$, H$_2$S \\ CS, CS$_2$}  & \makecell[l]{\textbf{MacMad17} Gray Cloud Deck \\ Parametric Rayleigh Scattering} \\ 
\hline \hline
\multicolumn{3}{c}{MultiNest Sampler Parameters}\\
\cline{1-3}
Live Points  & \multicolumn{2}{c}{1000}\\
\cline{2-3}
Evolution Tolerance & \multicolumn{2}{c}{1.0}\\
\cline{2-3}
Opacity Database Resolution  & \multicolumn{2}{c}{20000}
\enddata
\tablecomments{\textbf{MacMad17}: \citet{macdonald_hd_2017}. TLS prescription is described in \S\ref{subsection:coaaddposeidon} } 
\end{deluxetable*} 

\subsubsection{TauREx}\label{subsection:taurex}
To ensure the robustness of the measured abundances and planet parameters from our canonical model, we also performed free-chemistry retrievals on the individual visits described in \S\ref{subsection:visitbyvist} using the retrieval code \texttt{TauREx}. We used publicly available molecular opacities\footnote{\url{https://doi.org/10.5281/zenodo.15495830}} \citep{changeat_2025_15495830} compiled at a resolution of $R=50,000$ by \cite{taurexopacity} following the procedures described in \cite{exomolop}. We assumed an atmosphere composed mainly of hydrogen and helium with a solar He-to-H ratio (\ce{He/H2}=0.17) and trace molecular species. The retrievals included contributions from (i) the same species described in \S\ref{subsection:results}: \ce{H2O} \citep{taurexh2o}, \ce{CH4} \citep{taurexch4}, \ce{CO2} \citep{taurexco2}, \ce{CO} \citep{taurexco}, \ce{NH3} \citep{taurexnh3_1,taurexnh3_2}, \ce{H2S} \citep{taurexh2s_1,exomolop}, CS \citep{CS}, and \ce{CS2}, (ii) Rayleigh scattering, and (iii) collision-induced absorption for \ce{H2-CH4}, \ce{H2-H2}, \ce{H2-He}, \ce{CO2-CO2}, \ce{CO2-CH4}, and \ce{CO2-H2} pairs from HITRAN2024 \citep{hitrancia2024}. We generated absorption cross-sections for \ce{CS2} at $R=50,000$ with the \texttt{PyRaT Bay}\footnote{\url{https://github.com/pcubillos/pyratbay}} package \citep[Python Radiative Transfer in a Bayesian framework;][]{CubillosBlecic2021mnrasPyratBay}, assuming a hydrogen-helium dominated atmosphere (85\% \ce{H2} and 15\% \ce{He}) and using the line-by-line transitions from all isotopologues and the relevant total internal partition sums \citep{TIPS2024} available in HITRAN2024 \citep{HITRAN2024}.  The temperature-pressure profile was assumed to be isothermal and was modeled with 300 atmospheric levels that covered pressures of $10^{-7}-10^3$ bar. We adopted the same priors as the \texttt{POSEIDON} retrievals where possible (e.g., planet and stellar parameters, molecular abundances). To replicate the effects of the \cite{MacDonald2017} aerosol model, we included an opaque cloud deck along with a parameterized form of Mie scattering for small particles (fixed radii of 0.1 \textmu{}m) following the formalism of \cite{Leemie2013}. We performed the parameter estimation using the importance nested sampling algorithm \texttt{MultiNest} with an evidence tolerance of $\Delta \ln Z=1$ and $N=5000$ live points.

\subsubsection{Retrieval Results}\label{subsec:retrieval_results}

For our model selection criteria, we employed the metrics shown in \autoref{tab:retrieval_models}. The models for the co-added spectra that yielded the highest statistical evidence were M2.4 (Clouds+TLS) and M2.5 (Clouds+Haze). Despite the higher statistical evidence of the TLS model (M2.4), we retrieve spot temperatures that are physically too low. Spot temperatures that are $\sim$ 1000 K less than the stellar photosphere temperature are not expected for M-dwarfs \citep{mori_spot_temp} nor from our derived spot temperatures from our white light curve spot modeling. We used the median spot flux ratio derived from the white light curves (see \S\ref{subsec:whitelight}) along with PHOENIX spectra \citep{phoenix} to derive spot temperatures of $\sim3130$ K for visit 1 and $\sim3761$ K for visit 2. Yet, we are retrieving spot temperatures of 2395 K for visit 1 and 2679 K for visit 2. When comparing visit by visit results, model M2.4 (Clouds+TLS) had an unconstrained H$_2$O abundance for visit 1. \textbf{We also compare models on a visit by visit basis and find that aside from H$_2$O in visit 1, the other chemical abundances do not change significantly depending on the model choice (\autoref{appendix:model_comparison}).} In the end, we proceed with the most physically motivated model that also showed a high statistical evidence.

We select the M2.5 model (Clouds+Haze) as our final model due to the following: 

\begin{enumerate}[label=(\roman*)]
    \item uniformity in the derived C/O and metallicity across visits, since we physically do not expect these to change significantly in nearly consecutive visits
    \item challenges retrieving physical posteriors for TLS, including an nonphysical temperature for the spots \citep{mori_spot_temp} that did not match our spot temperatures from our white light curve, and very low spot coverage fractions
    \item the high retrieved bond albedo (assuming efficient uniform heat redistribution) from our consistently lower retrieved T$_{eq}$, which points to further evidence that clouds and hazes may be present in HATS-6 b. Previous GCM studies of this target also further point to an atmosphere dominated by aerosols \citep{kiefer_under_2024, kiefer_under_2024}.
\end{enumerate}

\begin{figure*}
\plotone{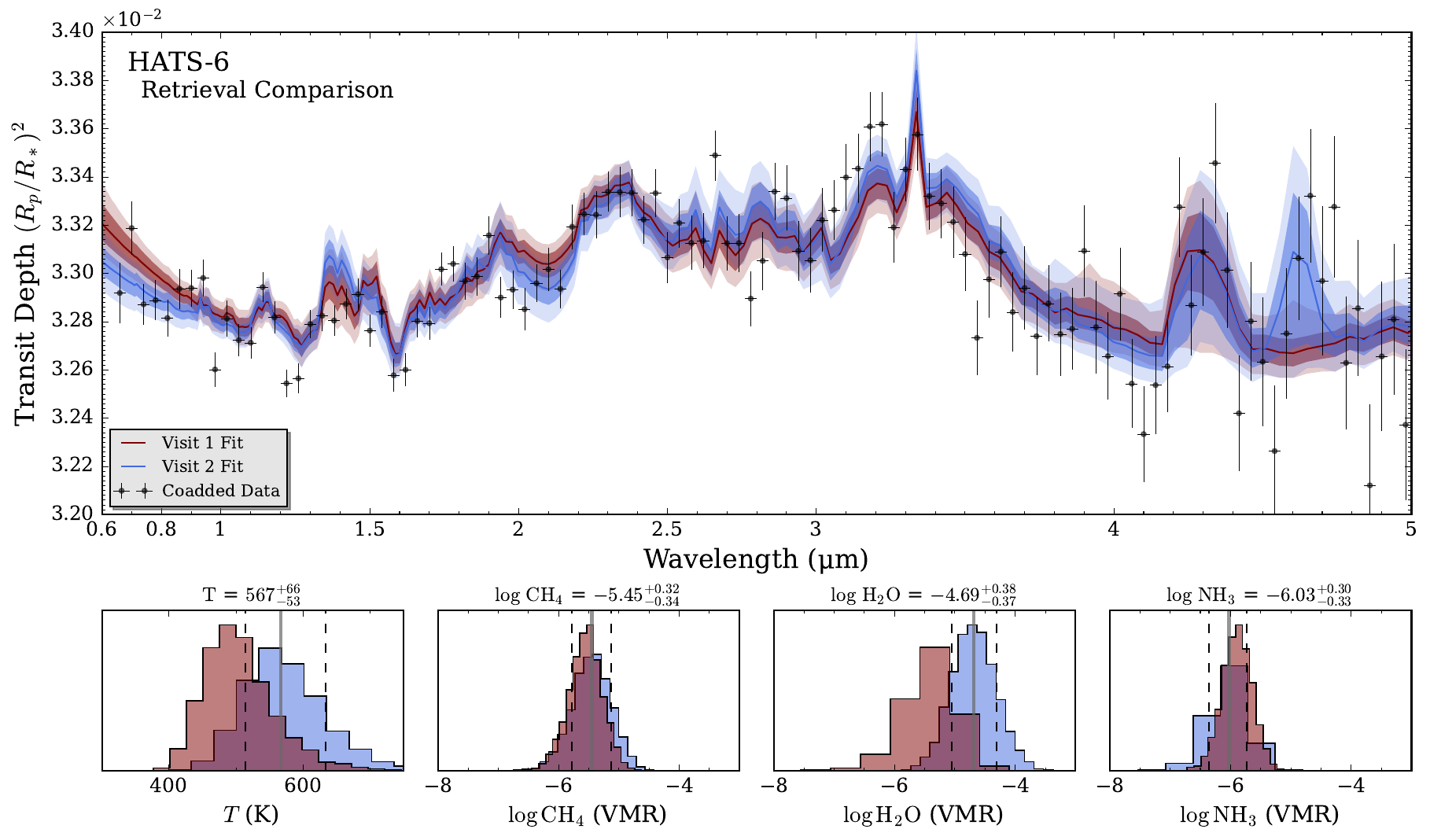}
\caption{Comparison of best-model fits per visit as well as posteriors for the most robustly detected species for our final \texttt{POSEIDON} model M2.5 (Clouds+Hazes). Note that the visit-by-visit retrievals were performed at pixel-level resolution (\S\ref{subsec:spectra}), but are plotted here against the co-added \textbf{Eureka1} spectra for easier visualization. For reference, \autoref{fig:spectral_contribution} shows the individual spectral contribution of each molecular species to the final model shown above.}
\label{fig:visitbyvisit_model}
\end{figure*}

\begin{figure*}
\epsscale{1.2}
\plotone{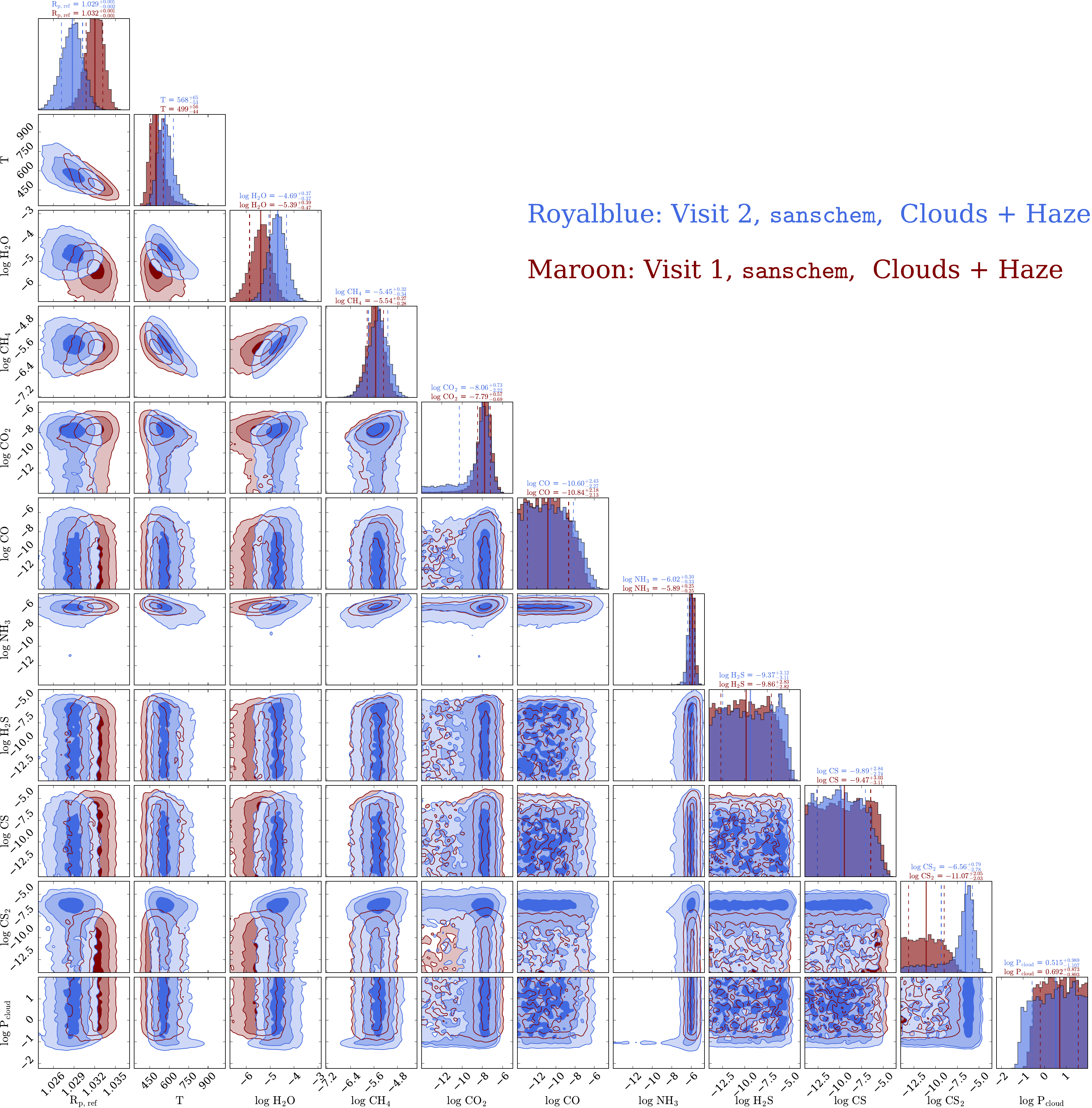}
\caption{Visit-by-visit posteriors for our final M2.5 (cloud+haze) model for our \texttt{POSEIDON} retrievals done with our \textbf{Eureka1} reduction. All chemical abundances retrieved are within 1$\sigma$ of each other, which was not the case for other models in our suite (i.e. cloud+TLS model M2.4).}
\label{fig:visitbyvisity_posteriors}
\end{figure*}

We absolutely do not rule out the presence of TLS, but believe that in the case of HATS-6 b, clouds and hazes are just as physically motivated (see also \S\ref{subsection:atm_discussion}) and produce more consistent visit-by-visit results. We also highlight the nonphysical spot parameters consistently retrieved for our TLS fits as a reason to not rely on our TLS fits, since statistical significance must be accompanied by a physically motivated result. Finally we stress that most of our final chemical abundances are not affected by our model selection (see \autoref{appendix:model_comparison}) and are within 1$\sigma$ of each other for both M2.4 and M2.5, meaning our final conclusions do not change. We therefore proceed with adopting M2.5 (Clouds+Haze) as our final model. We also performed retrievals of our final model on the co-added data in order to compare them with our retrieved abundances from our single-visit fits. These results are summarized in \autoref{tab:retrievals_final}, and we find general agreement between our co-added fit and our visit-by-visit fits. We summarize our final retrieved model in \autoref{fig:visitbyvisit_model} and the retrieved abundances in \autoref{fig:visitbyvisity_posteriors}. In general, we find strong constraints on the abundances of H$_2$O, CH$_4$, and NH$_3$ as well as a slightly lower \teq than expected. We also show the spectral contributions of each molecular species in our final model in \autoref{appendix:spectral_contributions}.

\begin{deluxetable*}{c|l|l|l|l|l|l|l}
\tablewidth{80pt}
\tabletypesize{\small}
\tablecaption{Model selection  \label{tab:retrieval_models}}
\tablehead{
\colhead{Name} & \colhead{Model Description} & \colhead{$\chi^2_{\nu}$} & \colhead{$\ln Z$} & \colhead{$\ln B_{2.1}$} & \colhead{Significance} & \colhead{$\log$[C/O]} & \colhead{$\log$[M/H]}}
\startdata
\multicolumn{6}{l}{Visit 1}  \\ \hline
M1 & Full Chemistry & 1.39 & 2586.34 & \nodata & \nodata & $0.63_{-0.42}^{+0.50}$ & $-2.04_{-0.31}^{+0.35}$ \\ \hline
M2.1 & Sans Chemistry & 1.42 & 2577.68 & 0.0 & \nodata & $0.36_{-0.53}^{+1.16}$ & $-2.60_{-0.20}^{+0.23}$\\
M2.2 & Clouds & 1.41 & 2579.83 & 2.15 & 2.6$\sigma$ & $0.34_{-0.60}^{+1.35}$ & $-2.89_{-0.41}^{+0.32}$\\
M2.3 & Haze & 1.43 & 2578.11 & 0.43 & 1.6$\sigma$ & $0.34_{-0.52}^{+1.00}$ & $-2.59_{-0.20}^{+0.22}$\\
M2.4 & Clouds + TLS & 1.3 & 2598.83 & 21.15 & 6.8$\sigma$ & $1.73_{-0.66}^{+0.78}$ & $-1.76_{-0.63}^{+0.78}$\\
\textbf{M2.5} & \textbf{Clouds + Haze} & 1.36 & 2587.52 & 9.84 &  4.8$\sigma$ & $-0.12_{-0.31}^{+0.41}$ & $-2.31_{-0.27}^{+0.29}$\\
\hline
\multicolumn{6}{l}{Visit 2} \\ \hline
M1 & Full Chemistry & 1.31 & 2590.96 & \nodata & \nodata & $-0.25_{-0.31}^{+0.34}$ & $-1.73_{-0.28}^{+0.31}$\\
\hline
M2.1 & Sans Chemistry & 1.33 & 2585.09 & 0.0 & \nodata & $-0.52_{-0.28}^{+0.30}$ & $-2.03_{-0.26}^{+0.28}$\\
M2.2 & Clouds & 1.29 & 2592.48 & 7.39 & 4.25$\sigma$ & $-0.79_{-0.21}^{+0.23}$ & $-2.24^{+0.36}_{-0.36}$\\
M2.3 & Haze & 1.34 & 2585.5 & 0.41 & 1.59$\sigma$ & $-0.53_{-0.26}^{+0.30}$ & $-2.04_{-0.25}^{+0.28}$\\
M2.4 & Clouds + TLS & 1.26 & 2597.95 & 12.86 & 5.42$\sigma$ & $-0.37_{-0.30}^{+0.38}$ & $-1.69_{-0.33}^{+0.39}$\\
\textbf{M2.5} & \textbf{Clouds + Haze} & 1.26 & 2595.19 & 10.1 & 4.87$\sigma$ & $-0.69_{-0.20}^{+0.23}$ & $-1.83_{-0.34}^{+0.36}$ \\
\enddata 
\tablecomments{The adopted model, M2.5 (see \S\ref{subsec:retrieval_results}), is in bold. The final model was selected both based on statistical evidence as well as physical evidence such as uniformity in C/O and metallicity across visits and the unphysically low retrieved spot temperatures for the models containing TLS (M2.4 Clouds + TLS).  Metallicity is calculated relative to solar. For reference, the Solar C/O in dex is $\log$[C/O] = -0.25.}
\end{deluxetable*}
\begin{deluxetable*}{r|l||l|l|l}
\tablewidth{0.98\textwidth}
\tabletypesize{\small}
\tablecaption{Atmospheric retrieval priors and posteriors for the M2.5 model with clouds and haze using \texttt{POSEDION}. The abundances are the volume mixing ratios for each species. \label{tab:retrievals_final}}
\tablehead{
\colhead{Parameters} & \colhead{Priors} & \colhead{Visit 1} & \colhead{Visit 2} & \colhead{Co-added}}
\startdata
$\mathrm{R}_{\mathrm{p, \, ref}}$ & $\mathcal{U}(0.85, 1.15)$ & $1.032^{+0.001}_{-0.001}$ & $1.029^{+0.001}_{-0.002}$ & $1.031^{+0.001}_{-0.001}$ \\
$\mathrm{T}$ & $\mathcal{U}(200.00, 1500.00)$ & $499^{+56}_{-44}$ & $568^{+65}_{-53}$ & $514^{+35}_{-27}$ \\
$\log \, \mathrm{H_2 O}$ & $\mathcal{U}(-14.00, -1.00)$ & $-5.39^{+0.39}_{-0.47}$ & $-4.69^{+0.37}_{-0.37}$ & $-4.88^{+0.25}_{-0.24}$ \\
$\log \, \mathrm{CH_4}$ & $\mathcal{U}(-14.00, -1.00)$ & $-5.54^{+0.27}_{-0.28}$ & $-5.45^{+0.32}_{-0.34}$ & $-5.38^{+0.18}_{-0.19}$ \\
$\log \, \mathrm{NH_3}$ & $\mathcal{U}(-14.00, -1.00)$ & $-5.89^{+0.25}_{-0.25}$ & $-6.02^{+0.30}_{-0.33}$ & $-6.03^{+0.18}_{-0.19}$ \\
$\log \, \mathrm{CO_2}$ & $\mathcal{U}(-14.00, -1.00)$ & $-7.79^{+0.57}_{-0.69}$ & $-8.06^{+0.73}_{-2.22}$ & $-7.59^{+0.41}_{-0.44}$ \\
$\log \, \mathrm{CO}$ & $\mathcal{U}(-14.00, -1.00)$ & $-10.84^{+2.18}_{-2.13}$ & $-10.60^{+2.43}_{-2.27}$ & $-9.92^{+2.81}_{-2.78}$  \\
$\log \, \mathrm{H_2 S}$ & $\mathcal{U}(-14.00, -1.00)$ & $-9.86^{+2.83}_{-2.82}$ & $-9.37^{+3.12}_{-3.11}$ & $-10.11^{+2.68}_{-2.56}$ \\
$\log \, \mathrm{CS_2}$ & $\mathcal{U}(-14.00, -1.00)$ & $-11.07^{+2.05}_{-2.03}$ & $-6.56^{+0.79}_{-2.78}$ & $-7.14^{+0.73}_{-3.33}$ \\
$\log \, \mathrm{CS}$ & $\mathcal{U}(-14.00, -1.00)$ & $-9.47_{-3.11}^{+3.03}$ & $-9.89_{-2.74}^{+2.84}$ & $-9.41_{-3.13}^{+2.98}$ \\
$\log \mathrm{a}$ & $\mathcal{U}(-4, 8)$ & $0.34_{-0.50}^{+0.58}$ & $-0.18_{-0.31}^{+0.38}$ & $0.43_{-0.38}^{+0.39}$  \\
$\mathrm{\gamma}$ & $\mathcal{U}(-20, 2)$ & $-3.49_{-1.47}^{+1.23}$ & $-1.76_{-0.97}^{+0.62}$ & $-3.66_{-0.95}^{+0.89}$ \\
$\log \mathrm{P}_{\mathrm{cloud}}$ & $\mathcal{U}(-4, 1)$ & $0.69_{-0.89}^{+0.87}$ & $0.51_{-1.11}^{+0.99}$ & $0.75_{-0.88}^{+0.85}$  \\
\hline
$\log[\mathrm{C/O}]$ & \nodata & $-0.131_{-0.323}^{+0.436}$  & $-0.690_{-0.205}^{+0.233}$ & $-0.460_{-0.169}^{+0.199}$ \\
$\log[\mathrm{M/H}]$ & \nodata & $-2.310_{-0.295}^{+0.303}$ & $-1.834_{-0.337}^{+0.356}$ & $-1.990_{-0.209}^{+0.219}$ \\
$\log[\mathrm{C/H}]$ & \nodata & $-2.191_{-0.286}^{+0.292}$ & $-2.048_{-0.323}^{+0.338}$ & $-2.013_{-0.192}^{+0.196}$  \\
$\log[\mathrm{O/H}]$ & \nodata & $-2.337_{-0.459}^{+0.39}$ & $-1.635_{-0.369}^{+0.374}$ & $-1.826_{-0.242}^{+0.246}$ \\
$\log[\mathrm{S/H}]$ & \nodata & $< 0.287$ & $< 0.433$ & $<0.053$ \\ 
$\log[\mathrm{N/H}]$ & \nodata & $-1.957_{-0.249}^{+0.246}$ & $-2.095_{-0.327}^{+0.295}$ & $-2.096_{-0.19}^{+0.18}$ \\
\hline
$\chi^2_{\nu}$ & \nodata & 1.36 & 1.26 & 1.73 \\
$\ln Z$ & \nodata & 2587.52 & 2595.19 & 827.38\\
\enddata 
\tablecomments{All chemical ratios are with respect to solar except C/O which is in native units. For reference, the Solar C/O in dex is $\log$[C/O] = -0.25.}
\end{deluxetable*}

Next, in order to explore whether differences in our data reduction were the culprit for the excess 3 \textmu{}m feature and low retrieved T$_{eq}$, we ran the final model fit on the following data reductions: (i) the \textbf{ExoticJEDI} reduction, and (i) the \textbf{Eureka1} reduction. For both we used the co-added versions, where we binned the spectral light curves down to our wavelength grid before fitting them (See \S\ref{subsection:final_reduction} and \autoref{appendix:bin}). The wavelength grid used for our binned versions was calculated for 40 nm bin widths between 0.6 and 5.3 \textmu{}m, resulting in 120 spectroscopic channels.

\begin{figure*}
\plotone{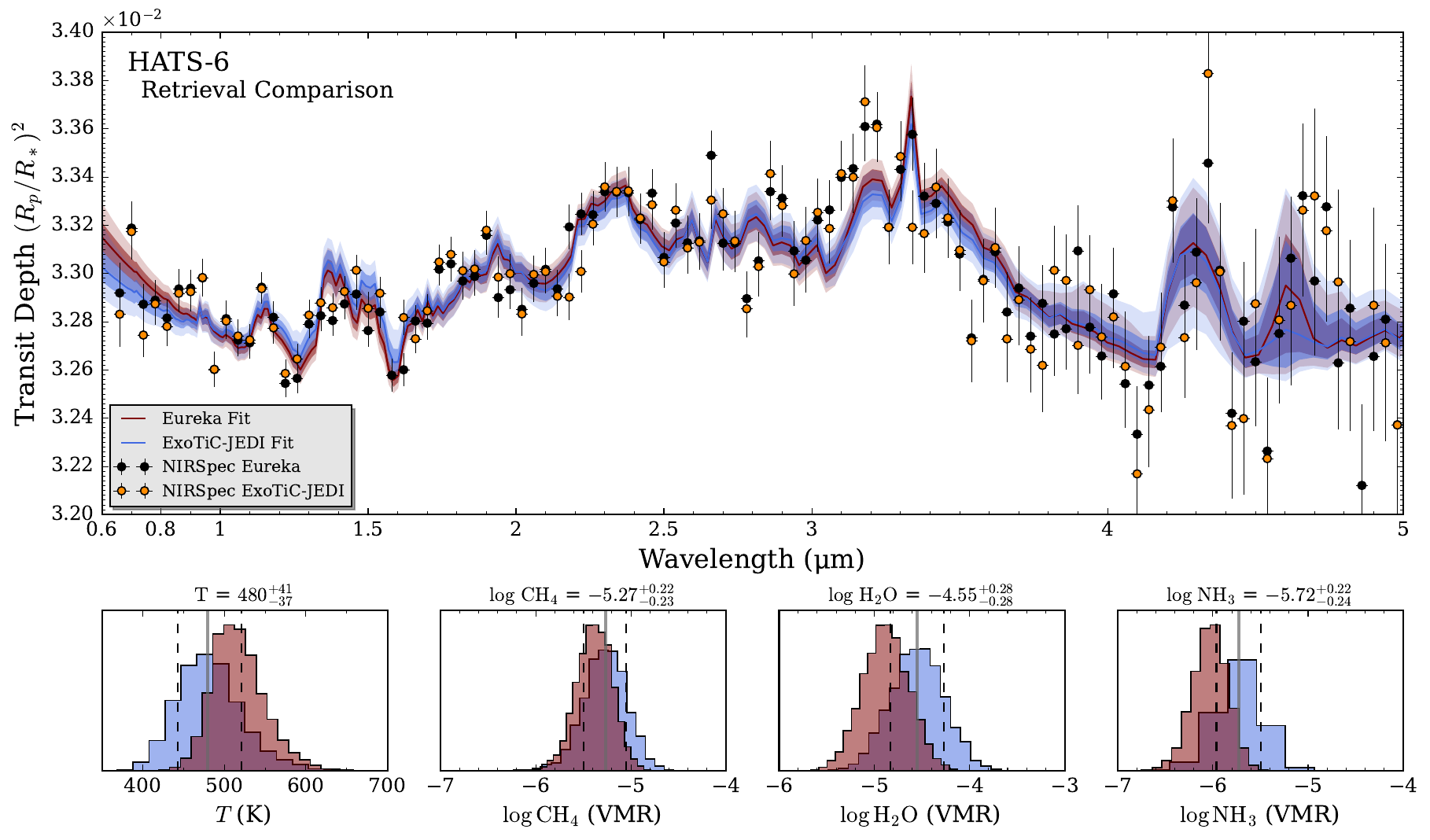}
\caption{Retrieved abundance comparisons of our most robustly detected species for two different data reductions: \textbf{Eureka1} and \textbf{ExoticJEDI}. Our results for H$_2$O, CH$_4$, and NH$_3$ as well as T$_{eq}$ are impervious to data reduction choices. These fits were performed with \texttt{POSEIDON} on the final selected model (M2.5) and the co-added data. For reference, \autoref{fig:spectral_contribution} shows the individual spectral contribution of each molecular species to the final model shown above.}
\label{fig:data_reduction_tests}
\end{figure*}

We compared the results from our co-added \textbf{ExoticJEDI} and \textbf{Eureka1} fit, and found general agreement between all fits. Despite some variation in goodness-of-fit criteria, in terms of retrieved parameter values there were minimal differences. The CO$_2$ abundance did produce tighter posterior constraints specifically in our \textbf{Eureka1} reduction. Yet, retrieved values for temperature, H$_2$O, CH$_4$, NH$_3$, were robustly similar across reductions. We summarize the reduction tests in \autoref{fig:data_reduction_tests}. We also show the spectral contributions of each molecular species in our final model in \autoref{appendix:spectral_contributions}. Ultimately, data reduction differences did not answer the question of why we consistently retrieved low \teq.

We also ran our final model with another retrieval suite \texttt{TauREx} (\S\ref{subsection:taurex}) in order to test the robustness of our results across retrieval suites. Each code has a different treatment for cloud and haze properties which did not allow for direct implementation of the exact same model. For \texttt{TauREx} a parametrization of the code's Mie scattering treatment was used to recreate a Rayleigh scattering slope as implemented in \texttt{POSEIDON} as well as a gray cloud deck. Furthermore, in the case of \texttt{TauREx}, a different reference pressure was used. Despite these differences in implementation, the general results are rather impervious to retrieval choices. Notably, all retrieved abundances are within 3$\sigma$ of each other between all three retrieval codes. The differences between the retrieved abundances can likely be attributed to the differences in the retrieved haze parameters. The fact that we are retrieving a generally lower \teq is completely impervious to retrieval suite choice. 

\begin{figure*}
\plotone{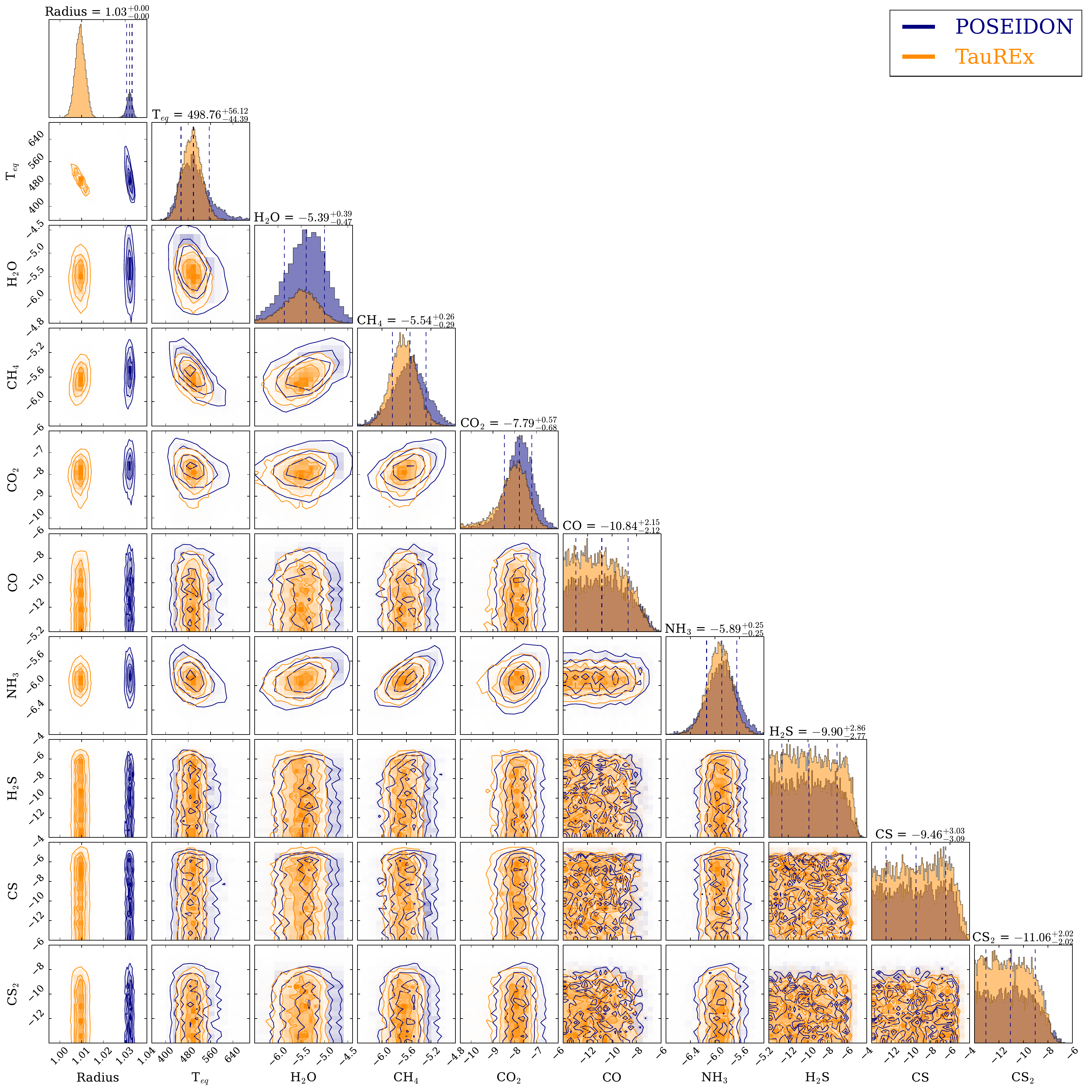}
\caption{Posteriors for our final M2.5 (cloud+haze model) on visit 1 by retrieval suite. The models show close agreement. The fact that we are retrieving a low \teq is completely impervious to retrieval suite choice.}
\label{fig:visit1_posteriors}
\end{figure*}

\begin{figure*}
\plotone{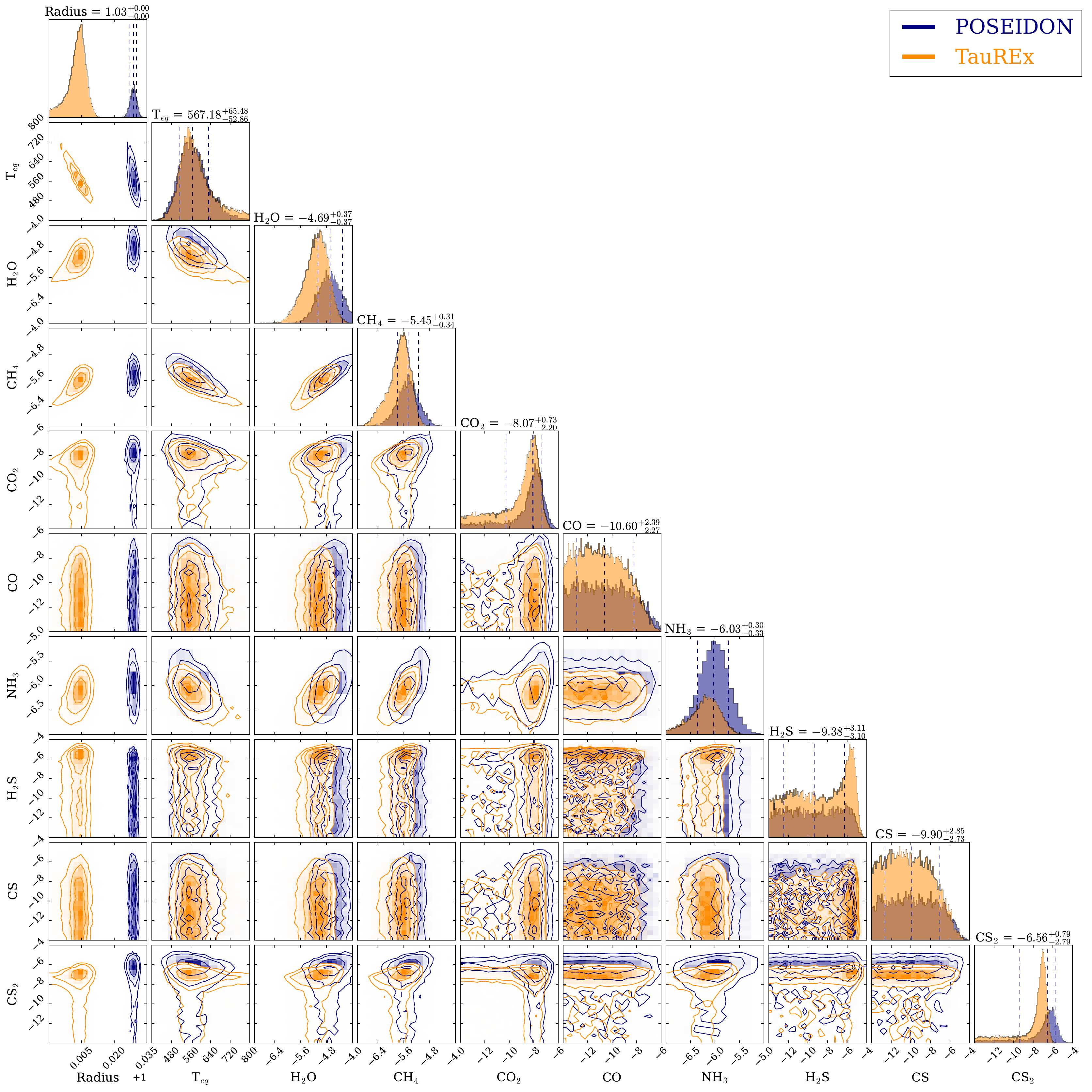}
\caption{Posteriors for our final M2.5 (cloud+haze) model on visit 2 by retrieval suite. The models show close agreement. The fact that we are retrieving a low \teq is completely impervious to retrieval suite choice.}
\label{fig:visit2_posteriors}
\end{figure*}

\subsubsection{Leave One Out Tests}\label{subsec:looretrieval_results}

In order to test the robustness of our best-constrained species, we also performed a leave one out test of Bayesian retrievals with H$_2$O, CH$_4$, CO$_2$, and NH$_3$ using the co-added \textbf{Eureka1} data and our final model (M2.5). In these tests, we re-ran our final model with everything held constant except for the removal of a single molecule in order to determine the statistical significance of each molecule when fitting the model to the data. We then compare the Bayesian evidence of each model against the original model containing all molecules (our M2.5 model) and derive an upper limit for the significance of each molecule. The results of this test are summarized in \autoref{tab:loo}. For each molecule the ln $B_{1,2}$ (with respect to our M2.5 model) was H$_2$O = 22.61, CH$_4$ = 79.24, NH$_3$ = 11.93, and CO$_2$ = 4.24. Following the convention from \citet{Thorngren_bayes}, we consider $\ln B_{1,2}$ $\geq$ 3.0 to be 'Minor Results', $\ln B_{1,2}$ $\geq$ 5.9 to be 'Detections', and $\ln B_{1,2}$ $\geq$ 14.4 as 'Major Discoveries'.

\begin{deluxetable*}{c|l|l|l|l}
\tablewidth{80pt}
\tabletypesize{\normalsize}
\tablecaption{Leave One Out Tests  \label{tab:loo}}
\tablehead{
\colhead{Molecule} & \colhead{$\chi^2_{\nu}$} & \colhead{$\ln Z$} & \colhead{$\ln B_{1,2}$} & \colhead{Significance}}
\startdata
\multicolumn{5}{l}{M2.5: Clouds + Haze} \\ \hline
\textbf{M2.5} & 1.73 & 827.38 & 0.0 & \nodata\\ \hline
H$_2$O & 2.21 & 804.77 & 22.61 & $<$ 7.03$\sigma$\\
CH$_4$ & 3.36 & 748.14 & 79.24 & $<$ 12.80$\sigma$ \\
NH$_3$ & 1.98 & 815.45 & 11.93 & $<$ 5.25$\sigma$\\
CO$_2$ & 1.83 & 823.14 & 4.24 & $<$ 3.38$\sigma$\\
\enddata 
\tablecomments{The reference model is described in \S\ref{subsec:retrieval_results}. The $\ln$ B$_{1,2}$ is calculated following \citet{Sellke2001} where model 1 is our M2.5 model with all molecules included. For reference, we follow the convention from \citet{Thorngren_bayes}, and consider $\ln B_{1,2}$ $\geq$ 3.0 to be 'Minor Results', $\ln B_{1,2}$ $\geq$ 5.9 to be 'Detections', and $\ln B_{1,2}$ $\geq$ 14.4 as 'Major Discoveries'.}
\end{deluxetable*}


\section{Discussion}\label{sec:discussion}

\subsection{A Low Retrieved Equilibrium Temperature}\label{section:low_teq}

In this paper, we perform an extensive comparison of both data reduction and retrieval choices for the atmosphere of HATS-6 b. In our initial free chemistry retrievals, we were retrieving T$_{eq}$ close to 300 K. By iterating through different retrieval tests and physically motivating a final set of chemical species using VULCAN, we were able to push our retrieved T$_{eq}$ closer to what was expected from photometric analysis. Based on our final canonical model (\S\ref{subsec:retrieval_results}), we retrieved an atmospheric \teq of around $514^{+37}_{-30}$ K which is still nearly 200 K lower than expected from the initial characterization \citep{Hartman2015} and follow-up observations of HATS-6 with SpeX \citep{tianjun_mdwarf}. The lower \teq retrieved was persistent across data reductions, visits, and retrieval suite choices. 

HATS-6 b is not the only planet to present unexpected differences in retrieved \teq \citep[e.g.,][]{gressier_jwst-tst_2025}. The T$_{eq}$ = 712 K measurement from \citet{Hartman2015} assumes a zero bond albedo with highly efficient heat transport. Keeping the latter assumption, the bond albedo for our retrieved T$_{eq}$ is A$_B$ = 0.71. This bond albedo is comparable to that of Venus (A$_B$ $\sim$ 0.75) in our own Solar System and higher than that of Jupiter \citep[A$_B$ $\sim$ 0.5;][]{2018NatCo...9.3709L}. The higher bond albedo of Jupiter and other gas giants in our Solar System is due to the presence of condensed molecules such as H$_2$O, CH$_4$, and NH$_3$ \citep{cowan_and_agol_bond_albedo}, therefore a high albedo could be due to the presence of clouds and hazes in the atmosphere of HATS-6 b. This would be consistent with our results from retrievals which generally preferred a cloudy and hazy model over a clear atmosphere. 

Despite this, relating the planet's bond albedo to any physical reflective properties in the atmosphere (i.e. so as to distinguish haze scattering from cloud presence) is a non-trivial matter that requires phase information from the planet \citep{heng_bond_albedo1, heng_bond_albedo2}. Previous statistical studies of Hot Jupiters typically favor bond albedo of $< 0.35$ \citep{cowan_and_agol_bond_albedo,2015MNRAS.449.4192S,2017ApJ...850..154S}. Furthermore, bond albedo is sensitive to the host stellar type \citep[e.g.,][]{marley_1999}, making it possible that GEMS as a population has very different bond albedo than the Hot Jupiter population.

Another explanation for such a significant difference in \teq is the potential presence of limb asymmetries \citep[e.g.,][]{macdonald_why_2020,2022ApJ...933...79W}, since applying 1-D retrieval techniques to data with asymmetric terminators can cause deviations from expected bulk properties and inferred abundances. Given the recent findings for WASP-107 b \citep{murphy_evidence_2024}, we considered limb asymmetries in our analysis, but upon further study we found limited evidence in our data of any limb asymmetries (See \S\ref{sec:datareduction}). Although no strong evidence of limb asymmetries was present, there are possible degeneracies with limb darkening that may be muddling any signs of limb asymmetries. Better stellar models for M-dwarfs would allow us to break these degeneracies in the future and test the presence of limb asymmetries with more certainty. 

We also considered how in-transit spot crossings could be affecting our data, but we also found no chromatic evidence that these produced any differences in our transmission spectra (See \S\ref{sec:datareduction}). Regardless, we acknowledge that HATS-6 b is orbiting an M-dwarf and there could be nuanced stellar activity occurring that we were unable to model with our current capabilities. The M2.4 Clouds+TLS fits had high statistical evidence (\autoref{tab:retrieval_models}), and we do not rule out the possibility that there is stellar activity at play for this planet. Despite this caveat, our TLS fits still yielded similar retrieved temperatures ($\sim$ 500 K) and chemical abundances to our other fits (see \autoref{appendix:model_comparison}). 

We believe that limitations in our understanding of M-dwarfs could potentially be driving some of these unexpected results. We are currently limited in our ability to untangle stellar effects in our spectra (i.e. spots, and limb darkening effects) from atmospheric signals (i.e. limb asymmetries). Therefore, more detailed modeling of M-dwarf activity could help constrain some of these effects for HATS-6 b. Our study highlights the importance of understanding the stellar environment of exoplanetary systems, and the potential value of high-signal observations for breaking some of these degeneracies.  We acknowledge that since this paper was written, NewEra stellar models \citep{newera_phoenix} have been released to the community, a potential step forward in improving constraints on the stellar environments of M-dwarf systems. Furthermore, the Pandora mission \citep{pandora} has launched with plans to constrain spot coverage fractions and the stellar environments of several systems, including M-dwarfs, to observe how the stellar environment affects our observations of exoplanetary atmospheres and how to mitigate these impacts in the infrared \citep{2024SPIE13092E..14Q,pandora}. 

\subsection{An Excess Feature around 3 Microns}\label{section:feature}

\begin{figure*}
\plotone{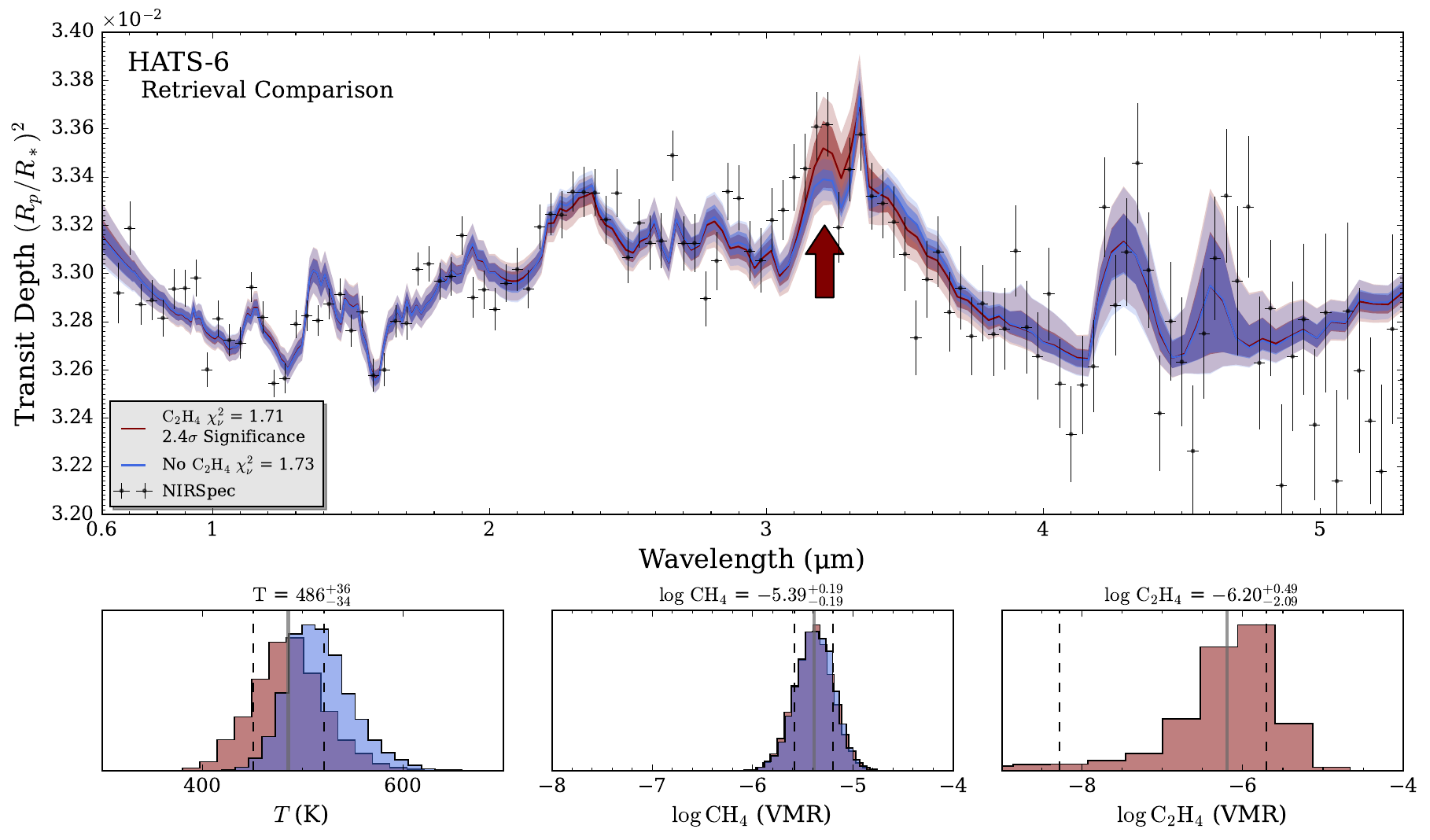}
\caption{Comparisons between models with and without C$_2$H$_4$, and the resulting retrieved values for methane and \teq for the \textbf{Eureka1} reductions. The significance noted is derived from the Bayesian evidence of each model. The presence of C$_2$H$_4$ drives the \teq further down from the expected \teq. We find a slight constraint on C$_2$H$_4$ at VMR abundances of $\sim$ 10$^{-6.20}$ and a slightly lower $\chi_\nu^2$ value of 1.71 in comparison to our model without C$_2$H$_4$ ($\chi_\nu^2$ = 1.73). We note the evidence for this feature is more significant in visit 2 than visit 1. For reference, \autoref{fig:spectral_contribution_c2h4} shows the individual spectral contribution of C$_2$H$_4$ to the model shown above, and the arrow in the plot above highlights the region of strongest contribution from C$_2$H$_4$.
}
\label{fig:c2h4}
\end{figure*}

When performing our initial exploratory retrievals, our full chemistry fits were displaying even lower \teq than our final informed chemistry fits. Through the wavelength cutoff tests (\S\ref{subsection:wavecut}) we found that this phenomenon was driven entirely by an excess absorption feature around 3 microns that was not explained by the CH$_4$ feature alone. This excess slope was consistently fitted by a C$_2$H$_4$ feature inferred by the full chemistry models in all of our fits. Previous studies have also shown that JWST has the capability to detect these hydrocarbons \citep[e.g.,][]{Gasman2022}. Despite the strong constraint on C$_2$H$_4$, our physical modeling (even when accounting for chemical disequilibrium) did not predict the presence of C$_2$H$_4$ or any aliphatic hydrocarbon in an atmosphere of $\sim$ 700 K (\autoref{fig:TPprofile}). The excess slope remained consistent in our fits and because of this, we ran our final retrieval model with C$_2$H$_4$ as a placeholder (see \autoref{fig:c2h4}). For our co-added run, we find a slight constraint on C$_2$H$_4$ at VMR abundances of $\log\mathrm{C_2H_4}=-6.19^{+0.49}_{-2.10}$ and a slightly lower $\chi_\nu^2$ value of 1.71 but also lower $\ln$ Z = 828.98 in comparison to our model without C$_2$H$_4$ ($\chi_\nu^2$ = 1.73; $\ln$ Z = 827.38), which does not point to a strong constrain on this feature in the coadded data, yet we note the evidence for this feature is more significant in visit 2 than visit 1. For visit 1, there is a slight preference for C$_2$H$_4$ (ln B$_\mathrm{ref, C_2H_4}$ of 3.5) which corresponds to a maximum detection significance of 3$\sigma$, at a more tightly constrained VMR of $\log\mathrm{C_2H_4}=-5.57^{+0.41}_{-0.43}$. For visit 2, there is a preference for C$_2$H$_4$ (ln B$_\mathrm{ref, C_2H_4}$ of 7.5) which corresponds to a maximum detection significance of 4.3$\sigma$, at a much less constrained VMR of $\log\mathrm{C_2H_4}=-0.08^{+2.65}_{-3.33}$. Regardless, the abundances retrieved for C$_2$H$_4$ for visit 1 are nearly similar to our retrieved CH$_4$ abundances, which is not expected when the inferred source of hydrocarbons like C$_2$H$_4$ is CH$_4$ photolysis \citep[e.g.][]{Morley2015,Venot2020,tsai_comparative_2021}. The retrieved median abundance for visit 2 is closer to what is expected based on our forward models.

\begin{figure*}
\plotone{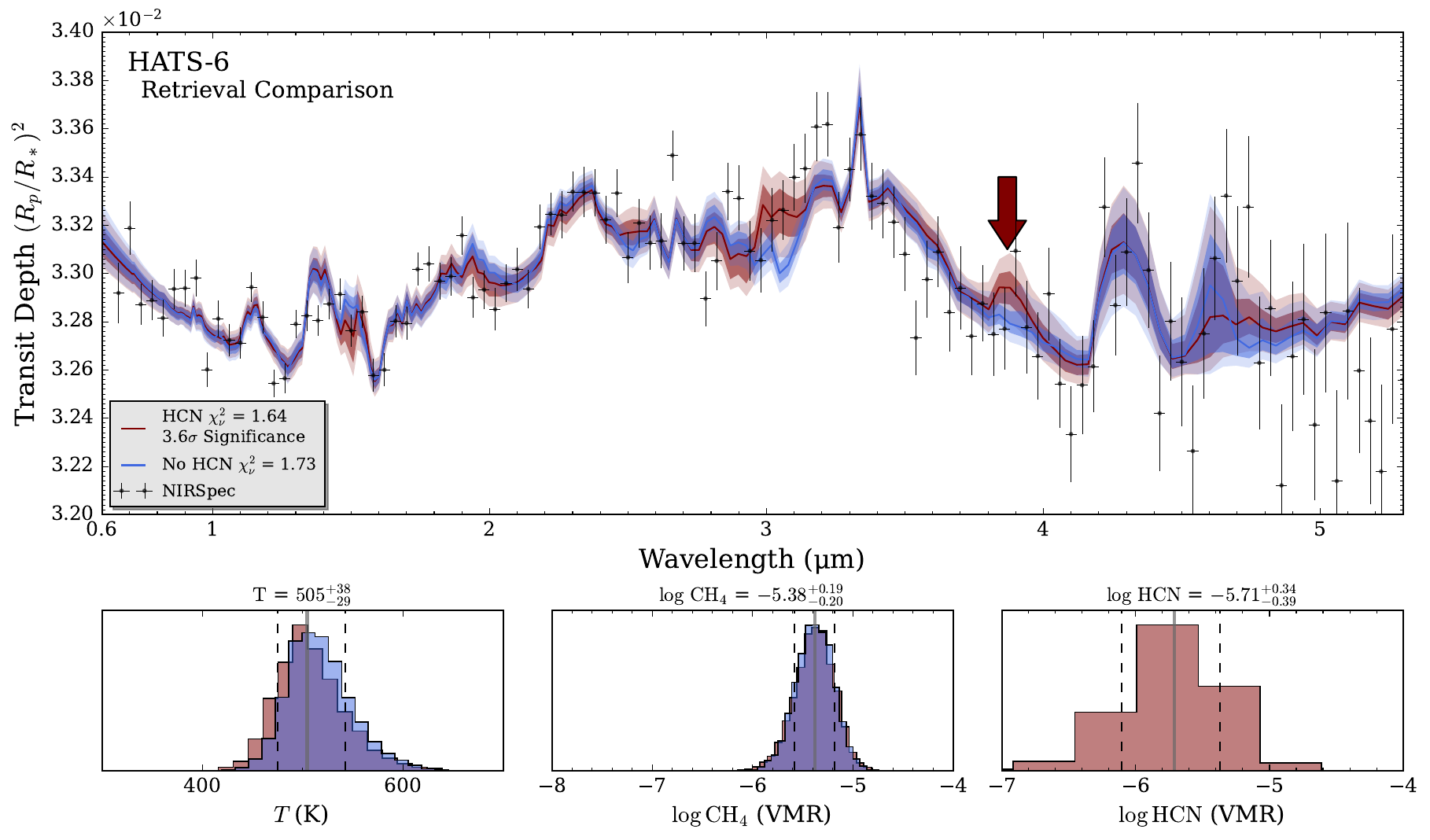}
\caption{Comparisons between models with and without HCN and the resulting retrieved values for methane and \teq for the \textbf{Eureka1} reductions. The maximum significance noted is derived from the Bayesian evidence of each model. HCN provides some amelioration to the discrepancy around the 3 \textmu{}m region, yet introduces a new feature around 3.8--4.0 \textmu{}m. Our retrieved abundance for HCN is consistent with the presence of HCN high up in the atmosphere in our chemical disequilibrium models \autoref{fig:TPprofile}, yet the expected HCN abundance drops abruptly around the pressures best probed by our spectra (\autoref{appendix:pressure}). For reference, \autoref{fig:spectral_contribution_hcn} shows the individual spectral contribution of HCN to the model shown above, and the arrow highlights the region where HCN contributes an excess feature near 4 \textmu{}m.}
\label{fig:hcn}
\end{figure*}

Due to the present degeneracies in the molecular features around the 3 \textmu{}m space, we also performed our final retrieval model with HCN as a placeholder for this excess slope (see \autoref{fig:hcn}). We find a constraint on HCN at VMR abundances of $\log\mathrm{HCN}=-5.71^{+0.34}_{-0.40}$ for our co-added fits and a lower $\chi_\nu^2$ value of 1.64 in comparison to the reference model ($\chi_\nu^2$ = 1.73) with an ln B$_\mathrm{ref, HCN}$ of $\sim$ 4.9 corresponding to a maximum detection significance of 3.58$\sigma$. This suggests that the presence of HCN mitigated the discrepancy between the model and data around the 3 \textmu{}m region as well, yet it introduced a new feature around 3.8--4.0 \textmu{}m that worsened the fit in that region. Our retrieved abundance for HCN is consistent with the presence of HCN high up (P $<$ 10$^{-5}$; see \autoref{fig:TPprofile}) in the atmosphere in our chemical disequilibrium models, yet the HCN abundance drops abruptly around the pressures best probed by our spectra (\autoref{appendix:pressure}).  The excess slope around 3 \textmu{}m being more pronounced in one visit over the other could be due to several effects: i) variability in the stellar environment, ii) variability in the atmosphere, iii) higher precision of the data in our first visit. 

Previous GCM studies of HATS-6 b \citep[e.g.,][]{kiefer_under_2024,kiefer_why_2024} have also shown a possible CH$_4$ enhancement of $\sim$ 200 ppm in the evening terminator, which could point towards limb asymmetries being the culprit of this excess. Although these tests showcase possible explanations for the excess slope in the 3 \textmu{}m region of our data, we are limited not only by the available chemistry of our standard opacity lists, but also by a degenerate wavelength region that hosts multiple hydrocarbon features \citep[e.g.,][]{niraula_hydrocarbons}. In the end, we do not draw a conclusion on the presence of a feature in this region, but rather present to the reader the complexities that arise in understanding the chemistry of a warm Saturn around an M-dwarf with our current chemical and stellar models. 

This unexplained excess artifact in the data motivates further additional observations of HATS-6, especially in wavelength ranges where there are fewer degeneracies between features of hydrocarbon species. MIRI wavelength ranges have the capacity to disentangle different hydrocarbon features \citep{Gasman2022} and offer more information regarding the possible presence of mineral clouds in the atmosphere of HATS-6 b \citep{kiefer_under_2024, kiefer_why_2024}, making multi-wavelength observations of this target a worthwhile endeavor. Additionally, due to the degeneracies that arise between limb darkening and the possible presence of limb asymmetries, further stellar modeling of M-dwarfs that allow us to draw upon more informed priors for our stellar modeling will be crucial in understanding the atmosphere of these GEMS moving forward. 

\subsection{A low-metallicity and sub-solar C/O atmosphere with methane and ammonia}\label{subsection:atm_discussion}

Our final model (M2.5 Clouds+Haze) presents a low-metallicity ($\log\mathrm{[M/H]}=-1.99^{+0.2}_{-0.2}$) and sub-solar C/O ($\log[\mathrm{C/O}]$ = $-0.46^{+0.2}_{-0.2}$) atmosphere with the presence of clouds and haze. The metallicity and C/O values were derived from our best-fit free retrieval (M2.5) using the function \texttt{volume\_mixing\_ratios2metallicity} in \texttt{petitRADTRANS} \citep{molliere_petitradtrans_2019} and the co-added results. C/O is largely unconstrained for most of our retrievals which makes it difficult to compare results in C/O across models. In contrast, the low metallicity was persistent across all runs (see \autoref{tab:retrieval_models}).  Sub-solar C/O has been identified in other low-density Saturn-mass targets such as WASP-39 b \citep{wasp39_co1,wasp39_co2}, which could be due to chemical disequilibrium, or in the case of HATS-6 b, also a muting of the carbon features due to the presence of hazes or clouds. Our cloud and haze parametrization approaches were rather simple, mostly allowing the retrieval flexibility when fitting the data. Yet, the possibility of clouds and hydrocarbon hazes in the atmosphere of HATS-6 b is physically motivated by a low retrieved temperature, which points to a high albedo, as well as previous GCM studies of this target \citep{kiefer_under_2024, kiefer_why_2024} that showcase the presence of complex mineral clouds. 

For our visit-by-visit fits, we find significantly different C/O ratios between visits. For all models except M2.5 (Clouds+Haze), we consistently find super-solar C/O for visit 1 and sub-solar C/O for visit 2. For model M2.5 we find C/O ratios that are 1$\sigma$ away from each other, which is what further motivated our choice for this model over any others. Our other best-fit model M2.4 (Clouds+TLS) points to the presence of TLS effects on our spectra. If TLS is indeed the cause between the visit-by-visit differences, then the presence of TLS could be biasing the C/O ratio to be higher in our first visit. This is due to the modeled effects of spot inhomogeneities on the surface of our star affecting the reliable measurement of water abundance. We show a direct comparison of model M2.4 and model M2.5 abundances as well as detail priors and posteriors for model M2.4 (clouds + TLS) in \autoref{appendix:model_comparison}. Water is our most significantly abundant oxygen-bearing species that contributes to our derived C/O and any lack of constraint on its features due to TLS is directly correlated to deriving much higher C/O ratios. 

Therefore, several lines of evidence point to a possible sub-solar C/O for this planet, but we caution that the differences in these values by visit also motivates further follow-up, especially in the context of the possible presence of TLS. One way to explore these visit-by-visit differences would be to perform multi-visit free retrievals. Stellar surfaces are prone to change, so fitting the TLS parameters independently per visit while jointly fitting the atmospheric abundances could be a way to glean more insight on the differences we are seeing in our visit-by-visit data. This also urges the need for tools that can reliably perform this kind of retrieval, as multi-visit observations of exoplanet atmospheres are becoming increasingly common in the era of JWST. 

When speaking of hazes, hydrocarbon aerosols are expected to dominate giant planet atmospheres with \teq $<$ 950 K \citep{Gao2020} due to the presence of CH$_4$ as a reservoir species \citep{Fortney2020}. Studies have shown that CH$_4$ photolysis could lead to the presence of hydrocarbons such as acetylene (C$_2$H$_2$) and ethene (C$_2$H$_4$) in a planet's atmosphere \citep[e.g.][]{Venot2015,Molaverdikhani2019}. These molecules are important precursors for the presence of complex haze \citep{Morley2015,Venot2015,tsai_comparative_2021}. HATS-6 b sits in a parameter space where the presence of haze cannot be ruled out, and given the excess feature (\S\ref{section:feature}) around 3 \textmu{}m, the low retrieved T$_{eq}$ which points to a high bond albedo (\S\ref{section:low_teq}), and the possibility of an active stellar environment, there is a chance that this planet may contain complex hydrocarbon haze. 

When testing for the effects of clouds in our spectra (see \S\ref{subsection:virga}), we were limited by the assumption of spherical particulates and the available species in \texttt{virga} 1.0, as well as our NIRSpec wavelength range (0.6 $-$ 5.3 \textmu{}m. Yet, previous literature has also modeled the presence and formation of clouds in the atmosphere of HATS-6 b \citep{kiefer_why_2024, kiefer_under_2024} and have found that HATS-6 b lies in a parameter space that is amenable to the formation of more complex cloud species, for which effects are strongest in wavelengths $>$ 8 \textmu{}m. For warm Saturns, atmospheric temperatures are expected to be more uniform in comparison to their hot Jupiter counterparts \citep{christie2022,helling2023}, and due to their lower global temperatures, efficient horizontal heat transport is expected for these planets \citep{kataria2016,komaceck2016}. These create amenable conditions for global homogeneous cloud coverage. GCM models \citep{kiefer_under_2024} found observable differences between the morning and evening terminators for cloud deck height, CH$_4$, and H$_2$O features on the order of 100--200 ppm. They found evidence for a cloud-driven temperature inversion at pressures lower than 10$^{3}$ bars, around the same pressure floor that our transmission spectrum stops probing. This kind of temperature inversion has been predicted for atmospheres with extended cloud-decks or photochemical haze  \citep{heng2012,morley2012,steinrueck2023}. Observing such a temperature inversion, especially if its cloud-deck induced, is beyond the constraints provided by transmission spectroscopy. Yet, \citet{kiefer_under_2024} mention that the presence of a visible CH$_4$ feature in the transmission spectra of HATS-6 b could indicate that there is strong vertical mixing and upwelling of CH$_4$ in the atmosphere. The GCM study by \citet{kiefer_why_2024} ultimately presents HATS-6 b as a mostly cloudy atmosphere, with a temperature inversion `anti-greenhouse' effect that causes lower temperatures in the deeper parts of the atmosphere, and a possibly observable limb asymmetry of the CH$_4$ feature around 3\textmu{}m. We note that this limb asymmetry could be a potential explanation for our excess feature (\S\ref{section:feature}). In order to verify this, improved stellar models will be needed in order to break degeneracies between stellar effects (i.e. limb darkening) and differences in the transmission spectra arising due to limb asymmetries. 

We ultimately placed reliable constraints for four gas species in the atmosphere of HATS-6 b: H$_2$O at a VMR of $\log\mathrm{H_2O}=-4.88^{+0.25}_{-0.24}$ and $\ln B_{1,2}$=22.61; CH$_4$ at a VMR of $\log\mathrm{CH_4}=-5.38^{+0.18}_{-0.19}$ and $\ln B_{1,2}$=79.24; NH$_3$ at a VMR of $\log\mathrm{NH_3}=-6.03^{+0.18}_{-0.19}$ and $\ln B_{1,2}$=11.93; and CO$_2$ at a VMR of $\log\mathrm{CO_2}=-7.59^{+0.41}_{-0.44}$ and $\ln B_{1,2}$=4.24. Following the convention from \citet{Thorngren_bayes}, we consider $\ln B_{1,2}$ $\geq$ 3.0 to be 'Minor Results', $\ln B_{1,2}$ $\geq$ 5.9 to be 'Detections', and $\ln B_{1,2}$ $\geq$ 14.4 as 'Major Discoveries'. Model comparison metrics for these tests are summarized in \autoref{tab:loo}. 

\subsection{Bulk Properties of HATS-6}

We aimed to constrain the bulk properties of HATS-6b using interior and thermal evolution models. Here, we used the open-source code \texttt{GASTLI} \citep{2024A&A...688A..60A}, which calculates adiabatic, one-dimensional hydrostatic models at a given intrinsic temperature. The thermal evolution is then constructed from the static models by integrating the luminosity equation, connecting the intrinsic temperature to the planetary age.

The common approach to constrain planetary bulk compositions employs thermal evolution models in conjunction with measured values of planetary mass, radius and age \citep[e.g.,][]{guillot_composition_2008,thorngren_mass-metallicity_2016}. However, this method is challenging if the planetary radius is larger than what is predicted from standard evolution models for a pure hydrogen-helium composition. 

\begin{figure}
    \centering
    \includegraphics[width=0.9\linewidth]{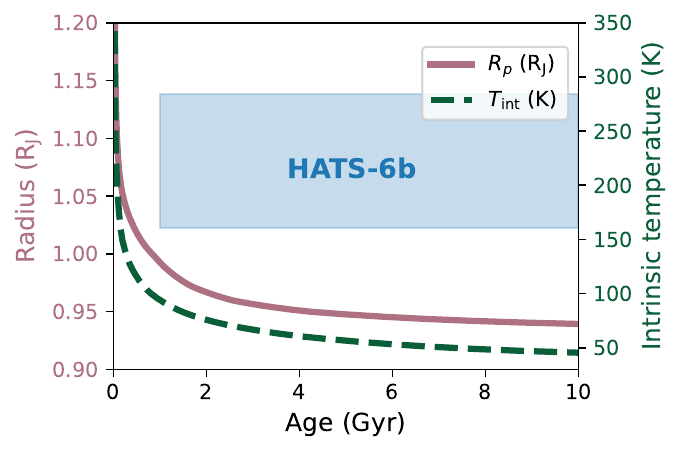}
    \caption{Radius (solid mauve line) and intrinsic temperature evolution (dashed dark green line) and the observed planet radius with $\pm1\sigma$ uncertainties (blue box). The cooling curve was calculated assuming a pure hydrogen-helium composition in proto-solar ratios without heavy elements. This demonstrates that HATS-6b is inflated beyond what standard cooling models predict. Explaining its observed radius requires an additional energy source.}
    \label{fig:thermal_evolution}
\end{figure}

To demonstrate this is the case for HATS-6b, we calculated the cooling of the planet using its measured mass, zero-albedo equilibrium temperature, and assumed a pure hydrogen-helium composition and no central heavy-element core. This yielded an upper bound for the predicted radius from cooling curves \footnote{This is a good estimate for the upper bound as long as there are no inverted heavy-element gradients in the envelope. A hypothetical heavy-element rich atmosphere sitting on top of a pure hydrogen-helium envelope would slow down the cooling due to the increased opacity in the atmosphere.}. The results for the planetary radius and intrinsic temperature together with the observational constraints on the planetary radius and age are shown in Fig. \ref{fig:thermal_evolution}. For the system age, we intentionally assume a large range from 1 to 10 Gyr to account for the likely large age uncertainty of previous estimates. Adopting the system age from \citet{Hartman2015} of 8.1 $\pm$ 4.3 Gyr would strengthen our conclusion. Matching the observed radius within $1\sigma$ is not possible even for ages at about 1 Gyr. This suggests that the planet is inflated, and that it likely has an additional energy source that is not accounted for in standard thermal evolution models.

The inferred intrinsic temperature from our atmospheric forward modeling (see \S\ref{sec:forwardmodel}) opens up another path to estimate the interior properties for inflated planets: Rather than using thermal evolution models, we can calculate static interior models at specific intrinsic temperatures and constrain the bulk metallicity without using the age \citep[see also][]{sing_warm_2024}. The atmospheric forward models in this work constrain the intrinsic temperature to about 50 K for best-fit chemical equilibrium and disequilibrium models. Given the inflated radius of HATS-6b, from an evolution point of view, a higher intrinsic temperature would be worth exploring as well. For estimating the bulk composition of HATS-6b, we therefore test both the intrinsic temperature retrieved with the forward models (50 K), as well as a hotter intrinsic temperature of 300 K.

We first calculated a four-dimensional grid of static interior models with \texttt{GASTLI} where the dimensions were the planetary mass ($M_p$), atmospheric metallicity (log[M/H]), intrinsic temperature ($T_{\rm{int}}$) and core mass fraction (CMF). Atmospheric models below log[M/H] = -2 were extrapolated. We used a constant C/O ratio of 0.10, which is the lower limit by the atmospheric grid.

The grid of the evolution models was then used to construct a linear four-dimensional interpolator $f(M_p, T_{\rm{int}}, \rm{CMF}, \log(\rm{M/H})$) that calculates the planetary radius. This was used as a forward model in a Markov Chain Monte Carlo approach to estimate posterior distributions of the model parameters. To constrain the parameters within the grid of thermal evolution models, we used uninformed uniform priors for the mass (0.08 to 0.6 M$_{\rm{J}}$), intrinsic temperature (30 to 400 K), core mass fraction (0 to 0.6) and atmospheric metallicity (-3 to 0). The statistical model was further constrained by the observed parameters by using products of Gaussian likelihoods for the planetary mass, radius, retrieved atmospheric metallicity and intrinsic temperature.

\begin{figure*}[t]
    \centering
    \includegraphics[width=\linewidth]{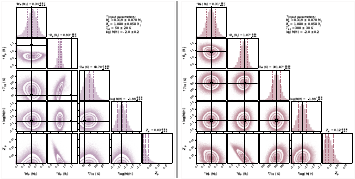}
    \caption{Estimates for bulk metallicity $Z_p$ using measurements of planet mass, radius, intrinsic temperature, and atmospheric metallicity. \textbf{Left)} The fits assumed a Gaussian prior on $T_{int}$ centered at 50 K based on best-fit chemical disequilibrium models and \textbf{Right)} 300 K, which best matched the observed radius. The input parameters for the statistical model are marked with a $^\dagger$ and listed in the each panel.}
    \label{fig:interior_retrieval}
\end{figure*}

The posterior distributions of the model parameters are shown in \autoref{fig:interior_retrieval}. Instead of the core mass fraction, we show the total planet metallicity $Z_p$ and note that since the atmospheric metallicity is small, $Z_p \simeq \rm{CMF}$. We find that HATS-6b contains $Z_p = 0.14^{+0.09}_{-0.08}$ of heavy elements, which is three orders of magnitude larger than the mean atmospheric metallicity of $Z_{\rm{atm}} \simeq 10^{-4}$ (corresponding to log[M/H] = -2). The extreme contrast between the interior and atmospheric metallicities is significant evidence that the planet is not fully mixed.

Previously reported bulk-metallicity estimates retrieved with the \texttt{planetsynth} thermal evolution models \citep{muller_synthetic_2021} were consistent with either no or very few heavy elements. However, on top of using a smaller planetary radius, the successful interior retrievals were likely related to samples with a young age and a radius that deviates more than 1$\sigma$ from the observed mean value. If HATS-6b is indeed inflated as Fig. \ref{fig:thermal_evolution} suggests, previous bulk-metallicity estimates are perhaps unreliable, and those presented in this work should be more realistic.

We note that there are clearly theoretical uncertainties attached to the evolution models used to infer the composition of giant exoplanets \citep[see, e.g.,][for a review]{muller_warm_2023}. One major uncertainty relates to the equations of state and opacities or atmospheric models \citep{thorngren_mass-metallicity_2016,poser_effect_2019,muller_theoretical_2020,howard_giant_2024}. Another important caveat is that the \texttt{GASTLI} models assume an adiabatic core-envelope structure and do not account for potential dilute-core structures in the interior \citep[e.g.,][]{helled_fuzziness_2017,wahl_comparing_2017,debras_new_2019,muller_challenge_2020,miguel_jupiters_2022,howard_jupiters_2023} or inverted heavy-element gradients in the atmosphere \citep{howard_exploring_2023,muller_can_2024} that could affect the thermal evolution. While these uncertainties would affect the inferred bulk metallicity of HATS-6b, they would not alter our main conclusion that there is a gigantic divide between the bulk and atmospheric metallicities.

\subsection{HATS-6 b in the context of GEMS}

HATS-6 b is part of a survey that aims to constrain how atmospheric and bulk metallicity of GEMS compare to giant planets around FGK stars. TOI-5205 b \citep{canas_gems_2025}, HATS-75 b \citep{2026AJ....171..294A}, and TOI-5293A b \citep{toi5293} have all shown significant signs of stellar contamination in their transmission spectra. In contrast, although HATS-6 b presents some evidence of stellar heterogeneity in the white light curve, we find that the transmission spectra for HATS-6 b is not as deeply affected by the TLS as the previous two planets. Despite stellar contamination, TOI-5205 b, HATS-75 b, and TOI-5293A b are sub-solar metallicity and super-solar C/O atmospheres. We note that our HATS-6 b results are consistent with the low atmospheric metallicity and high bulk metallicity measurements for these other three planets. Although our derived C/O ratios are hard to consolidate across visits, our final model yields a $\log[\mathrm{C/O}]$ of $\sim -0.46$, which is sub-solar in contrast to that of the other targets (TOI-5205 b: 3$\sigma$ lower limit $\log[\mathrm{C/O}] \gtrsim 0.09$ ;  HATS-75 b: $\log[\mathrm{C/O}] \sim 1.04$; TOI-5293A b: $\log[\mathrm{C/O}] \sim 1.23$). In general, the derived C/O for HATS-6 b trends lower than its GEMS counterparts; this could possibly be due to better constraints on oxygen-bearing species (namely H$_2$O), due to the lack of strong stellar contamination in HATS-6 b observations, but also due to a general muting of the CH$_4$ feature due to clouds. For all planets, robust detections of CH$_4$ have been made. 

For HATS-6 b, in comparison to other GEMS survey targets, uncertainties and degeneracies remain that are much harder to untangle since we do not have strong constraints on the TLS effect. The clearest example of this is the excess feature seen around 3 \textmu{}m, which could be due to unprecedented chemistry, limb asymmetries, haze, or changes in the stellar environment itself. 


\section{Conclusion}\label{sec:conclusion}

In this paper, we present the NIRSpec PRISM transmission spectra of HATS-6 b, a warm Saturn-mass and Jupiter-size planet orbiting an M-dwarf. We reduced our data with two different pipeline implementations and tested our sensitivity to in-transit spot crossings and limb asymmetries. We find no strong evidence for limb asymmetries, and find that our transmission spectra is not affected by chromatic effects from in-transit spots. We then generated forward models of chemical equilibrium and disequilibrium expectations for HATS-6 b and used those to inform our priors for free-chemistry retrievals. We also ran chemical equilibrium retrievals which failed to capture features in the data and yielded atmospheric metallicities that hit the lowest prior bounds, pointing towards the presence of disequilibrium chemistry in the atmosphere of HATS-6 b.

\begin{itemize}
\item[$\bigstar$] From informed free-chemistry retrievals, we infer a sub-solar metallicity($\log\mathrm{[M/H]}=-1.99^{+0.2}_{-0.2}$) and sub-solar C/O ($\log[\mathrm{C/O}]$ = $-0.46^{+0.2}_{-0.2}$; where solar $\log[\mathrm{C/O}]$ is -0.25) atmosphere with the presence of clouds and haze. We find a lower than expected T$_{eq}$ ($514^{+35}_{-27}$K) for HATS-6 b which permeated all retrieval tests. 
\end{itemize}

The presence of clouds and haze is physically consistent with previous GCM studies of HATS-6 b \citep{kiefer_under_2024, kiefer_why_2024} as well as with the high bond albedo derived from the consistently low T$_{eq}$ retrieved for this planet. Assuming our retrieved T$_{eq}$ and efficient heat transport yields a bond albedo of A$_B$ of 0.71, close to that of Venus in our own atmosphere, and higher than Jupiter (A$_B$ = 0.5). The higher bond albedo of Jupiter and other gas giants in our Solar System are due to the presence of condensed molecules such as H$_2$O, CH$_4$, and NH$_3$ \citep{cowan_and_agol_bond_albedo}, therefore a high albedo could be due to a presence of clouds and hazes in the atmosphere of HATS-6 b.
 
 \begin{itemize}
 \item[$\bigstar$] We confidently constrain four chemical species in the atmosphere of HATS-6 b: H$_2$O ($\log\mathrm{H_2O}=-4.88^{+0.25}_{-0.24}$), CH$_4$ ($\log\mathrm{CH_4}=-5.38^{+0.18}_{-0.19}$), NH$_3$ ($\log\mathrm{NH_3}=-6.03^{+0.18}_{-0.19}$), and CO$_2$ from the 4.3 \textmu{}m feature ($\log\mathrm{CO_2}=-7.59^{+0.41}_{-0.44}$).
 \end{itemize}
 
Metrics for the leave one out tests for these molecules are summarized in \autoref{tab:loo}.
 
 \begin{itemize}
 \item[$\bigstar$] This is one of the few NH$_3$ detections in an exoplanetary atmosphere to date \citep{ammonia_direct, ammonia_groundbased1, ammonia_groundbased2}, and only the second detection of ammonia from transmission spectroscopy. 
 \end{itemize}
 
 Constraining NH$_3$ could offer potential insights into the bulk nitrogen content of a planet from which we could derive an N/O ratio \citep{nitrogen_bulk}. Deriving this ratio could be a potential tracer for planetary formation alongside measured C/O, since N/O has been shown to monotonically increase with radial distance \citep[e.g.,][]{cridland_nitrogen,piso_nitrogen}. The first detection of ammonia through transmission spectroscopy was in the atmosphere of WASP-107 b \citep{welbanks_wasp107}. WASP-107 b is a Neptune-mass but Jupiter-size exoplanet with an inflated atmosphere and a slightly higher T$_{eq}$ of 770 K which has presented evidence for limb asymmetries and high internal temperatures ($>$ 345 K) suggestive of tidally driven inflation.

\begin{itemize}
\item[$\bigstar$] While constraining the bulk properties of HATS-6 b, we find that HATS-6 b likely has an inflated radius relative to what is expected from thermal evolution models, which would require an additional energy source to explain its observed radius, similar to the case of WASP-107 b. 
\end{itemize}

We also find a strong discrepancy between the observed atmospheric metallicity and the derived bulk metallicity (which is three orders of magnitude higher). This would suggest that HATS-6 b, similar to the other GEMS TOI-5205 b and HATS-75 b, is not fully mixed, analogous to gas giants in our own Solar System.

We note an excess feature in the data around 3 \textmu{}m and note that the strength of this excess feature is visit-dependent. Although free chemistry retrievals attributed C$_2$H$_4$ to this excess feature, we also explore HCN due to its chemical plausibility in the atmosphere of HATS-6 b.

\begin{itemize}
\item[$\bigstar$] We retrieve the C$_2$H$_4$ feature at a VMR of $\log\mathrm{C_2H_4}=-6.19^{+0.49}_{-2.10}$ with a maximum detection significance of 2.38$\sigma$, and HCN at a VMR of $\log\mathrm{HCN}=-5.71^{+0.34}_{-0.40}$ with a maximum detection significance of 3.59$\sigma$ as possible explanations for the excess feature around 3 \textmu{}m. Due to the plethora of hydrocarbon features in this wavelength region \citep[i.e.][]{niraula_hydrocarbons}, we do not draw any conclusions from this analysis. Rather, we urge follow-up observations of this target at wavelength regions that can help break this degeneracy.
\end{itemize} 
 
 Finally, we stress the need for better constraints on M-dwarf stellar environments (e.g. stellar age and rotation periods, spot coverage fractions, etc.), in order to break degeneracies in the transmission spectra of high-signal targets such as GEMS. This need is especially crucial for targets that do not present obvious signs of stellar contamination in the transmission spectra, such as the huge TLS slopes seen in other GEMS targets \citep[e.g. TOI-5205][]{canas_gems_2025} which have helped constrain the stellar photosphere. We also recommend the development of spectral retrieval tools that allow for multi-visit fits in order to better coalesce differences across visits in the potential presence of TLS and aerosol contamination, and recommend HATS-6 b for a case study in order to test the usefulness of these tools in the future. 


\section*{Acknowledgments}
We thank the anonymous referee for valuable feedback which has improved the quality of this manuscript. 
We thank Adriana Kuehnel for her assistance in reducing JWST data as part of the GEMS JWST survey. 

This work is based on observations made with the NASA/ESA/CSA James Webb Space Telescope. The data were obtained from MAST at STScI, which is operated by the Association of Universities for Research in Astronomy, Inc., under NASA contract NAS 5-03127 for JWST. These observations are associated with program \#3171. Support for program \#3171 was provided by NASA through a grant from the Space Telescope Science Institute, which is operated by the Association of Universities for Research in Astronomy, Inc., under NASA contract NAS 5-03127.

The JWST data presented in this paper were obtained from MAST at STScI. The specific observations analyzed can be accessed via \dataset[10.17909/22d4-8136]{https://doi.org/10.17909/22d4-8136}. Support for MAST for non-HST data is provided by the NASA Office of Space Science via grant NNX09AF08G and by other grants and contracts.

CIC acknowledges support by NASA Headquarters through (i) an appointment to the NASA Postdoctoral Program at the Goddard Space Flight Center, administered by ORAU through a contract with NASA and (ii) under award number 80GSFC24M0006. S-.M. T. is supported by the National Science and Technology Council (grants 114-2112-M-001-065-MY3) and an Academia Sinica Career Development Award (AS-CDA-115-M03). S-.M. T. acknowledges the access to high-performance computing facilities provided by Academia Sinica Institute of Astronomy and Astrophysics (ASIAA).

Resources supporting this work were provided by the NASA Scientific Computing project through the NASA Center for Climate Simulation (NCCS) at Goddard Space Flight Center. This content is solely the responsibility of the authors and does not necessarily represent the views of the NCCS.

Goddard affiliates acknowledge support from the GSFC Sellers Exoplanet Environments Collaboration (SEEC), which is supported by NASA's Planetary, Astrophysics and Heliophysics Science Divisions’ Research Program.

\vspace{5mm}
\facilities{JWST}

\software{
\texttt{astroquery} \citep{Ginsburg2019},
\texttt{astropy} \citep{AstropyCollaboration2018},
\texttt{catwoman} \citep{jones_catwoman_2020},
\texttt{dynesty} \citep{Speagle2020},
\texttt{emcee} \citep{emcee2013},
\texttt{Eureka!} \citep{Bell2022},
\texttt{ExoTiC-JEDI} \citep{Alderson2022},
\texttt{ExoTiC-LD} \citep{Grant2024_exotic-ld},
\texttt{fleck} \citep{Morris2022},
\texttt{GASTLI} \citep{2024A&A...688A..60A},
\texttt{GGChem} \citep{ggchem},
\texttt{harmonica} \citep{grant2022transmission},
\texttt{jwst} \citep{bushouse2023},
\texttt{juliet} \citep{espinoza_juliet_2019},
\texttt{lightkurve} \citep{LightkurveCollaboration2018},
\texttt{matplotlib} \citep{Hunter2007},
\texttt{numpy} \citep{vanderWalt2011},
\texttt{pandas} \citep{McKinney2010},
\texttt{PICASO} \citep{mukherjee_picaso_2023},
\texttt{petitRADTRANS} \citep{molliere_petitradtrans_2019},
\texttt{planetsynth} \citep{muller_synthetic_2021}, 
\texttt{PLATON} \citep{Zhang2019},
\texttt{POSEIDON} \citep{MacDonald2017},
\texttt{pyMSG} \citep{2023JOSS....8.4602T},
\texttt{PyMultiNest} \citep{2016ascl.soft06005B},
\texttt{PyRaT Bay} \citep{CubillosBlecic2021mnrasPyratBay},
\texttt{pysynphot} \citep{2013ascl.soft03023S},
\texttt{scipy} \citep{Virtanen2020},
\texttt{spotrod} \citep{spotrod2014},
\texttt{TauREx} \citep{taurex1,taurex2},
\texttt{VULCAN} \citep{tsai_vulcan_2017,tsai_comparative_2021},
}
\bibliography{references}{}
\bibliographystyle{aasjournal}



\appendix

\section{Limb-Darkening Comparison}\label{appendix:limbdark}

\begin{figure*}[!htb]
\epsscale{1}
\plotone{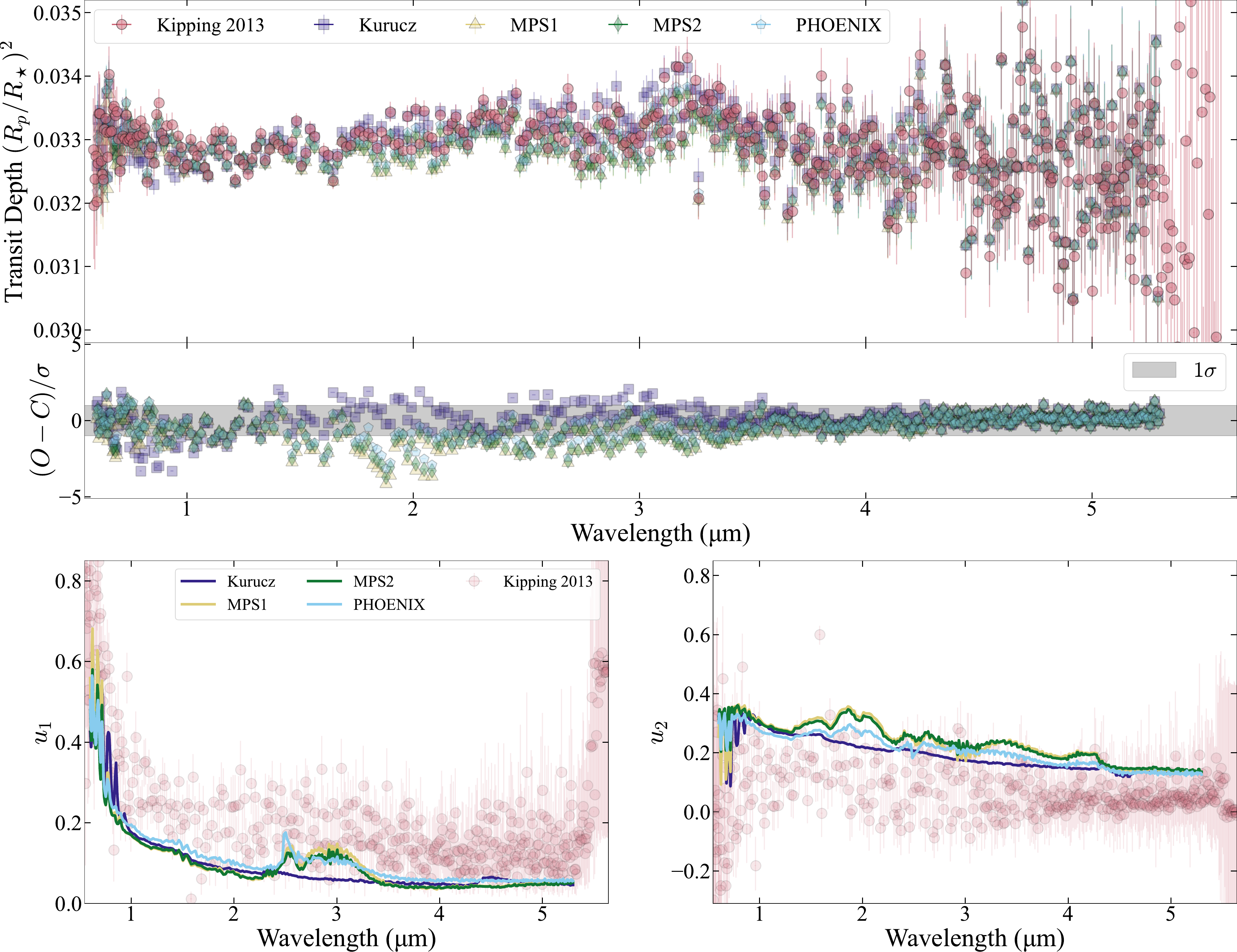}
\caption{\textbf{Top:} HATS-6 b transmission spectra for visit 1. The spectra shown are derived using fixed quadratic limb-darkening coefficients (u$_1$ and u$_2$) from ExoTiC-LD or fitting for the quadratic limb-darkening coefficients following the parametrization from \citet{kipping_efficient_2013} (q$_1$ and q$_2$) and labeled as such. \textbf{Middle:} The residuals (O-C) with respect to the free limb-darkening spectra (labeled ``Kipping 2013''). The residuals are divided by the observed error of the ``Kipping 2013'' spectrum. \textbf{Bottom:} The resulting limb-darkening coefficients, u1 and u2, for each implementation}
\label{fig:limb-darkening}
\end{figure*}

For the \textbf{Eureka1} reduction, we allowed the limb darkening parameters to float and implemented wide uniform priors using the Kipping 2013 \citep{kipping_efficient_2013} prescription for a quadratic limb darkening law. Limb darkening prescriptions have been extensively studied for the effect they have on extracted abundances \citep[e.g.,][]{coulombe_biases_2024, keers_reliable_2024}. As a test, we adopted different limb darkening prescriptions by also implementing different stellar grids from ExoTiC-LD for one of our reductions in order to observe what effects they had on our resulting spectra. We find that the transmission spectrum is most affected by limb-darkening choices blueward of 3 \textmu{}m (\autoref{fig:limb-darkening}). We note that none of the stellar grids were able to fully capture the complexities that arose in our spectra, although the stellar grids were much more representative of the data than is the case for other GEMS \citep{kanodia_toi-5205b_2023}. In order to avoid biasing our data with a pre-computed stellar grid, we proceed with Kipping 2013 \citep{kipping_efficient_2013} free limb darkening as our final limb-darkening prescription.

\pagebreak

\section{Bin-then-fit vs Fit-then-bin}\label{appendix:bin}

\begin{figure*}[!htb]
\epsscale{1}
\plotone{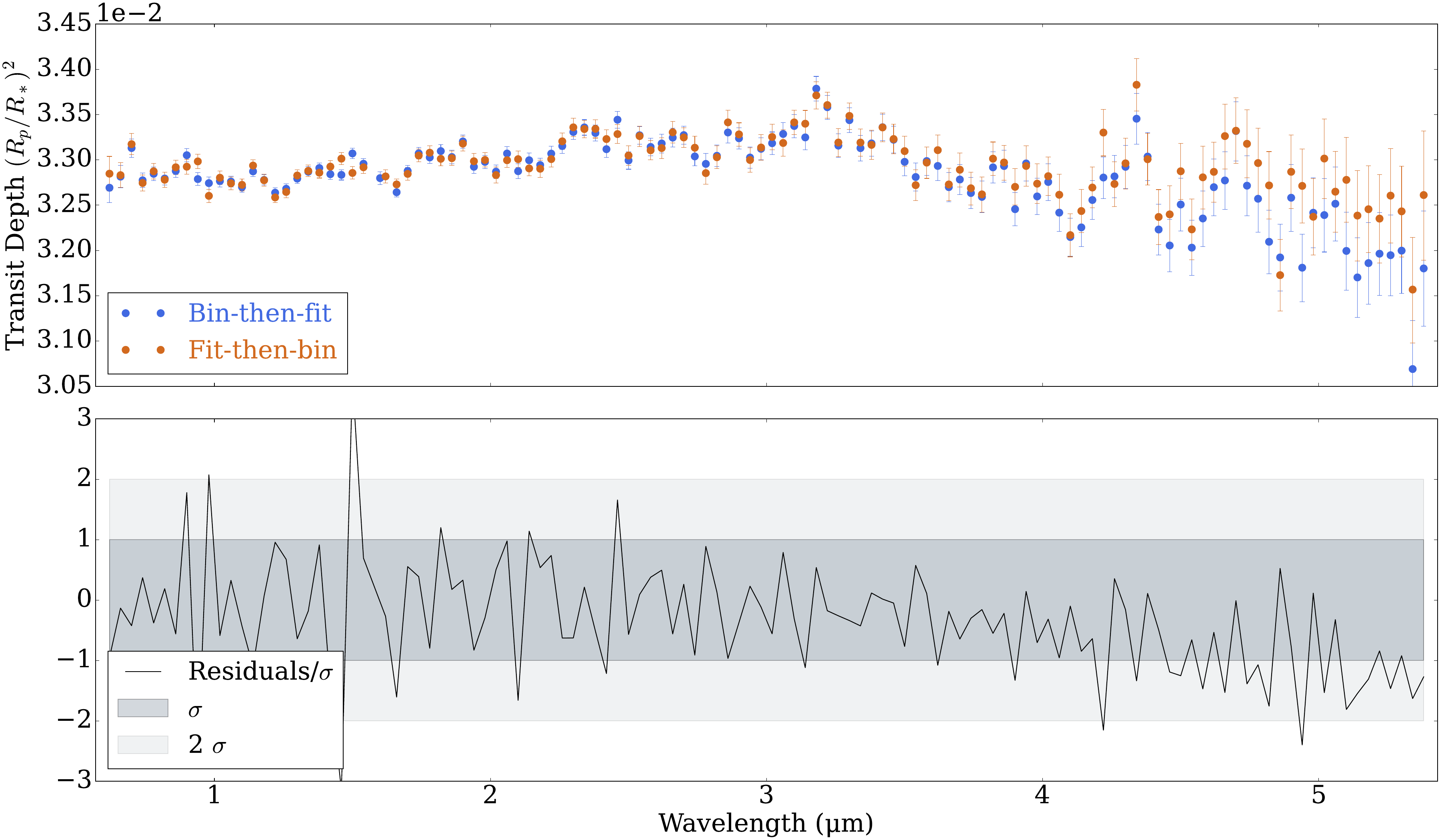}
\caption{\textbf{Top:} The transmission spectra resulting from two different binning approaches. For one approach, we fit the light curves at the pixel-level, resulting in $\sim$ 470 channels that we then bin down to a binned wavelength grid of 120 channels that are 40 nm wide. For the other, we employ the 120 channel wavelength grid at the light curve fitting stage, and bin our light curves before fitting them. The high residual difference around 1.5 \textmu{}m is directly caused by a hot pixel in the trace that created high correlated noise in one of the light curves. This hot pixel was easily removed at the pixel-level resolution, but was not identified by the light-curve binning routine and in turn biased that data point for the ``bin-then-fit'' strategy, highlighting the dangers of binning at the light-curve stage. \textbf{Bottom}: Residuals with respect to the uncertainties in the first approach. The binning approach affects the data redder than 4 \textmu{}m most.}
\label{fig:bin}
\end{figure*}

During our reductions, we employ two different binning strategies: ``bin-then-fit'' v.s. ``fit-then-bin''. In order to test the effect of different binning strategies in the data, we tested were the following:  ``bin-then-fit'' v.s. ``fit-then-bin''. Binning before spectroscopic fitting leads to less computationally intensive fitting, but it may also hide important outliers that bias the resulting spectra. Fitting the light curves in native resolution first may also lead to smaller uncertainties in the extracted spectra \citep{coulombe_biases_2024}. Testing these approaches with the same reduction (\textbf{ExoticJEDI}) we find that the ``fit-then-bin'' v.s. ``bin-then-fit'' approaches led to some differences in the resulting spectra. Especially redder than 4 \textmu{}m and when a light curve with high correlated noise was not masked in the binning stage (\autoref{fig:bin}). In order to avoid any biases from outliers and maximize the amount of information we can extract from the spectra, we proceed with a pixel-level resolution spectral fit.

\pagebreak

\section{Chromatic Spot Fitting}\label{appendix:spot_chrome}

\begin{figure*}[!htb]
\epsscale{1}
\plotone{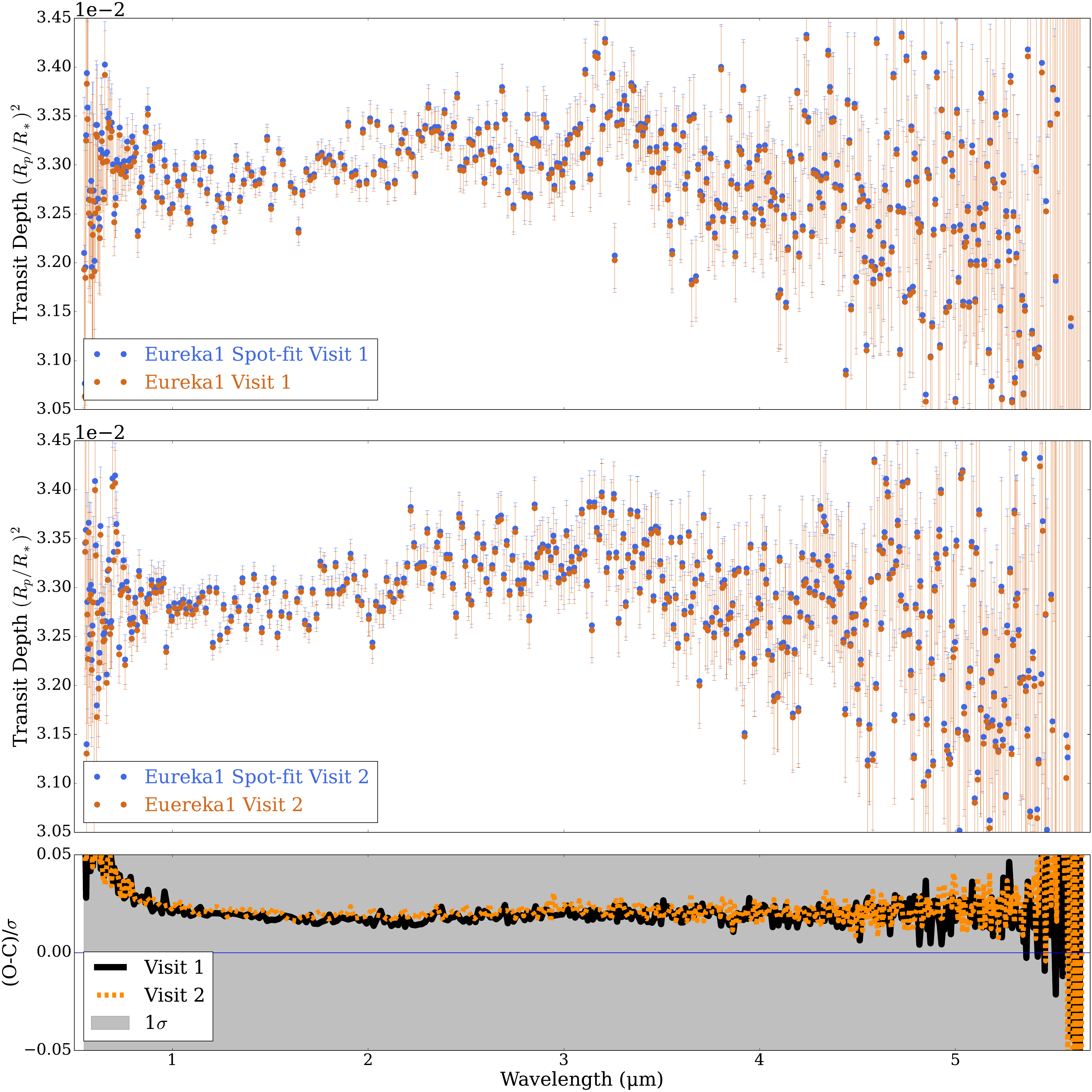}
\caption{\textbf{Top:} Transmission spectra for visit 1 with and without spot contrast fitted. \textbf{Middle:} Transmission spectra for visit 2 with and without spot contrast fitted. \textbf{Bottom:} Residuals (O-C) with respect to the uncertainties of the transmission spectra without spot fitting. We stress the fact that our residuals are zoomed in to a $\pm$ 0.05 $\sigma$ axis in order to show any structure, highlighting the minimal difference between these fits.}
\label{fig:spot_data}
\end{figure*}

We also investigated the chromatic effect of spot crossings with \texttt{spotrod} by running spectral fits for our best-fit spot configurations (see \S\ref{subsection:spots}). We do this by holding all spot parameters constant except the spot contrast. Despite our white light curves highly favoring a spot crossing configuration, we do not find notable differences in the resulting spectra between our non-spotted and spotted configurations (\autoref{fig:spot_data}). The same spots that were evident in the white light curve were not resolved spectroscopically with confidence. Because the wavelength-dependent transit depths generally demonstrated no sensitivity to spot crossings, we adopted the simplest model without spots. This decision was made in order to avoid any overfitting due to higher model complexity and any degeneracies that might arise from limb darkening since one of our spots is in the limb. The white light curve fits themselves presented some signs of over-fitting, especially for visit 2 (see \autoref{fig:wlc}), and because limb-darkening is fitted freely, having a spot on the limb can be highly degenerate with limb-darkening effects. 

\pagebreak

\pagebreak

\section{Limb Asymmetries}\label{appendix:limb_diff}

\begin{figure*}[!htb]
\epsscale{1}
\plotone{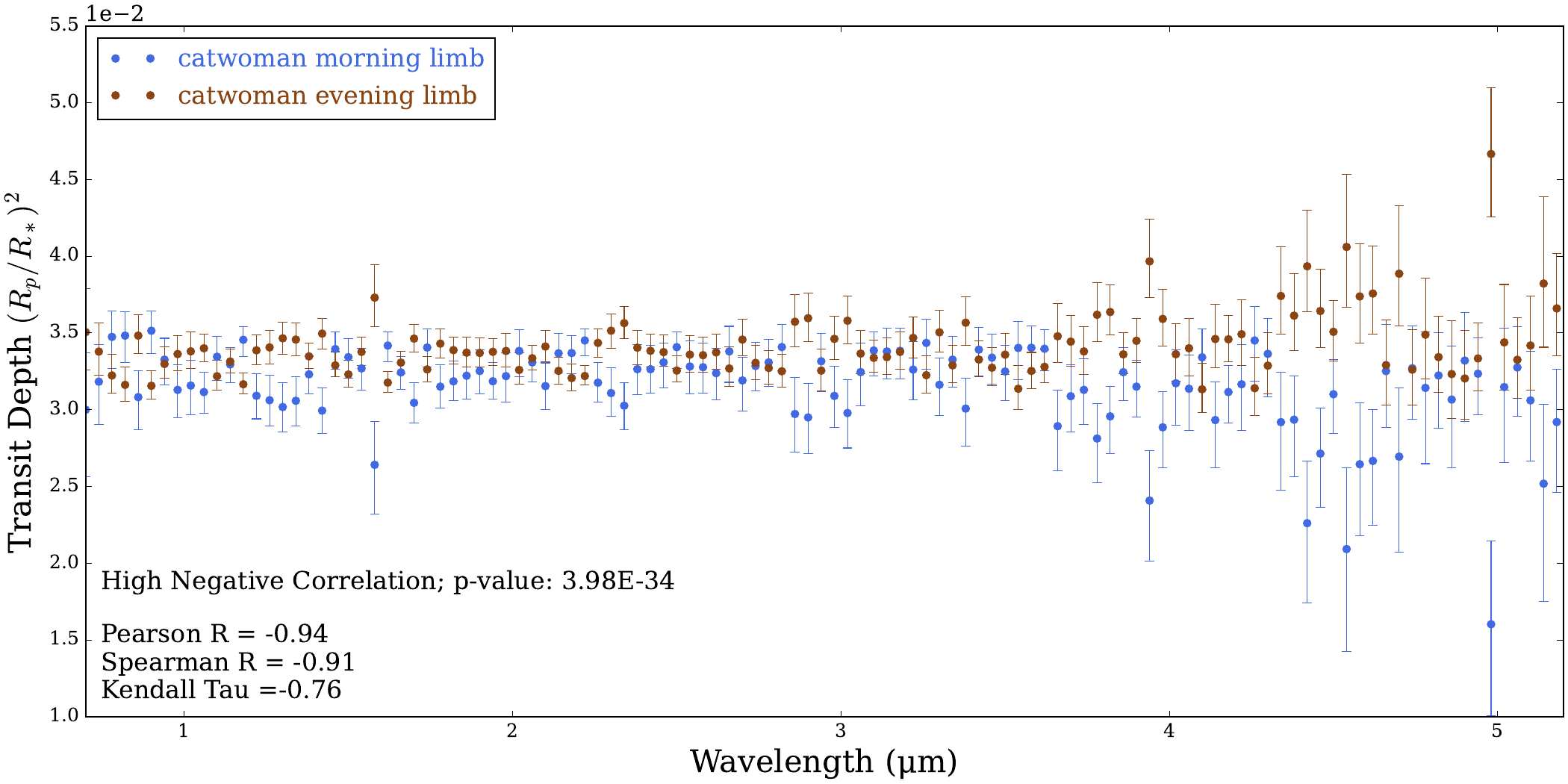}
\caption{The resulting transmission spectra for each limb with \texttt{catwoman} for visit 1, the correlation factors show that the resulting limb spectra are highly correlated.}
\label{fig:catwoman1}
\end{figure*}

\begin{figure*}[!htb]
\epsscale{1}
\plotone{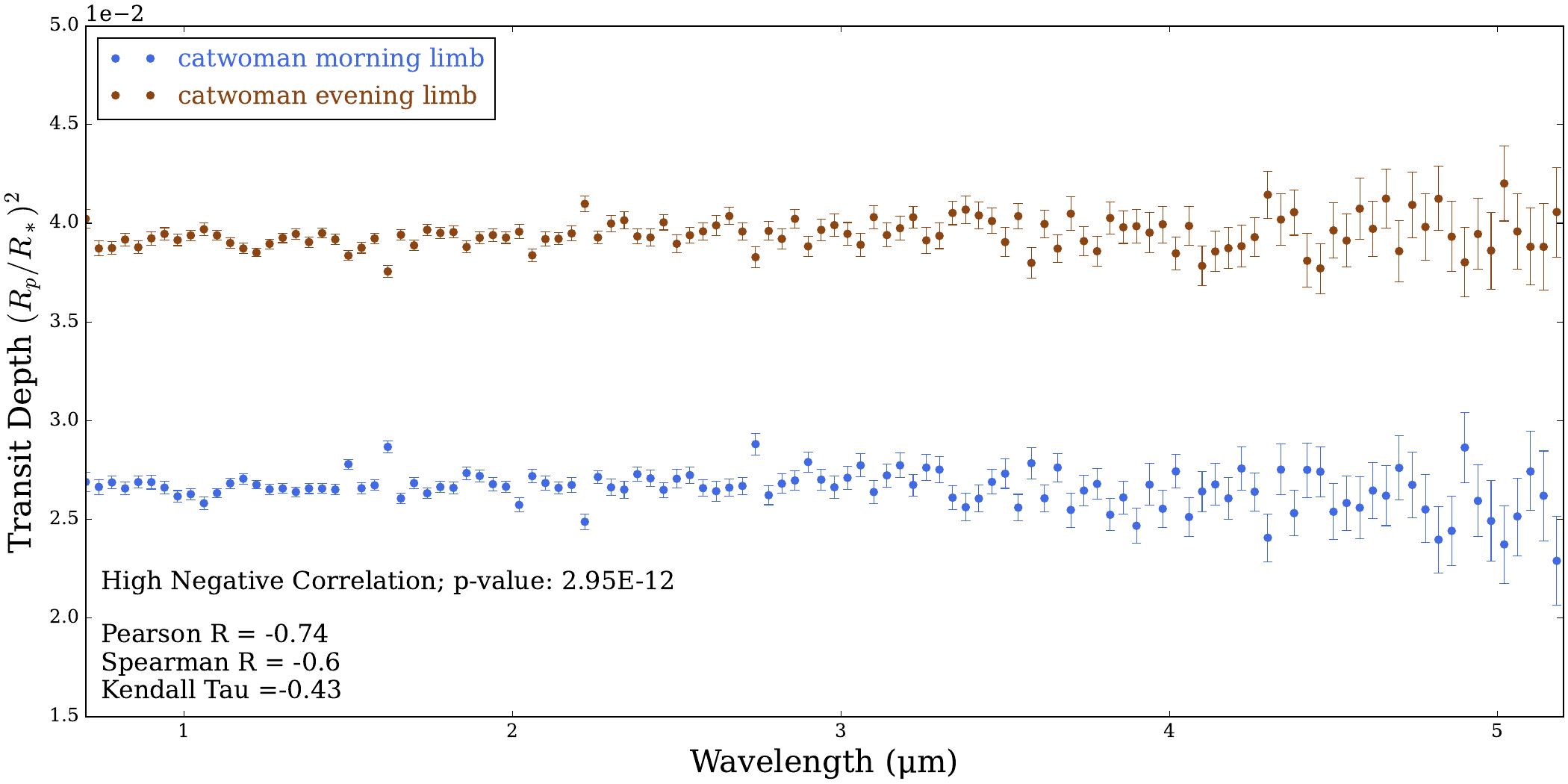}
\caption{The resulting transmission spectra for each limb with \texttt{catwoman} for visit 2, the correlation factors show that the resulting limb spectra are highly correlated.}
\label{fig:catwoman2}
\end{figure*}

\begin{figure*}[!htb]
\epsscale{1}
\plotone{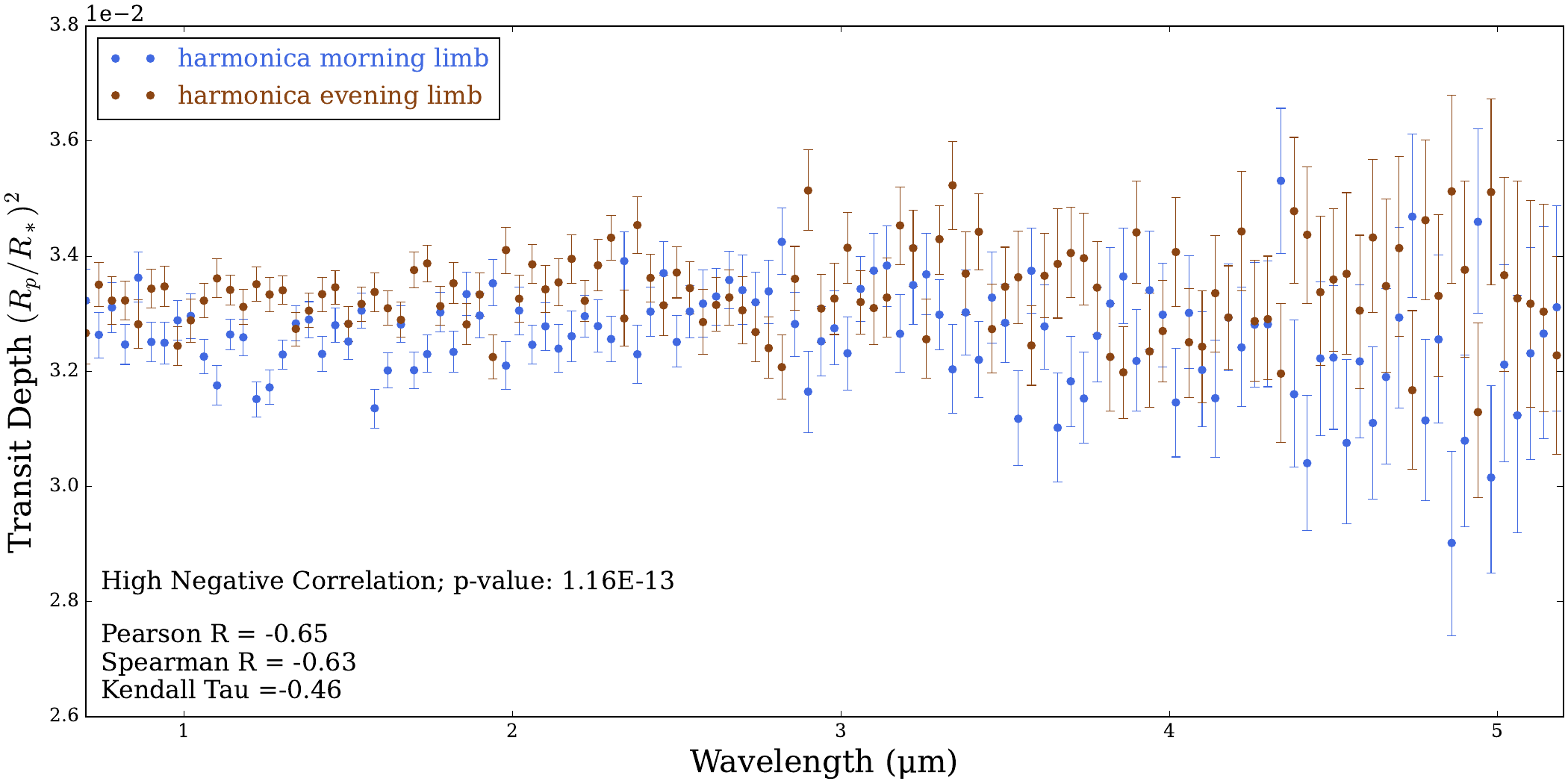}
\caption{The resulting transmission spectra for each limb with \texttt{harmonica} for visit 1, the correlation factors show that the resulting limb spectra are highly correlated.}
\label{fig:harmonica_visit1}
\end{figure*}

\begin{figure*}[!htb]
\epsscale{1}
\plotone{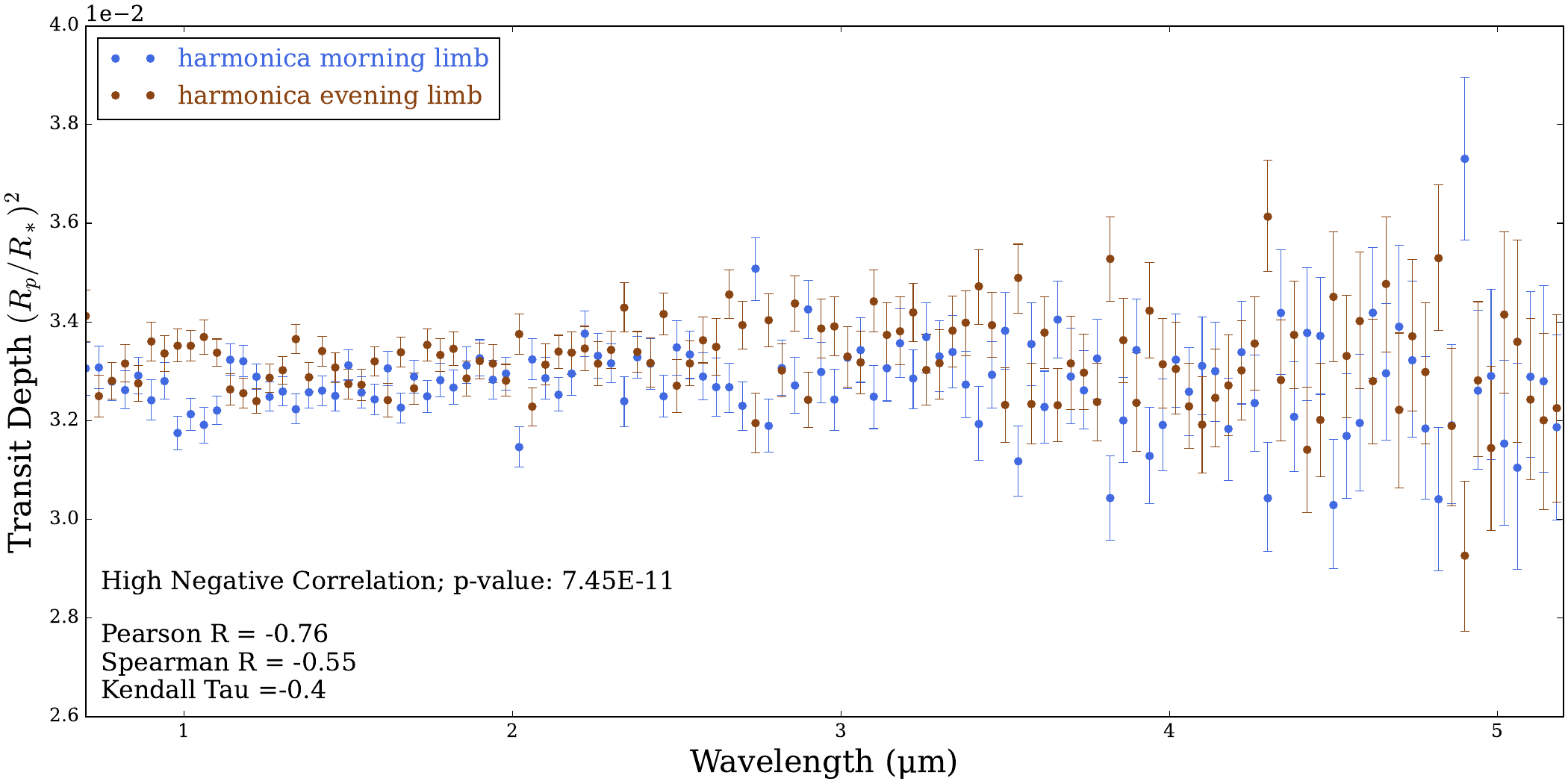}
\caption{The resulting transmission spectra for each limb with \texttt{harmonica} for visit 2, the correlation factors show that the resulting limb spectra are highly correlated.}
\label{fig:harmonica_visit2}
\end{figure*}

We also investigated the possibility of limb asymmetries in our data using both \texttt{catwoman} \citep{jones_catwoman_2020, espinoza_constraining_2021} and \texttt{harmonica} \citep{grant2022transmission} for our analysis. In both cases, \texttt{catwoman} and \texttt{harmonica}, we found high negative correlation between the resulting limb spectra  (see \autoref{fig:catwoman1}-\autoref{fig:harmonica_visit2}). We ran these tests both on the pixel-level and binned wavelength grid. The conclusions were consistent for both resolutions. Therefore, we once again proceed with our simplest model, assuming that no limb asymmetries are present.

\pagebreak

\pagebreak

\section{Comparison of the transmission spectra by reduction} \label{appendix:compare_visits}
\begin{figure*}[!htb]
\epsscale{1.2}
\plotone{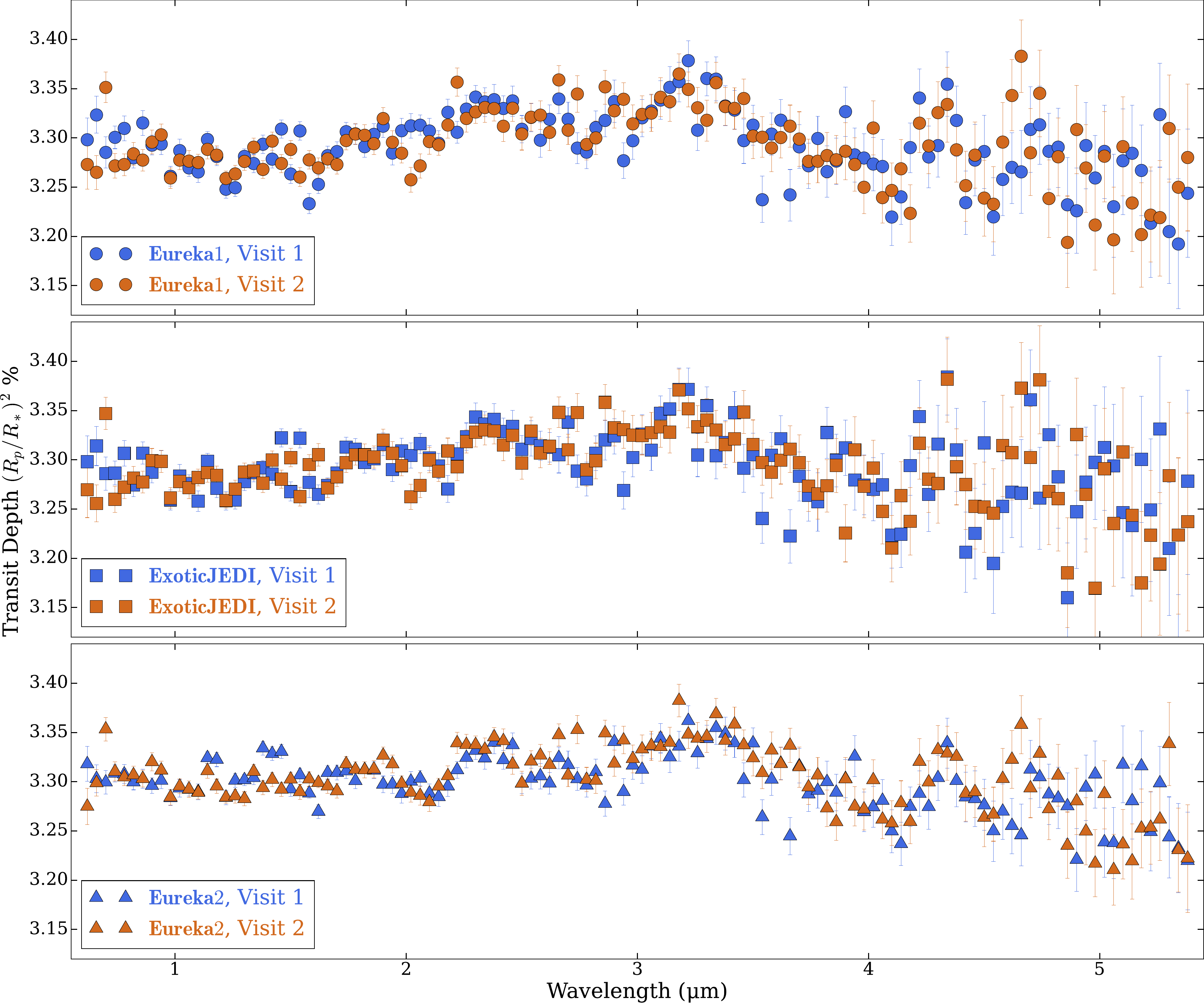}
\caption{The finalized, binned transmission spectra, shown by code, using \textbf{Eureka1}, \textbf{ExoticJEDI}, and \textbf{Eureka2} reductions. We find a generally good agreement for the visits between both visits, on average within 2 $\sigma$ of each other. The binned spectra have 120 spectral channels with 40nm bins.}
\label{fig:transmission_by_code}
\end{figure*}

\autoref{fig:transmission_by_code} presents a comparison of both visits for all three reductions described in \S\ref{sec:datareduction}. For all reductions, we consistently observe a slight difference ($<2\sigma$) between both visits (e.g., near $2.8-3.5$ \textmu{}m or 4.2 \textmu{}m), suggesting that the change in visits is independent of the choices made during the data reduction and may reflect a change in the surface of HATS-6. 

\pagebreak 

\section{Full Chemistry Fits}\label{appendix:full_chemistry}

\begin{figure*}[!htb]
\plotone{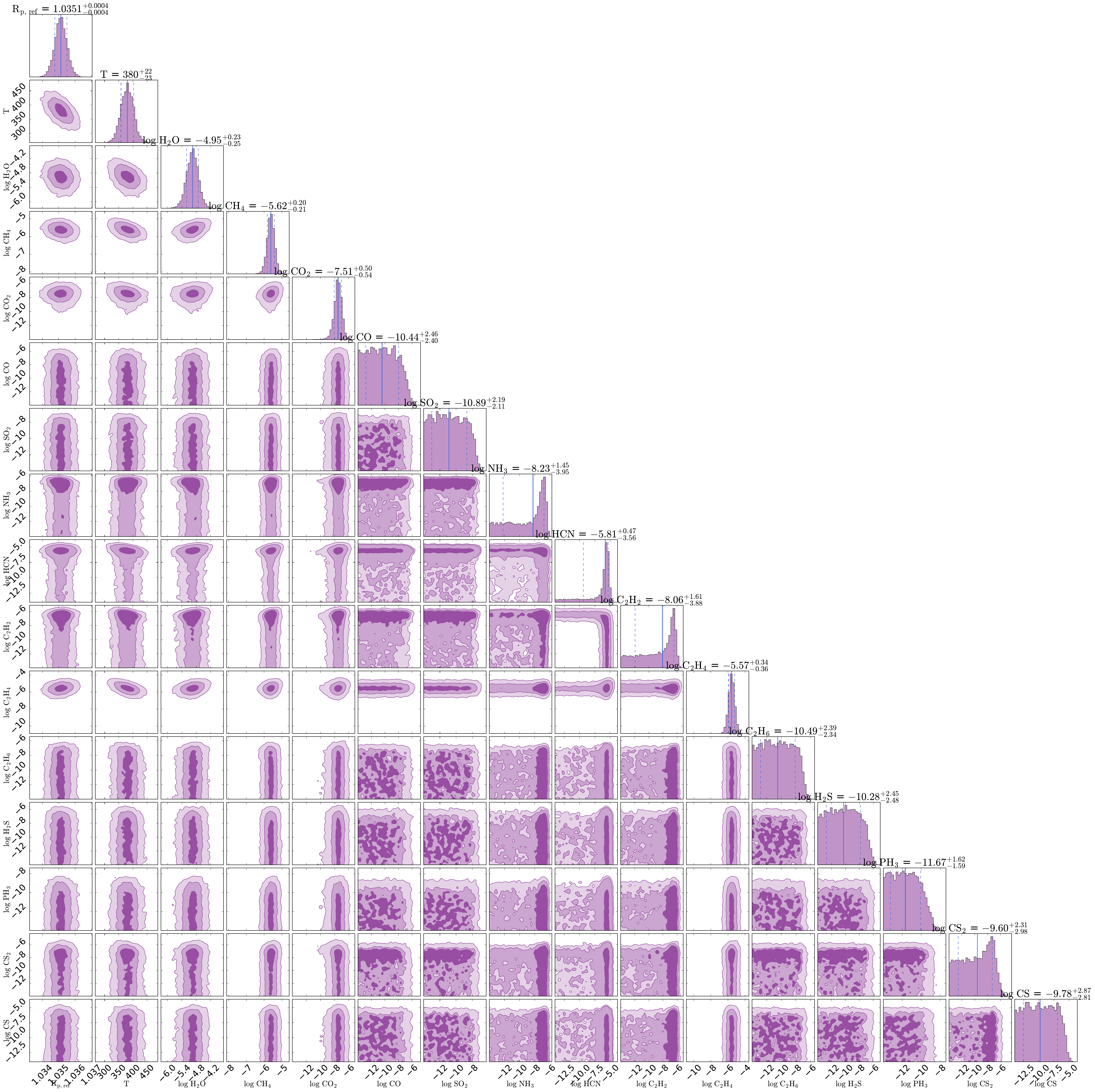}
\caption{Retrieved parameters and abundances for our best-fit free chemistry runs.}
\label{fig:fullchem}
\end{figure*}

We performed our first exploratory retrievals using the co-added \textbf{ExoticJEDI} reductions as well as a relaxed convergence criterion ($\Delta\ln Z$ = 10). The full chemistry fits included opacities for: H$_2$O, CH$_4$, CO$_2$, CO, SO$_2$, NH$_3$, C$_2$H$_2$, H$_2$S, PH$_3$, C$_2$H$_4$, and C$_2$H$_6$, as well as CIA from H$_2$–H$_2$, H$_2$–He, H$_2$–CH$_4$, CO$_2$–H$_2$, CO$_2$–CO$_2$, and CO$_2$–CH$_4$. We allowed the logarithmic mixing ratio of each of the listed species to float freely within prior bounds [$\mathcal{U}(-14.00, -1.00)$] while maintaining evenly mixed vertical gas abundances. In \autoref{fig:fullchem} we showcase the corner plot results from our full chemistry run with the binned coadded data (ln Z = 821.36) from this preliminary set of retrieval fits. We highlight the considerably low T$_{eq}$ of 380 K and the high retrieved abundance of C$_2$H$_4$. These findings drove us to further explore and constrain our retrievals (see \S\ref{subsection:wavecut} - \S\ref{subsec:retrieval_results}) using physically motivated chemical species from our VULCAN forward models (\autoref{fig:TPprofile}).

\pagebreak

\section{Testing Different Pressure Temperature Profile}\label{appendix:guillot}

\begin{figure*}[!htb]
\epsscale{0.8}
\plotone{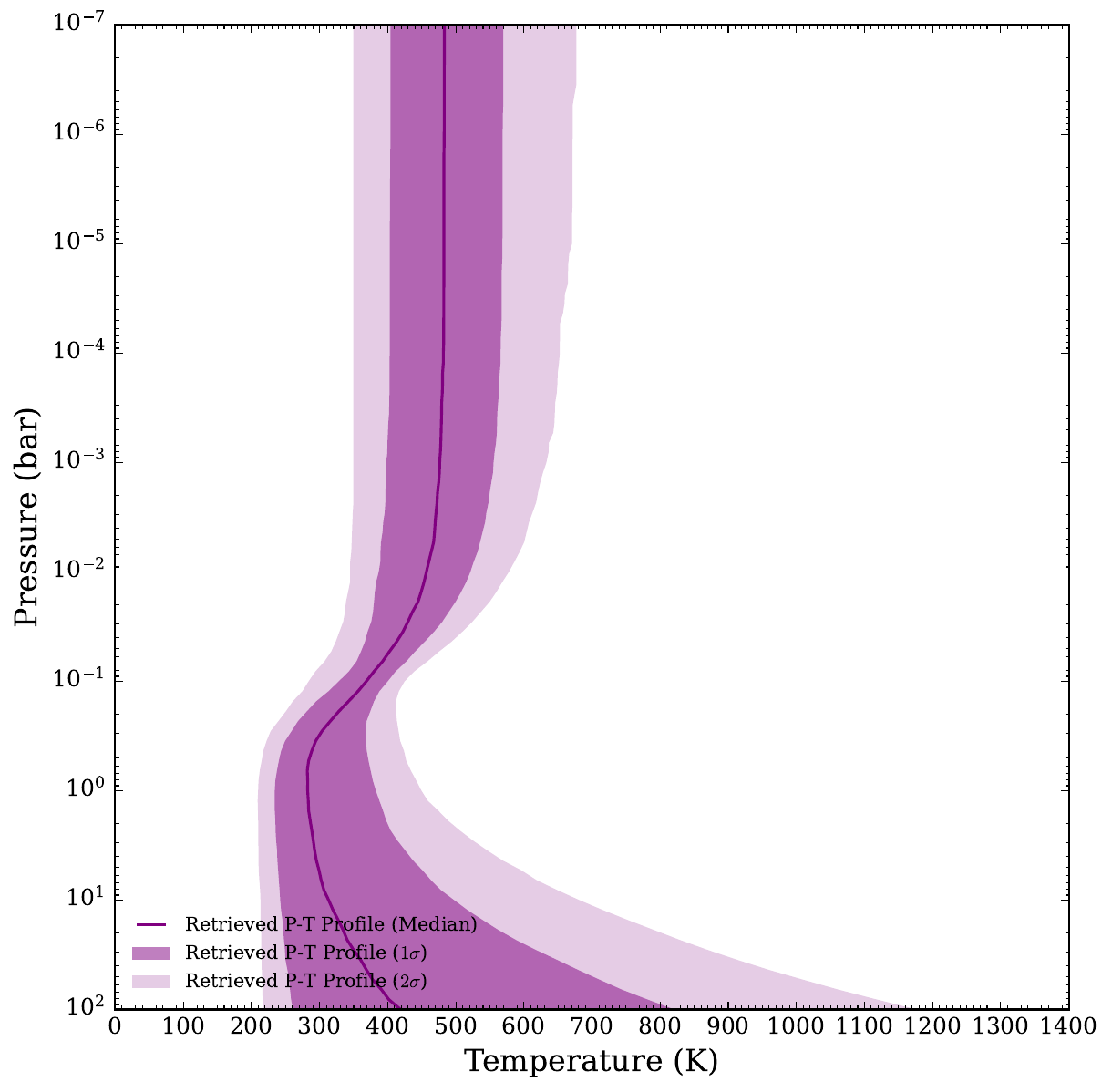}
\caption{The retrieved Guillot pressure-temperature profile and its 2$\sigma$ constraints. The profile remains mostly isothermal at our temperatures probed. The upper 2$\sigma$ uncertainties get close to our photometric T$_{eq}$ (712 K), yet the derived abundances are nearly indistinguishable from our isothermal retrieval.}
\label{fig:guillot}
\end{figure*}

\begin{figure*}[!htb]
\epsscale{0.8}
\plotone{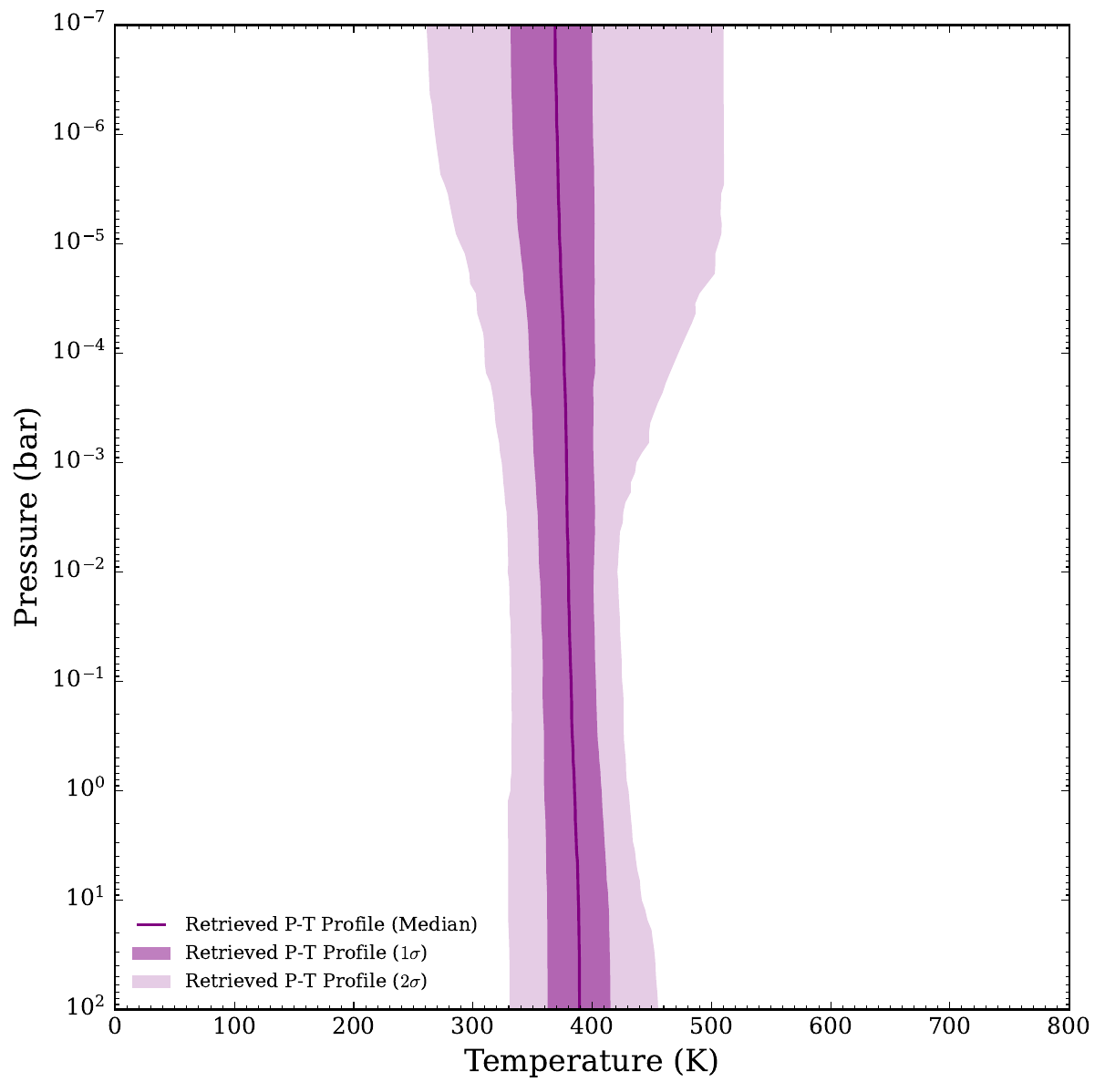}
\caption{The retrieved Madhusudhan and Seager pressure-temperature profile and its 2$\sigma$ constraints. The profile remains isothermal at our temperatures probed. The upper 2$\sigma$ uncertainties still do not encompass our photometric T$_{eq}$ (712 K).}
\label{fig:madhu}
\end{figure*}

For our retrievals, we tested fitting for a Guillot \citep{guillot_radiative_2010} temperature-pressure profile; $\chi^2_\nu$ = 1.94 and ln Z = 822.07. We ran these tests assuming a clear atmosphere with \emph{full chemistry}. The test yielded slightly better statistical evidence than the original fit ($\chi^2_\nu$ = 1.97, ln Z = 821.36). Aside from yielding a \teq of 255 K and a T$_{\rm{int}}$ of 138 K, the abundances derived from the Guillot profile fit were indistinguishable from the original fit. This was because the best-fit temperature-pressure profile was mostly isothermal in the regions of the atmosphere probed by our transmission spectra (\autoref{fig:guillot}). We had less successful results while employing the Madhusudhan and Seager pressure-temperature profile \citep{madandseager}. We retrieved an isotherm (see \autoref{fig:madhu}) and produced a worse fit to the data ($\chi^2_\nu$ = 2.08) than the isothermal full chemistry fit. In the end, from these tests we concluded that including a more complex and flexible thermal profile did not resolve the low T$_{eq}$ retrieved nor did it add any information to our retrievals. 

\pagebreak
\section{Pressure Contributions}\label{appendix:pressure}

\begin{figure*}[!htb]
\epsscale{1.2}
\plotone{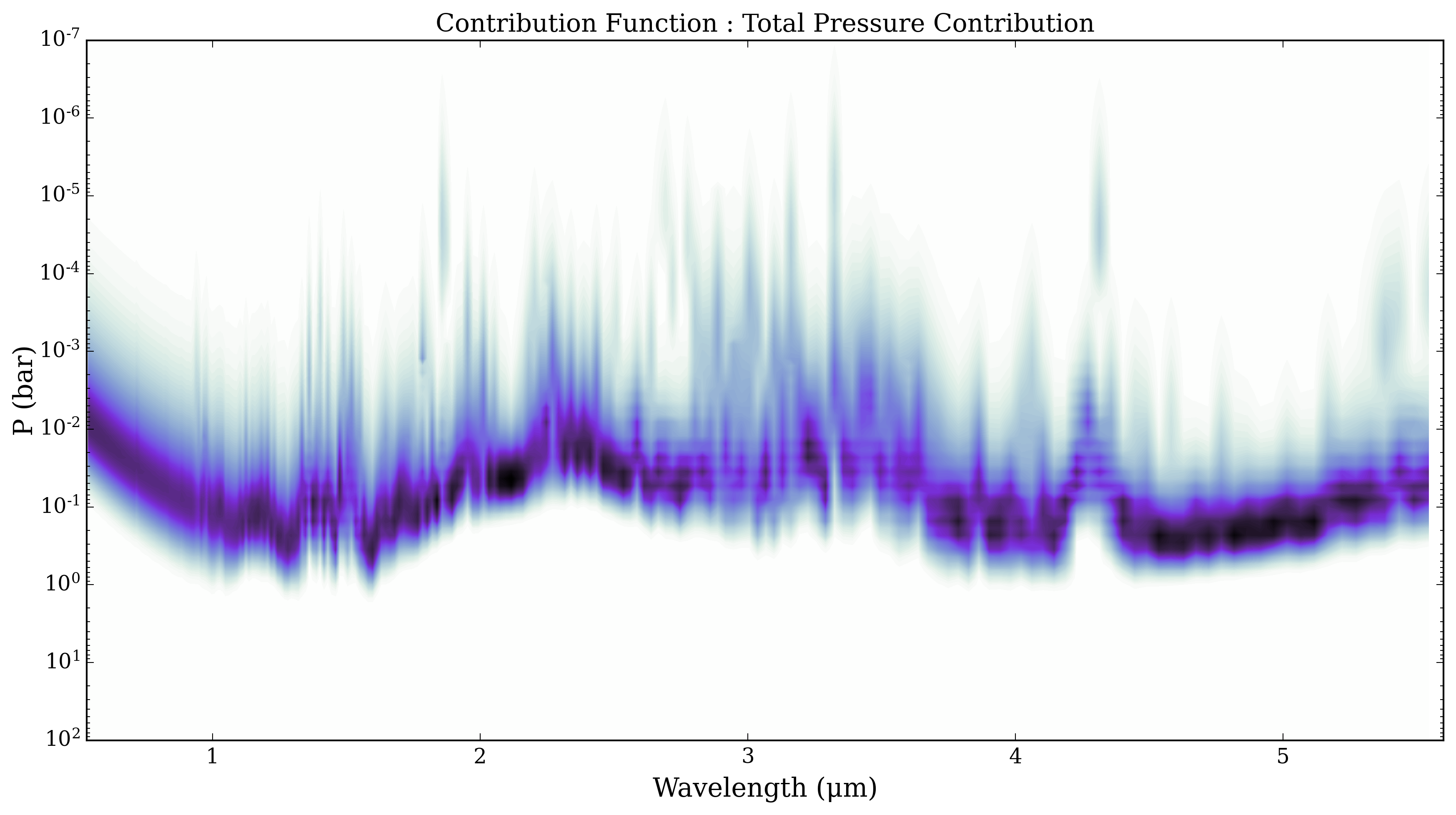}
\caption{Pressure contribution plot for our final model M2.5. The darker regions signify that the contribution from all opacity sources in the transmission spectra is at its highest in those regions.}
\label{fig:pressure}
\end{figure*}

In order to relate our free chemistry retrievals back to our chemically consistent forward models, we generate total pressure contribution plots that highlight the general strength of spectral contributions throughout our pressure grid (\autoref{fig:pressure}). We iteratively generated pressure contribution plots for most of our retrievals as an analysis tool. 

\pagebreak

\section{Model Comparisons}\label{appendix:model_comparison}

\begin{figure*}[!htb]
\epsscale{1.2}
\plotone{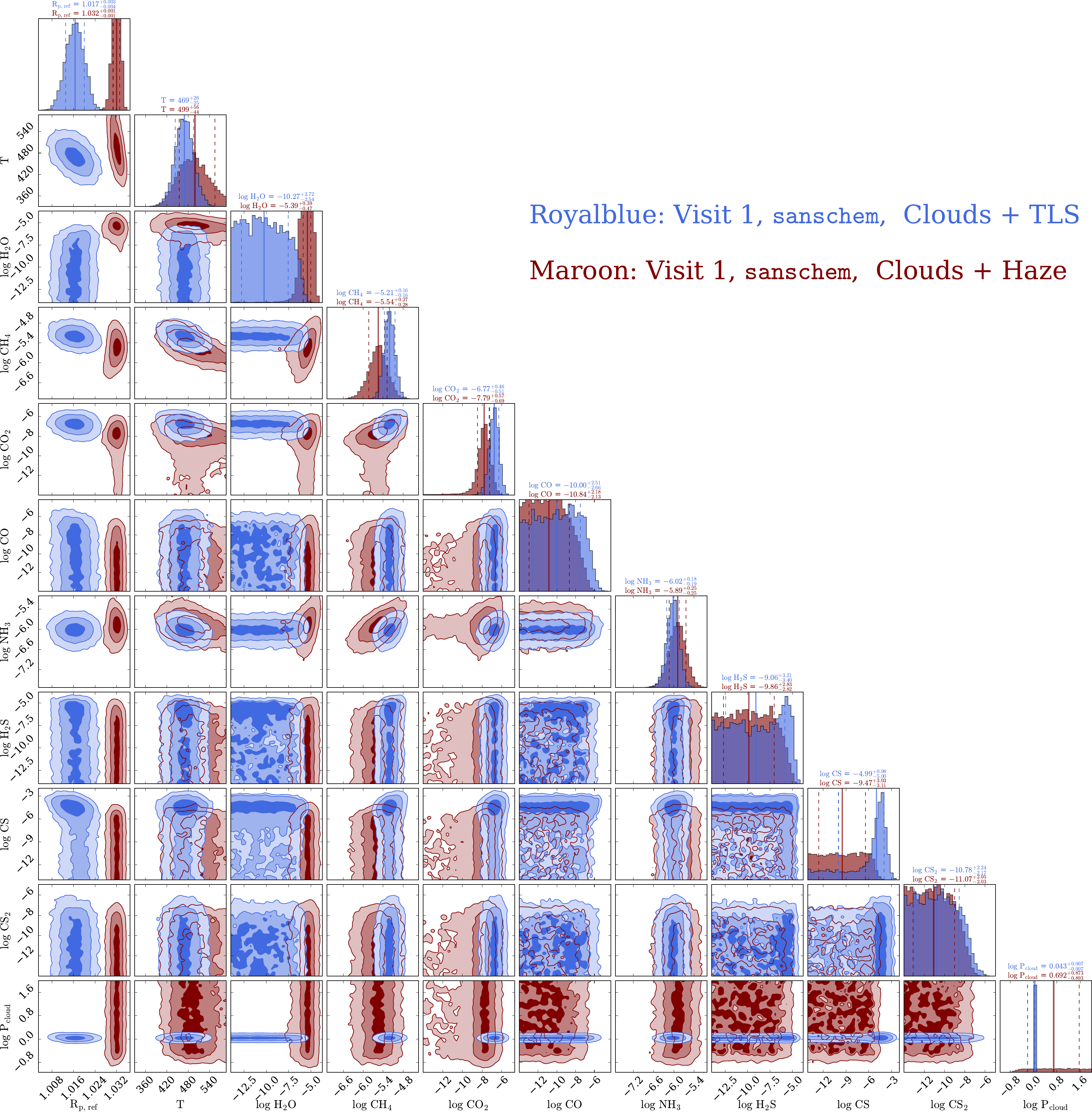}
\caption{Comparison for Models M2.4 (Clouds+TLS) and M2.5 (Clouds+Haze) for visit 1. We only show the shared parameters between the two models in order to show that the chemical abundances between these models do not change drastically. Most of the difference is absorbed by the haze and TLS parameters.}
\label{fig:visit1_comp}
\end{figure*}

\begin{figure*}[!htb]
\epsscale{1.2}
\plotone{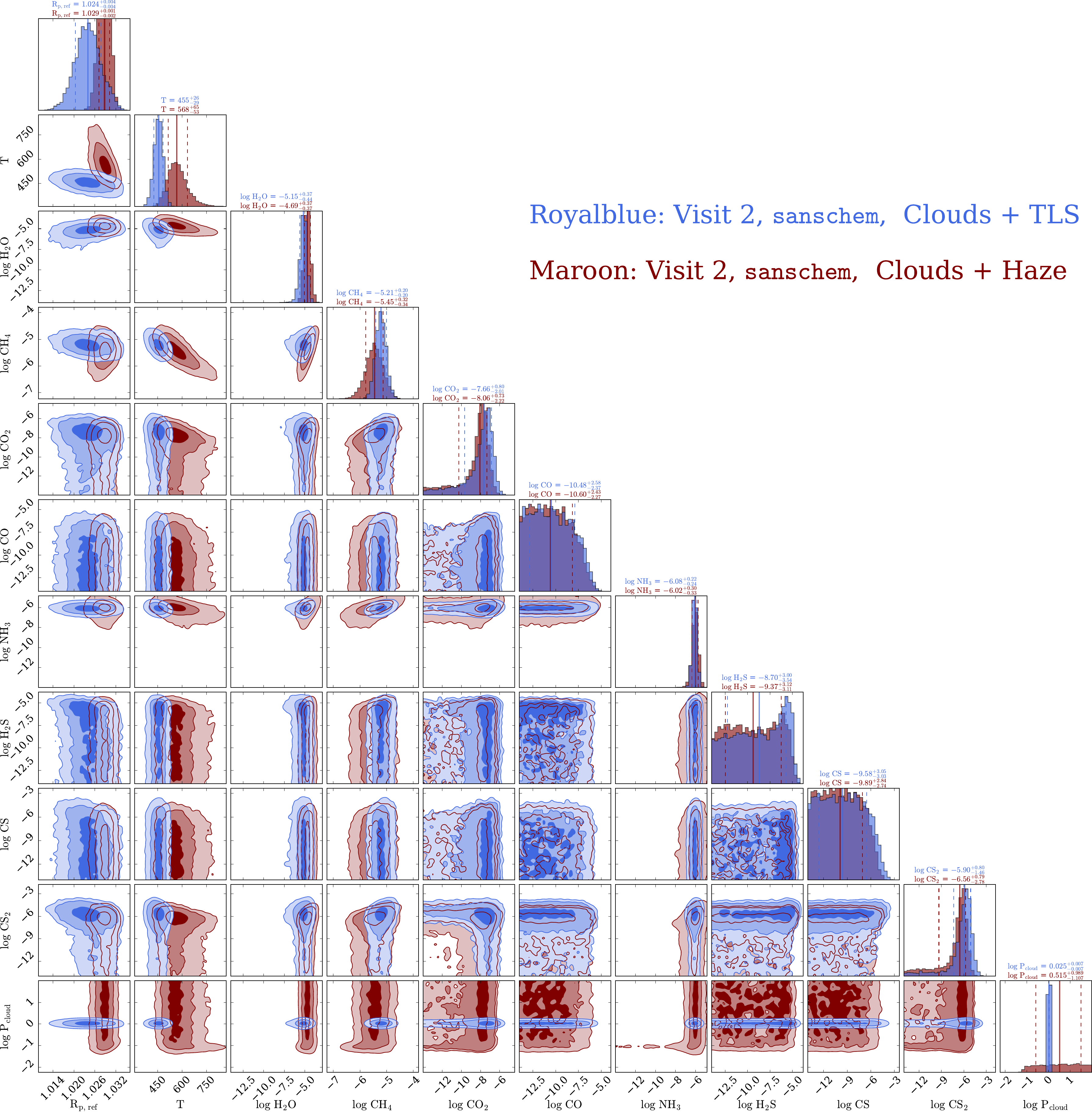}
\caption{Comparison for Models M2.4 (Clouds+TLS) and M2.5 (Clouds+Haze) for visit 2. We only show the shared parameters between the two models in order to show that the chemical abundances between these models do not change drastically. Most of the difference is absorbed by the haze and TLS parameters.}
\label{fig:visit2_comp}
\end{figure*}

During model comparison, we had two high-evidence models to choose between. The first model (M2.4) parametrized clouds as a gray cloud deck with cloud-top pressure $\log$(P) and TLS as a stellar surface with a spot coverage fraction, photosphere temperature, spot temperature, photosphere $\log$(g), and spot $\log$(g). The second model (M2.5) also parametrized a gray cloud deck, along with a haze slope ($\log~a$ and $\gamma$). This meant that M2.4 had two free parameters more than M2.5, and we took this into consideration during our model selection since higher model complexity might make it easier for the model to fit the data in ways that might not be physically motivated. Due to this, we also analyzed the M2.4 posteriors and found that the spot temperature to be consistently unphysically low for an M-dwarf (1000 K less than the stellar photosphere) \citep{mori_spot_temp}. By integrating the spot contrast in our white light curve spot fits, we derive spot temperatures of 3130 K for visit 1 and 3761 K for visit 2. Yet, we are retrieving spot temperatures of 2395 K for visit 1 and 2679 K for visit 2. Adjusting priors for the spot temperature did not alleviate this issue, causing the posterior to constantly hit lower prior bounds, or produce spot coverage fractions that were near-negligible (less than 3$\%$). Both the presence of TLS and haze are physically motivated for this planet, but our M2.5 model yielded more consistent results across visits and more consistent metallicity and C/O as a result. In \autoref{fig:visit1_comp} and \autoref{fig:visit2_comp} we show the full posteriors of M2.4 and M2.5 overplotted by visit to show that the retrieved abundances are near equal for both models except for H$_2$O which was degenerate with the TLS parameters for one of our visit 1. In the end, our conclusion that HATS-6 b is low metallicity is impervious to model selection, while the C/O is sensitive to the unconstrained water feature in visit 1 if TLS is employed in the model. We show the full priors and posteriors for M2.4 (Clouds+TLS) in \autoref{tab:TLS_results}.

\begin{deluxetable*}{r|l||l|l|l}
\tablewidth{0.98\textwidth}
\tabletypesize{\small}
\tablecaption{Atmospheric retrieval priors and posteriors for the M2.4 model with clouds and TLS using \texttt{POSEDION}. The abundances are the volume mixing ratios for each species. \label{tab:TLS_results}}
\tablehead{
\colhead{Parameters} & \colhead{Priors} & \colhead{Visit 1} & \colhead{Visit 2} & \colhead{Co-added}}
\startdata
$\mathrm{R}_{\mathrm{p, \, ref}}$ & $\mathcal{U}(0.85, 1.15)$ & $1.016_{-0.004}^{+0.004}$ & $1.021_{-0.004}^{+0.003}$ & $1.025_{-0.005}^{+0.003}$ \\
$\mathrm{T}$ & $\mathcal{U}(200.00, 1500.00)$ & $499_{-35}^{+39}$ & $570_{-43}^{+52}$ & $524_{-35}^{+43}$ \\
$\log \, \mathrm{H_2 O}$ & $\mathcal{U}(-14.00, -1.00)$ & $-9.74_{-2.86}^{+2.66}$ & $-4.69_{-0.36}^{+0.38}$ & $-5.18_{-0.34}^{+0.28}$ \\
$\log \, \mathrm{CH_4}$ & $\mathcal{U}(-14.00, -1.00)$ & $-5.23_{-0.19}^{+0.18}$ & $-5.22_{-0.31}^{+0.29}$ & $-5.30_{-0.19}^{+0.18}$ \\
$\log \, \mathrm{NH_3}$ & $\mathcal{U}(-14.00, -1.00)$ & $-6.05_{-0.22}^{+0.20}$ & $-6.11_{-0.38}^{+0.31}$ & $-6.22_{-0.24}^{+0.21}$ \\
$\log \, \mathrm{CO_2}$ & $\mathcal{U}(-14.00, -1.00)$ & $-6.85_{-0.50}^{+0.49}$ & $-7.50_{-1.89}^{+0.74}$ & $-7.26_{-0.42}^{+0.39}$ \\
$\log \, \mathrm{CO}$ & $\mathcal{U}(-14.00, -1.00)$ & $-10.10_{-2.60}^{+2.62}$ & $-10.35_{-2.46}^{+2.60}$ & $-9.10_{-3.23}^{+2.73}$  \\
$\log \, \mathrm{H_2 S}$ & $\mathcal{U}(-14.00, -1.00)$ & $-8.87_{-3.49}^{+3.13}$ & $-8.90_{-3.28}^{+3.09}$ & $-10.12_{-2.57}^{+2.73}$ \\
$\log \, \mathrm{CS_2}$ & $\mathcal{U}(-14.00, -1.00)$ & $-10.70_{-2.22}^{+2.23}$ & $-5.84_{-1.22}^{+0.77}$ & $-7.30_{-4.44}^{+2.38}$ \\
$\log \, \mathrm{CS}$ & $\mathcal{U}(-14.00, -1.00)$ & $-4.93_{-4.55}^{+0.93}$ & $-9.46_{-3.06}^{+3.26}$ & $-7.30_{-4.44}^{+2.38}$ \\
$\log \mathrm{P}_{\mathrm{cloud}}$ & $\mathcal{U}(-4, 1)$ & $-0.58_{-0.19}^{+0.60}$ & $-1.11_{-0.16}^{+0.14}$ & $-0.77_{-0.12}^{+0.14}$  \\
$\mathrm{f}_{\mathrm{spot}}$ & $\mathcal{U}(0, 1)$ & $0.04_{-0.01}^{+0.01}$ & $0.02_{-0.01}^{+0.01}$ & $0.03_{-0.00}^{+0.01}$  \\
$\mathrm{T}_{\mathrm{spot}}$ & $\mathcal{U}(2300, 1.2*T_*)$ & $2394.3_{-67.4}^{+112.3}$ & $2697.1_{-267.9}^{+443.7}$ & $3194.9_{-714.6}^{+286.8}$  \\
$\mathrm{T}_{\mathrm{phot}}$ & $\mathcal{N}(T_*, 54)$ & $3830.0_{-26.3}^{+38.2}$ & $3833.3_{-28.2}^{+39.8}$ & $3829.6_{-26.1}^{+38.2}$  \\
$\log \mathrm{g}_{\mathrm{spot}}$ & $\mathcal{U}(\log g_{phot}-0.5, \log g_{phot}+0.5)$ & $4.68_{-0.33}^{+0.32}$ & $4.71_{-0.35}^{+0.30}$ & $4.50_{-0.26}^{+0.45}$  \\
$\log \mathrm{g}_{\mathrm{phot}}$ & $\mathcal{U}(\log g_{phot}-0.5, \log g_{phot}+0.5)$ & $4.75_{-0.37}^{+0.28}$ & $4.63_{-0.32}^{+0.34}$ & $4.91_{-0.30}^{+0.17}$  \\
\hline
$\log[\mathrm{C/O}]$ & \nodata & $1.73_{-0.664}^{+0.783}$  & $-1.996_{-0.24}^{+0.344}$ & $0.006_{-0.291}^{+0.56}$ \\
$\log[\mathrm{M/H}]$ & \nodata & $-1.759_{-0.633}^{+0.78}$ & $-1.69_{-0.333}^{+0.385}$ & $-2.0^{+0.2}_{-0.2}$ \\
$\log[\mathrm{C/H}]$ & \nodata & $-1.429_{-0.51}^{+0.76}$ & $-1.722_{-0.348}^{+0.403}$ & $-1.807_{-0.24}^{+0.401}$  \\
$\log[\mathrm{O/H}]$ & \nodata & $-3.357_{-0.496}^{+0.48}$ & $-1.637_{-0.359}^{+0.376}$ & $-2.094_{-0.327}^{+0.275}$ \\
$\log[\mathrm{S/H}]$ & \nodata & $< 1.95$ & $< 1.68$ & $< 1.00$ \\ 
$\log[\mathrm{N/H}]$ & \nodata & $-2.122_{-0.221}^{+0.204}$ & $-2.179_{-0.37}^{+0.312}$ & $-2.285_{-0.24}^{+0.208}$ \\
\hline
$\chi^2_{\nu}$ & \nodata & 1.30 & 1.26 & 1.75 \\
$\ln Z$ & \nodata & 2598.83 & 2597.95 & 827.12\\
\enddata 
\tablecomments{All chemical ratios are with respect to solar except C/O which is in native units. For reference, the Solar C/O in dex is $\log$[C/O] = -0.25.}
\end{deluxetable*}

\clearpage
\pagebreak

\section{Spectral Decomposition}\label{appendix:spectral_contributions}

\begin{figure*}[!htb]
\epsscale{1.2}
\plotone{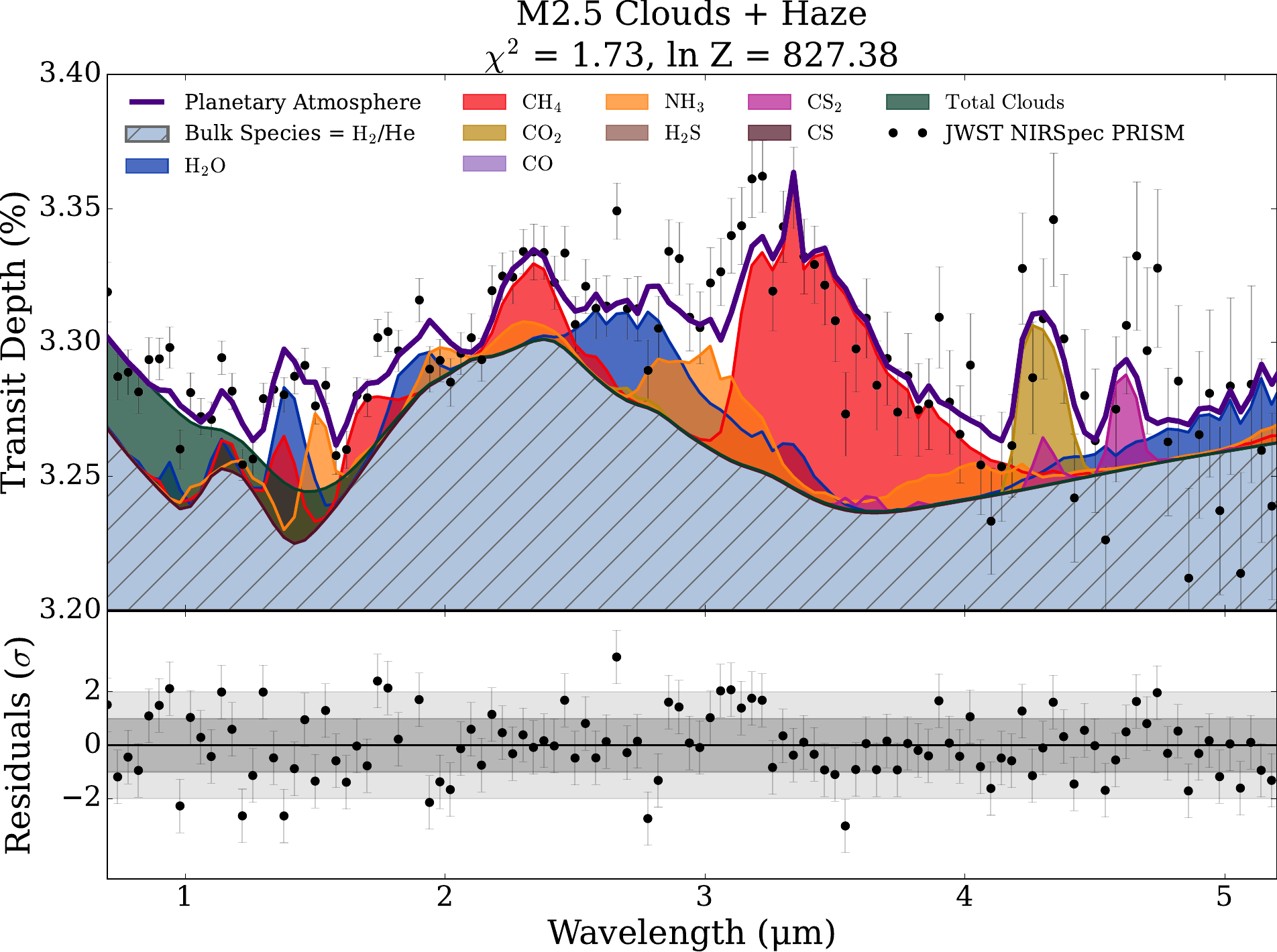}
\caption{Spectral decomposition plot for our final M2.5 Clouds + Haze model. Contributions for each species to the final planetary atmosphere model are highlighted, including the contribution of our cloud + haze treatment (``Total Clouds'')}
\label{fig:spectral_contribution}
\end{figure*}

\begin{figure*}[!htb]
\epsscale{1.2}
\plotone{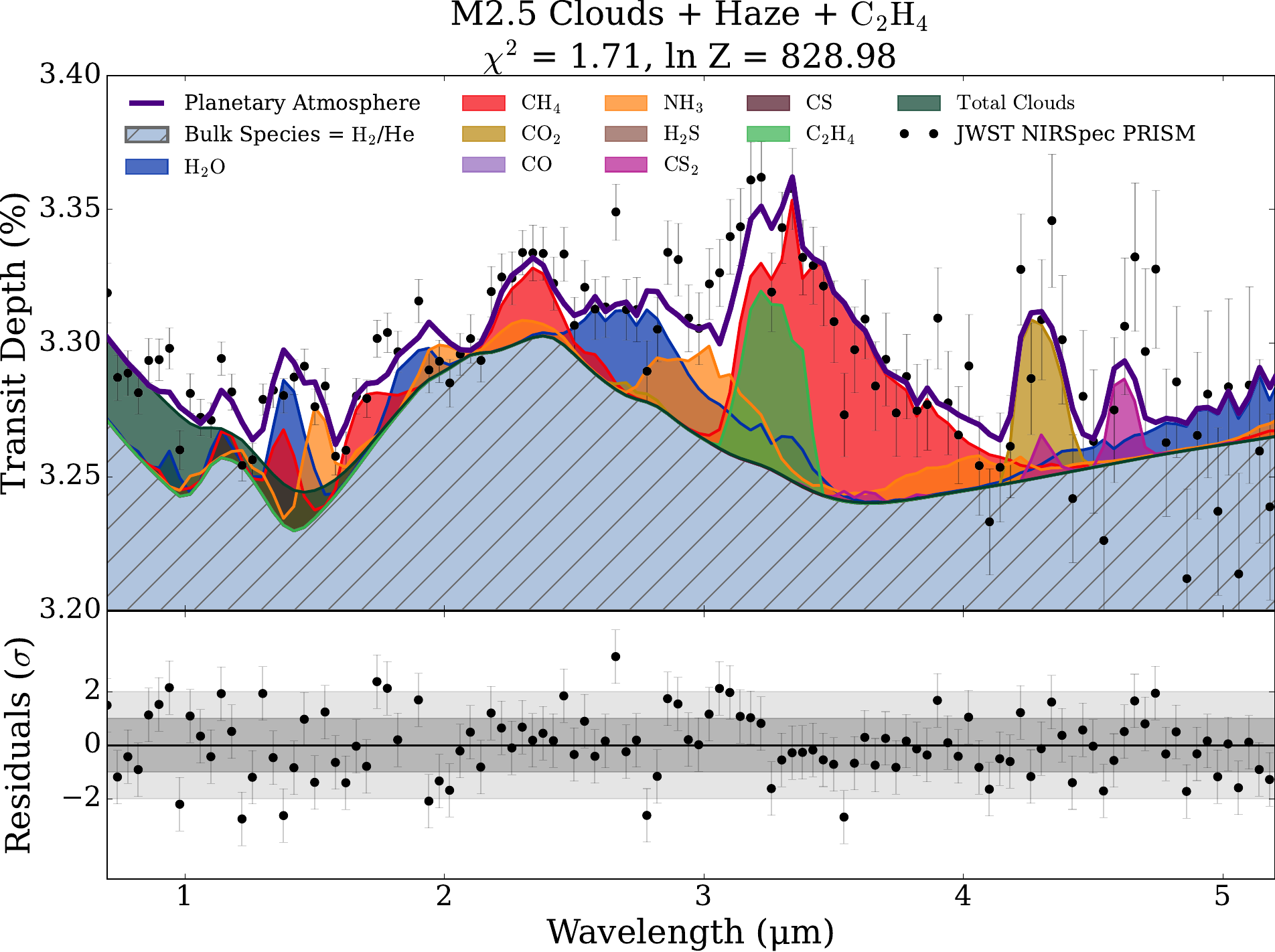}
\caption{Spectral decomposition plot for our final M2.5 Clouds + Haze model including C$_2$H$_4$. Contributions for each species to the final planetary atmosphere model are highlighted, including the contribution of our cloud + haze treatment (``Total Clouds'')}
\label{fig:spectral_contribution_c2h4}
\end{figure*}

\begin{figure*}[!htb]
\epsscale{1.2}
\plotone{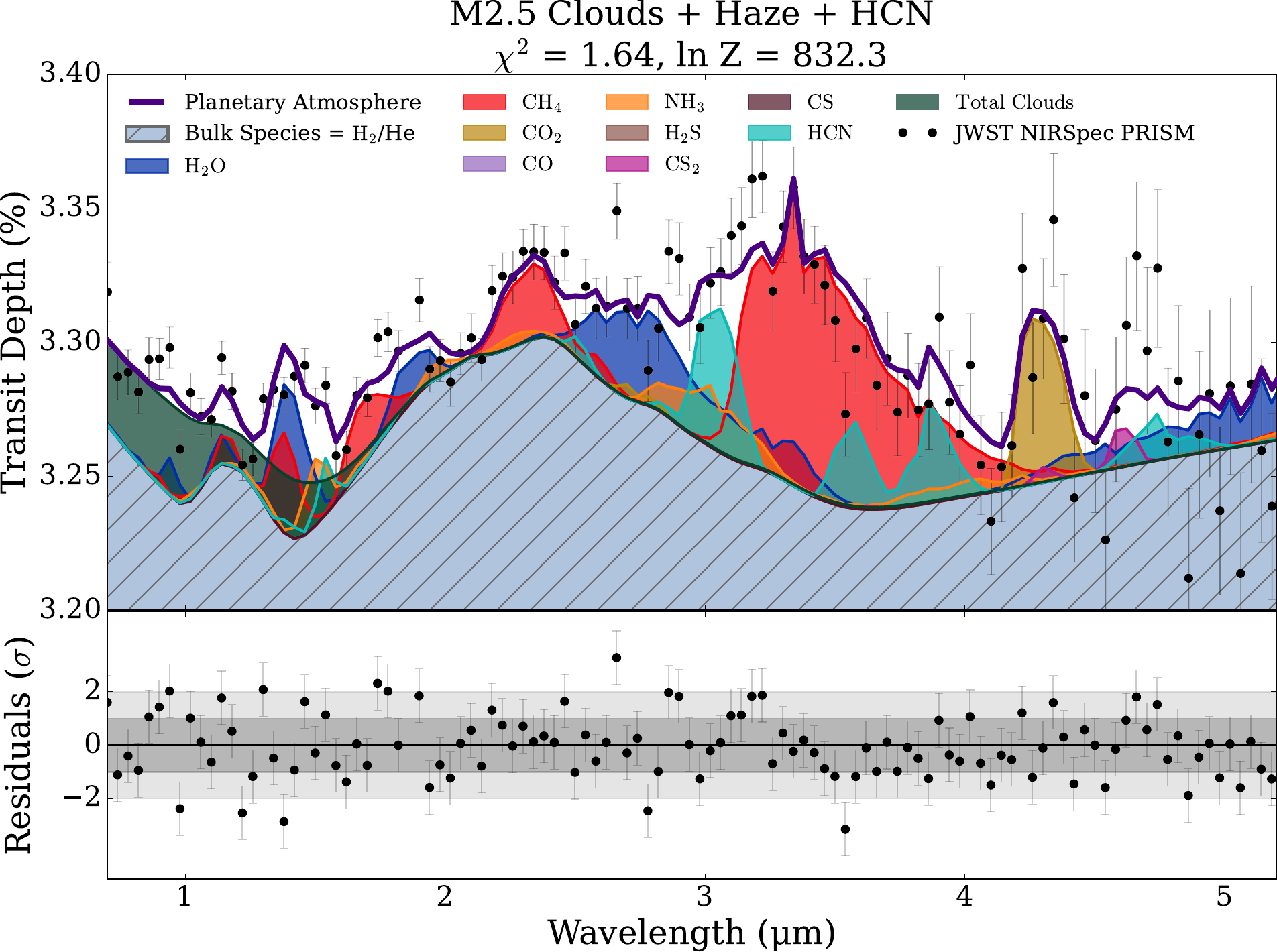}
\caption{Spectral decomposition plot for our final M2.5 Clouds + Haze model including HCN. Contributions for each species to the final planetary atmosphere model are highlighted, including the contribution of our cloud + haze treatment (``Total Clouds'')}
\label{fig:spectral_contribution_hcn}
\end{figure*}

For our best-fit model (\S\ref{subsec:retrieval_results}) as well as our HCN and C$_2$H$_4$ models (\S\ref{section:feature}), we run a spectral contribution function in which we forward model the planetary atmosphere and then remove opacities one by one in order to isolate each species. The results is a display of the spectral contribution of each species to the final best-fit planetary atmosphere model, seen in \autoref{fig:spectral_contribution} - \autoref{fig:spectral_contribution_hcn}.

\pagebreak

\end{document}